\documentclass[aps,prb,longbibliography,showpacs,floatfix,superscriptaddress,12pt]{revtex4-2}

\usepackage{geometry}
\usepackage[english]{babel}
\usepackage{blindtext}  
\usepackage{amsmath,amsfonts,amssymb}
\usepackage[dvipsnames]{xcolor}

\usepackage{graphicx}
\usepackage{float}
\usepackage{times}
\usepackage{color}
\usepackage{subfigure}
\usepackage{indentfirst}
\usepackage{setspace}
\usepackage{bm}

\usepackage[colorlinks,linkcolor=blue,anchorcolor=blue,citecolor=green]{hyperref}
\usepackage{CJKutf8}
\usepackage[ruled,vlined,linesnumbered]{algorithm2e}
\usepackage{algorithmic}

\graphicspath{{pdf}}

\newcounter{yc}

\allowdisplaybreaks

\begin{document}

\title{
	Destruction and recovery of the entanglement entropy of a many-body quantum
system after a single measurement}
\author{Bo Fan(\begin{CJK*}{UTF8}{gbsn}范波\end{CJK*})
}
\email{bo.fan@sjtu.edu.cn}
\affiliation{Shanghai Center for Complex Physics, School of Physics and Astronomy,
	\\Shanghai Jiao Tong University, Shanghai 200240, China}
	\author{\begin{CJK*}{UTF8}{gbsn}
			Can Yin (殷灿)
	\end{CJK*}}
\email{yin\_can@sjtu.edu.cn}
\affiliation{Shanghai Center for Complex Physics, School of Physics and Astronomy,
	\\Shanghai Jiao Tong University, Shanghai 200240, China}
\author{Antonio M. Garc\'ia-Garc\'ia}
\email{amgg@sjtu.edu.cn}
\affiliation{Shanghai Center for Complex Physics, School of Physics and Astronomy,
	\\Shanghai Jiao Tong University, Shanghai 200240, China}
\vspace{0.2cm}
\date{\today}
\vspace{0.3cm}

\begin{abstract}
	For one-dimensional non-interacting complex fermions, we compute numerically the probability distribution of the change in the entanglement entropy (EE) after saturation, resulting from a single measurement of the occupation number by using different measurements protocols. In the thermodynamic limit, the system is in the area-law phase for any monitoring strength, however we can also study the distribution in the critical phase characterized by a logarithmic scaling of the EE with system size by considering a sufficiently weak monitoring strength.
	For a quantum state diffusion protocol, where the change in EE is defined between two consecutive time steps, the distribution, Gaussian for weak monitoring, gradually develops symmetric exponential tails. For strong monitoring, the core turns from Gaussian to strongly peaked at zero suggesting the dominance of quantum Zeno effect. 
	 For the quantum jump and the projective measurement protocols, we observe clear deviations from Gaussianity characterized by broader and asymmetric tails, exponential for positive values of the change, and a peak at zero, likely a precursor of Zeno effect, that increases with the system size and the monitoring strength. 
	Intriguingly, the distribution is spatially inhomogeneous. For sites around the boundary separating the subsystems defining the EE, the distribution is close to Gaussian with a broad support while for the rest of sites has asymmetric exponential tails and a much narrower support. As the monitoring strength increases, the full distribution is controlled by the boundary sites.

\end{abstract}

\maketitle
\newpage

\section{Introduction}
The description of the impact of measurements on the quantum dynamics has been a recurrent research topic from the early days \cite{heisenberg1927} of quantum theory. Indeed, one of the accepted postulates of quantum mechanics \cite{von1955} is that, after a measurement, the wavefunction {\it collapses} to one eigenstate of the operator represented by the observable being measured with a probability given by the Born's rule, namely, the module squared of the overlap between the state before the measurement and the mentioned eigenstate. However, these so-called projective measurements are only a small subset of the experimental protocols \cite{schlosshauer_2005,wiseman2014} that are currently available to gain information on a quantum system.
Experimental \cite{minev2019} and theoretical developments \cite{dum1992,molmer93,dalibard1992,dum1992,daley2014} in quantum optics in the last three decades has broadened enormously the horizon of quantum measurements. Typical examples beyond projective measurement are continuous measurements based on photodetection, where the continuous monitoring is interrupted by quantum jumps. \cite{warren1986,zoller1987,gleyzes2007,minev2019,plenio1998}. This quantum jump protocol led to the development of the quantum trajectories formalism \cite{molmer93,dalibard1992,dum1992,daley2014} to model theoretically the experimental results which is numerically more efficient, and has a sharper physical interpretation, than the use of the full Lindbladian formalism \cite{lindblad1975,gorini1976} typical of open quantum systems.
Another popular continuous measurement protocol is the one based on homodyne detection \cite{wiseman1993,collett1987,wiseman2014,fuwa2015}, where at each time step the system is perturbed by a weak measurement characterized by a Gaussian noise \cite{cao2019a,alberton2021a,carisch2023,ladewig2022}.

The quantum measurement problem has gained new impetus in recent times not only because the availability of the mentioned experimental measurement set-ups with a high level of control but also because for quantum computation, and other quantum technologies, it is fundamental to extract information from quantum states with a minimal level of disturbance on the system so that quantum coherence is preserved. This is a challenging task as repeated measurements can trigger \cite{Li2018a,Skinner2019a,nahum2021a,ippoliti2021a,Li2019a,carisch2023,Szyniszewski2019a,Szyniszewski2020,Turkeshi2020a,turkeshi2021,Turkeshi2022a,legal2023,Soares2024} a measurement induced phase transitions (MIPT) in the entanglement entropy, and other quantum information observables non-linear in the density matrix, separating phases with different scaling with respect to the system size. Therefore, a high monitoring rate has the potential of destroying quantum features that may be necessary for practical applications. Although challenging because of the post-selection problem \cite{aaronson2005,guhne2009,delmonte2024}, the existence of MIPT has already \cite{Noel2022a,Koh2022} been confirmed experimentally.

The precise characterization of the MIPT in interacting many-body quantum systems are still being debated. However, for one dimensional free Dirac fermions, namely, non-relativistic complex fermions, with no disorder and subjected to either continuous \cite{cao2019a} or projective \cite{poboiko2023} measurements there is now growing consensus, despite some initial contradicting results \cite{buchhold2021a,alberton2021a}, that the system stays in the area-law phase for any monitoring strength. For projective measurements \cite{poboiko2023}, see also \cite{fava2025,Jian2023}, this conclusion was reached analytically by using the replica trick and mapping the problem onto a non-linear sigma model \cite{wegner1979the,efetov1980,efetov1983supersymmetry,EfetovBook}. However, a MIPT has been reported for Majorana fermions in one dimension \cite{fava2023}, for Dirac fermions with long-range hopping \cite{minato2022,mueller2022,fuji2020}, in higher dimensions \cite{poboiko2023a} or including interactions \cite{fazio2024,poboiko2024,fuji2020,Li2024,muller2025,guo2025,lumia2024}.
Conditions for the existence of MIPT \cite{pengfei2021a} has also been studied in the context of the Sachdev-Ye-Kitaev model \cite{kitaev2015,bohigas1971,french1971,maldacena2016,sachdev1993} and also \cite{milekhin2024,garcia2021,garcia2022c} its gravity dual which illustrates the broad appeal of these problems in rather different communities.

All the above findings are based on the study of averaged (over quantum trajectories) quantities. However, it is unclear whether the average is a good indicator of the observable because of the importance of rare events or other features leading to a non-Gaussian distribution with sufficiently broad tails. However, the study of the full distribution of probability of observables like the entanglement entropy \cite{schiro2024} or the local particle density and current \cite{turkeshi2024} has just started. Even in the simplest case of a non-interacting fermionic system in one dimension, studied in Refs.~\cite{schiro2024,turkeshi2024}, it is not yet known the precise form of these distributions for the different measurement protocols, its system size dependence or its role in the characterization of the transitions. 

Here, we address some of these questions through a detailed numerical study of the distribution function of the entanglement entropy for the three protocols mentioned earlier: projective measurements, quantum jumps and weak measurements. We shall focus on the distribution of the change of EE after a single measurement, though the global distribution of the EE will be studied as well, in the long time limit corresponding to the saturation of the EE. We start our analysis with the definition of the non-interacting fermionic model, the measurement protocol, and the method employed in the numerical calculation of the EE.

\section{Model and Method}\label{sec:model}

We investigate the dynamical effects of continuous and projective measurements in the dynamics of non-interacting complex spinless fermions in a one-dimensional chain with hopping to nearest neighbors described by the Hamiltonian,
\begin{equation}
	\hat{H} = J \sum_{i=1}^{L} \left[ \hat{c}_{i}^\dagger \hat{c}_{i+1} + \hat{c}_{i+1}^\dagger \hat{c}_{i} \right],
	\label{eq:Ham}
\end{equation}
where $ \hat{c}_i$ and $\hat{c}_i^\dagger$, $i=1,2,\cdots L$ are the annihilation and creation operators respectively for fermions at site $i$. $J$ is the hopping strength which for convenience is set to $J=1$. We impose periodic boundary conditions $\hat{c}_{i}=\hat{c}_{i+L}, \hat{c}_{i}^\dag=\hat{c}_{i+L}^\dag$ with $L$ the lattice size which we set to be even. The Hamiltonian commutes with the particle number operator $\hat{N}=\sum_{i=1}^{L}\hat{n}_i$ where $\hat{n}_i\equiv \hat{c}^\dag_i \hat{c}_i$ is the occupation number operator at site $i$ which has eigenvalues $0,1$. The eigenvalues of $\hat{N}$ are the total number of fermions $N = 1, 2, \ldots$. We consider the system at half-filling $N = L/2$. The initial state, termed N\'{e}el state, in an occupation number basis is $|\psi(t=0)\rangle=|101010\cdots\rangle$.

We consider the measurement of the occupation number at site $i$, $\hat{n}_i$, by three different protocols:

(1). Quantum state diffusion (QSD) protocol \cite{cao2019a,alberton2021a,carisch2023,ladewig2022}, also termed homodyne detection \cite{wiseman1993,collett1987,wiseman2014,fuwa2015} or weak measurements, is characterized by coupling the quantum system to a stochastic environment. The time evolution of the wave-function is governed by an Itô-type stochastic Schr\"{o}dinger equation with a weak Gaussian noise, parametrized by $\gamma$. Further details of this protocol are provided in Appendix ~\ref{app:QSD}.

(2). Quantum jump (QJ) protocol \cite{warren1986,zoller1987,gleyzes2007,minev2019,plenio1998}, a continuous monitoring protocol usually referred to as photo-detection in quantum optics where the measurement apparatus is always active and the detection, usually termed a quantum jump, occurs with a certain probability that depends on the monitoring strength $\gamma$. In our case, what is detected is an occupied site, $n_i = 1$. We model the time evolution of the state by the quantum trajectory method \cite{molmer93,dalibard1992,dum1992,daley2014} characterized by the jump operator $\hat{L}_i = \sqrt{\gamma}\hat{n}_i$. Further details are in Appendix ~\ref{app:QJ}.

(3). Projective measurement (PM) protocol refers to the standard description of measurements in quantum mechanics \cite{von1955} where the observable to be measured is represented by a Hermitian operator. As a consequence of the measurement, the wavefunction of the system {\it collapses} to one eigenstate of this operator with a probability given by the Born's rule, namely, the absolute value of the overlap between the eigenstate and the wavefunction before the measurement.
In our case, we measure the occupation number at site $i$ represented by the operator $\hat{n}_i$ which could have two outcomes $0,1$. The measuring rate per site, controlled by a Poisson distribution, is given by $\gamma$. Further details are found in Appendix ~\ref{app:PM}.

A crucial feature that simplifies enormously the numerical treatment of the model for all these protocols is that since $\hat{n}_i^2 = \hat{n}_i$, the operator controlling the  state evolution, even if subjected to measurements of the occupation number, will still be quadratic and preserve the particle number. As a consequence, generic states $|\psi(t)\rangle$ are Gaussian so they can be  written down as \cite{cao2019a},
\begin{equation}
	|\psi(t)\rangle = \prod_{k=1}^{N} \left[\sum_{j=1}^L U_{jk}(t) \hat{c}_j^\dagger\right] | \text{vac} \rangle.
	\label{eq:def_state}
\end{equation}
The information of the wavefunction $|\psi(t)\rangle$ is captured by the $L\times N$  matrix $U(t)$ that depends on the protocol. By an appropriate normalization, the determinant of $U$ is nothing but a Slater's determinant of single-particle wavefunctions, so $U^\dag U=\mathbf{1}_{N}$, with $\mathbf{1}_{N}$ the $N\times N$ identity matrix. The initial N\'{e}el state $|101010\cdots\rangle$ is represented in terms of the coefficient matrix $U$ as
\begin{equation}
	U_{ij}(t=0)=\delta_{2i-1,j}
	\label{eq:Neel}
\end{equation}
Another consequence of the Gaussianity of the evolved states is that the Wick's theorem allows to write a generic $2n-$point in terms of the two-point correlation function
\begin{equation}
	D_{ij}\equiv \langle \psi(t)|\hat{c}_i^\dag \hat{c}_j|\psi(t)\rangle 
    \label{eq:dij}
\end{equation}
which can be expressed in terms of $U(t)$ as follows.  
We label the $N$ particles as $1,2,\ldots,N$ and assign to each a site index $j_1, j_2, \ldots, j_N$. Here, the index $j$ indicates that we consider all possible combinations of $N$ occupied sites chosen from a total of $L$ sites. We denote the summation over all sets $\{j\}$ (with each $j_k\in\{1,2,\ldots,L\}$ for $k=1,\ldots,N$) by $\sum_{\{j\}}$. Consequently, the wavefunction in Eq.~\eqref{eq:def_state} is expressed as:
\begin{equation}
    |\psi(t)\rangle =\sum_{\{j\}}U_{j_1,1}(t)U_{j_2,2}(t)\cdots U_{j_N,N}(t)\hat{c}^\dag_{j_1}\hat{c}^\dag_{j_2}\cdots \hat{c}^\dag_{j_N} | \text{vac} \rangle=\sum_{\{j\}}\left[\prod_{i=1}^N U_{j_i,i}(t)\hat{c}^\dag_{j_i}\right] | \text{vac} \rangle
    \label{eq:def_state2}
\end{equation}
then we substitute Eq.~\eqref{eq:def_state2} into Eq.~\eqref{eq:dij},
\begin{equation}
	D_{mn}= \sum_{\{i\}, \{j\}}
	\biggl(\prod_{k=1}^N U_{i_k,k}^*\biggr)\,
	\biggl(\prod_{l=1}^N U_{j_l,l}\biggr)\,
	\langle \text{vac}|\hat{c}_{i_N}\cdots \hat{c}_{i_1}\hat{c}_m^\dag \hat{c}_n
	\hat{c}_{j_1}^\dagger \cdots \hat{c}_{j_N}^\dagger
	|\text{vac}\rangle=\sum_{k=1}^N U^*_{mk}U_{nk}
	\label{eq:CorM}
\end{equation}
where we contract the operators and use the orthogonality condition in the second equality.

Our primary focus is the entanglement entropy, $S = -\operatorname{Tr} (\hat{\rho}_A \ln \hat{\rho}_A)$, where $\hat{\rho}_A$ is the reduced density matrix for the subsystem $A = \{1,2,\dots,\ell\}$ with $\ell \in \{1,2,\dots,L\}$. The correlation matrix $D^A$ in sub-system $A$ is given by the $\ell \times \ell$ upper-left block of $D$, with elements defined as $D^A_{ij} = \operatorname{Tr} (\hat{\rho}_A\, \hat{c}_i^\dagger \hat{c}_j)$. Since $\hat{\rho}_A$ is Gaussian, Wick's theorem allows us to express it in terms of the eigenvalues of $D^A$ \cite{calabrese2005,fagotti2008,vidal2003,Mbeng2024},
\begin{equation}
	S(\ell, t) = -\sum_{i=1}^{\ell} \left( \lambda_i(t) \ln(\lambda_i(t)) + (1 - \lambda_i(t)) \ln(1 - \lambda_i(t)) \right)
	\label{eq:EE}
\end{equation}
where $\lambda_i \in [0,1]$, $i=1,2,\cdots, \ell$ are the eigenvalues of $D_{ij}^A$.

In order to study the entanglement dynamics, we evaluate $S(\ell,t)$ for a sufficiently large number of quantum trajectories so that the resulting distribution function of $S(\ell,t)$ is smooth which is necessary for a quantitatively analysis of its form, especially its tails.

\begin{figure}[!htbp]
	\begin{center}
		\subfigure[]{ \label{fig.QSD_L512_EE_vs_t}
			\includegraphics[width=5.4cm]{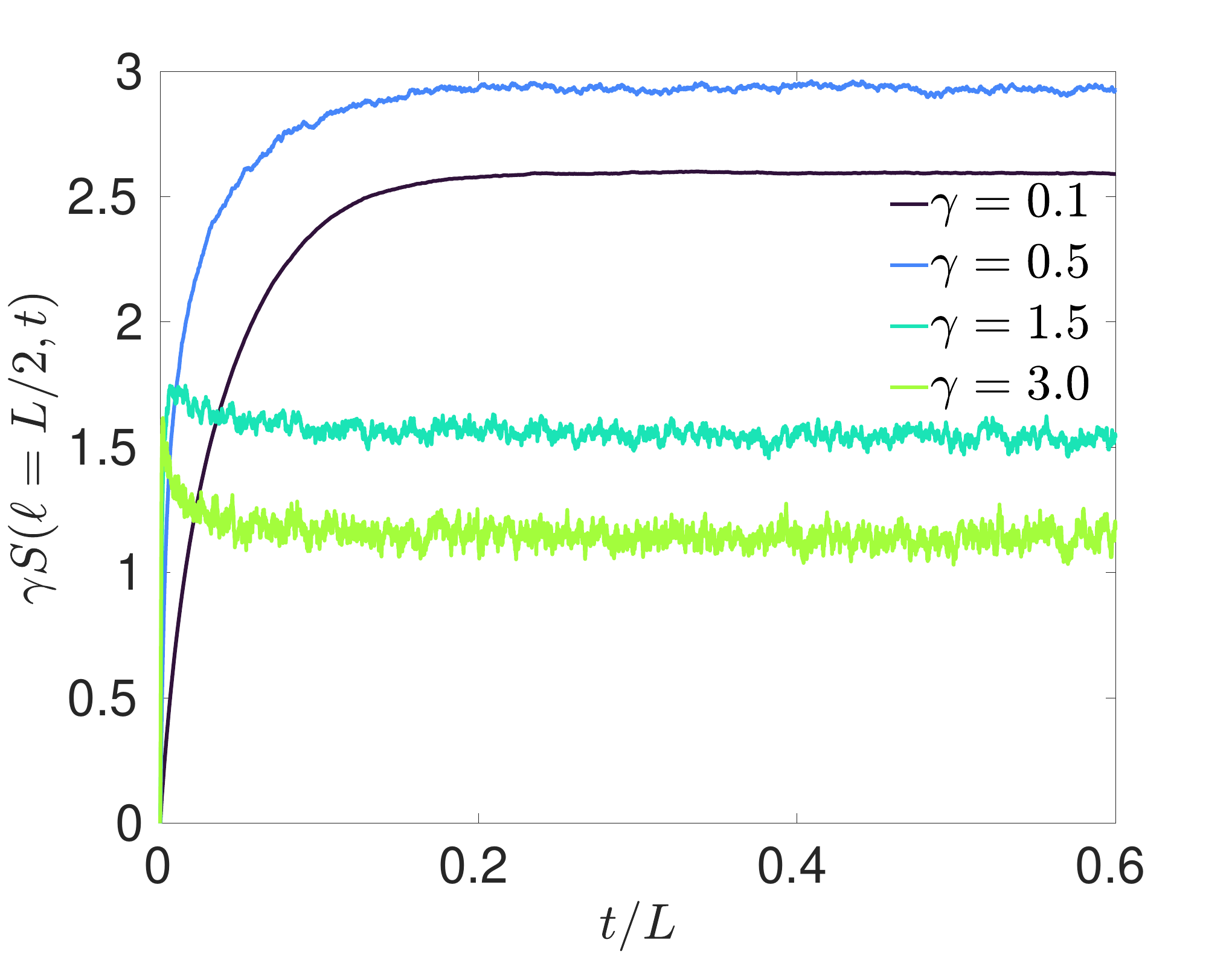} }\hspace{-0.2cm}
		\subfigure[]{ \label{fig.QJ_L512_EE_vs_t}
			\includegraphics[width=5.4cm]{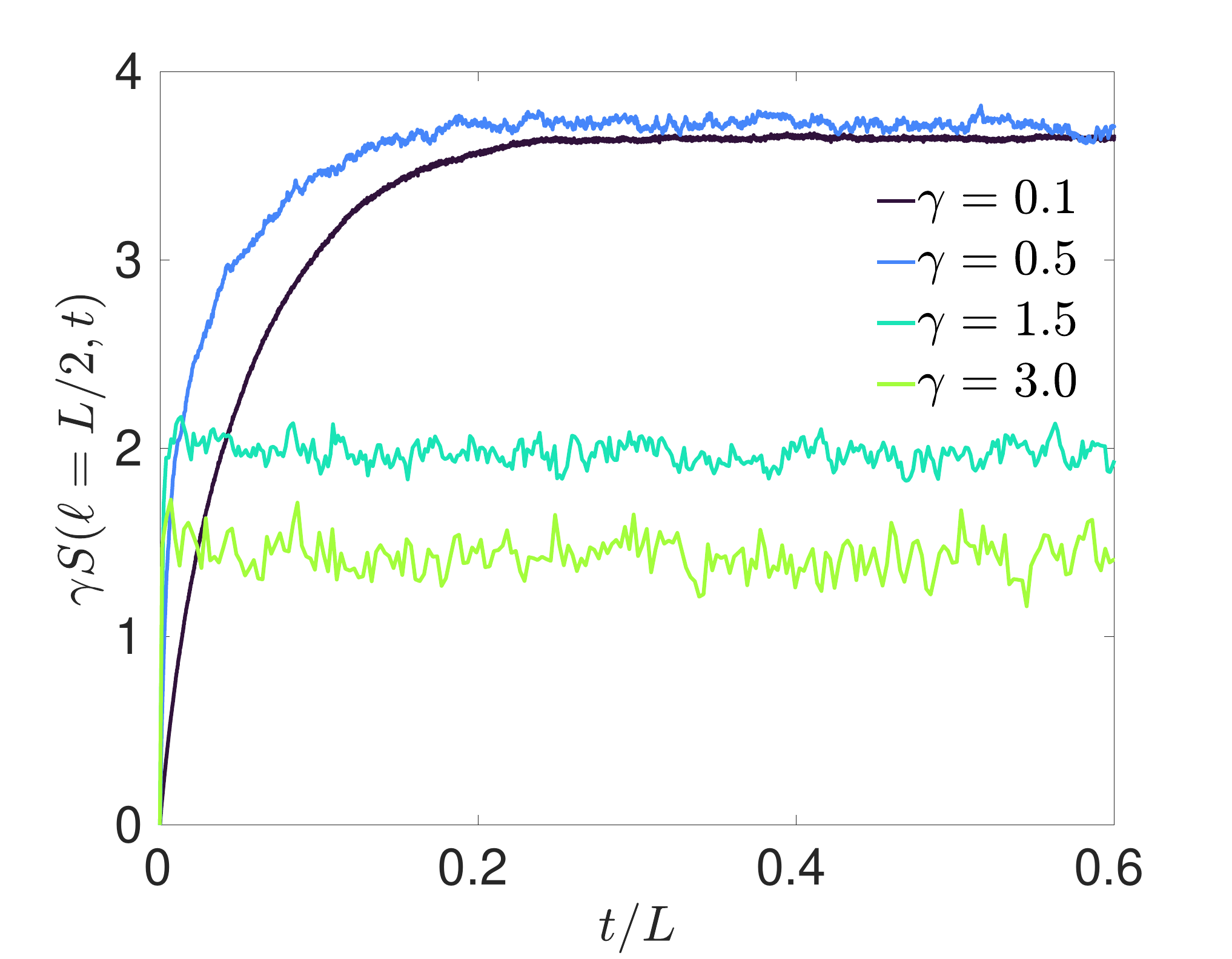} }\hspace{-0.2cm}
		\subfigure[]{ \label{fig.PM_L512_EE_vs_t}
			\includegraphics[width=5.4cm]{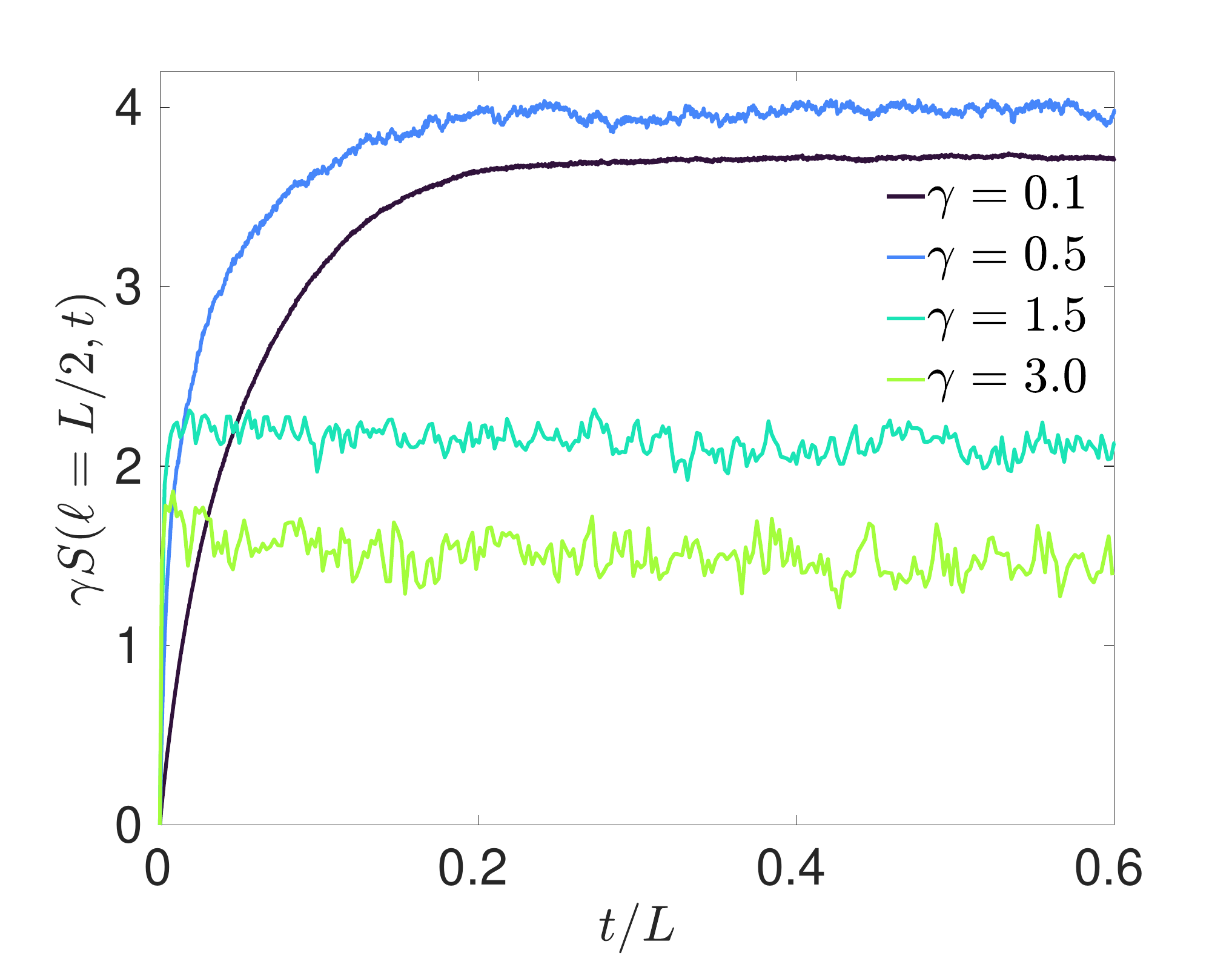} }
		\caption{
			The evolution of the entanglement entropy $S$ under different measurement protocols: \subref{fig.QSD_L512_EE_vs_t} QSD, \subref{fig.QJ_L512_EE_vs_t} QJ and \subref{fig.PM_L512_EE_vs_t} PM. The system size is $L=512$ and the subsystem size is $\ell = L/2$.
		}\label{Fig:Pro_EE_vs_t}
	\end{center}
\end{figure}

Across the paper, we fix the subsystem size to $\ell = L/2$ in order to systematically investigate changes in the EE under the three different measurement protocols, QSD, QJ and PM, mentioned above.
Here, we are interested in the probability distribution of the EE for sufficiently long times when it has reached its saturation value. Before we embark in this calculation, for reference, we depict in Fig.~\ref{Fig:Pro_EE_vs_t} the time dependence of the EE for the three protocols considered after an average over quantum trajectories.
As expected, the initial growth of the EE eventually stops for all protocols and monitoring strengths $\gamma$. The time at which the EE becomes flat depends on the monitoring strength and, if the strength is not too strong, on the system size.

\begin{figure}[!htbp]
	\begin{center}
		\subfigure[]{ \label{fig.QSD_L512_Pro_S_vs_g_fit}
			\includegraphics[width=5.4cm]{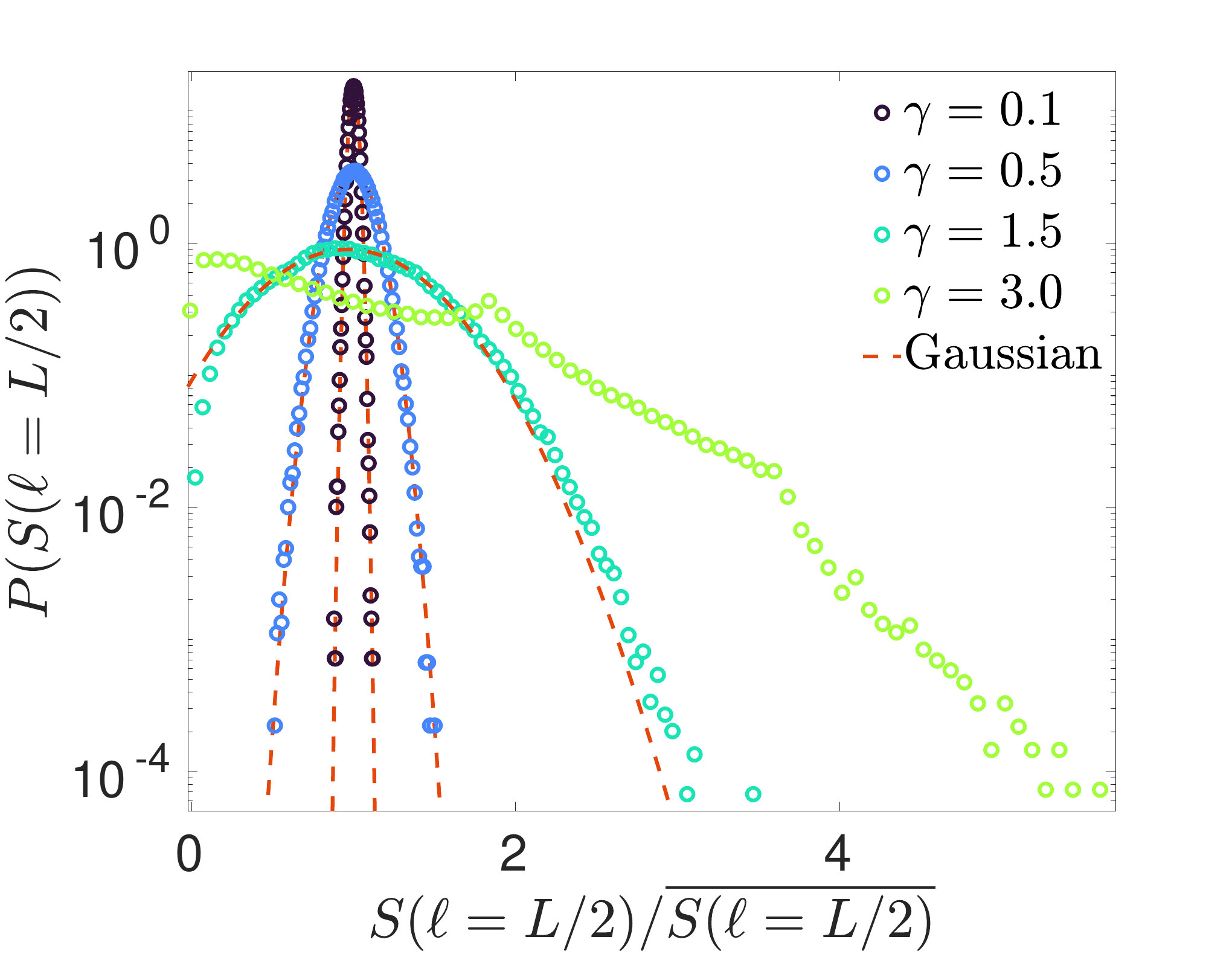} }\hspace{-0.2cm}
		\subfigure[]{ \label{fig.QJ_L512_Pro_EE_vs_g_fit}
			\includegraphics[width=5.4cm]{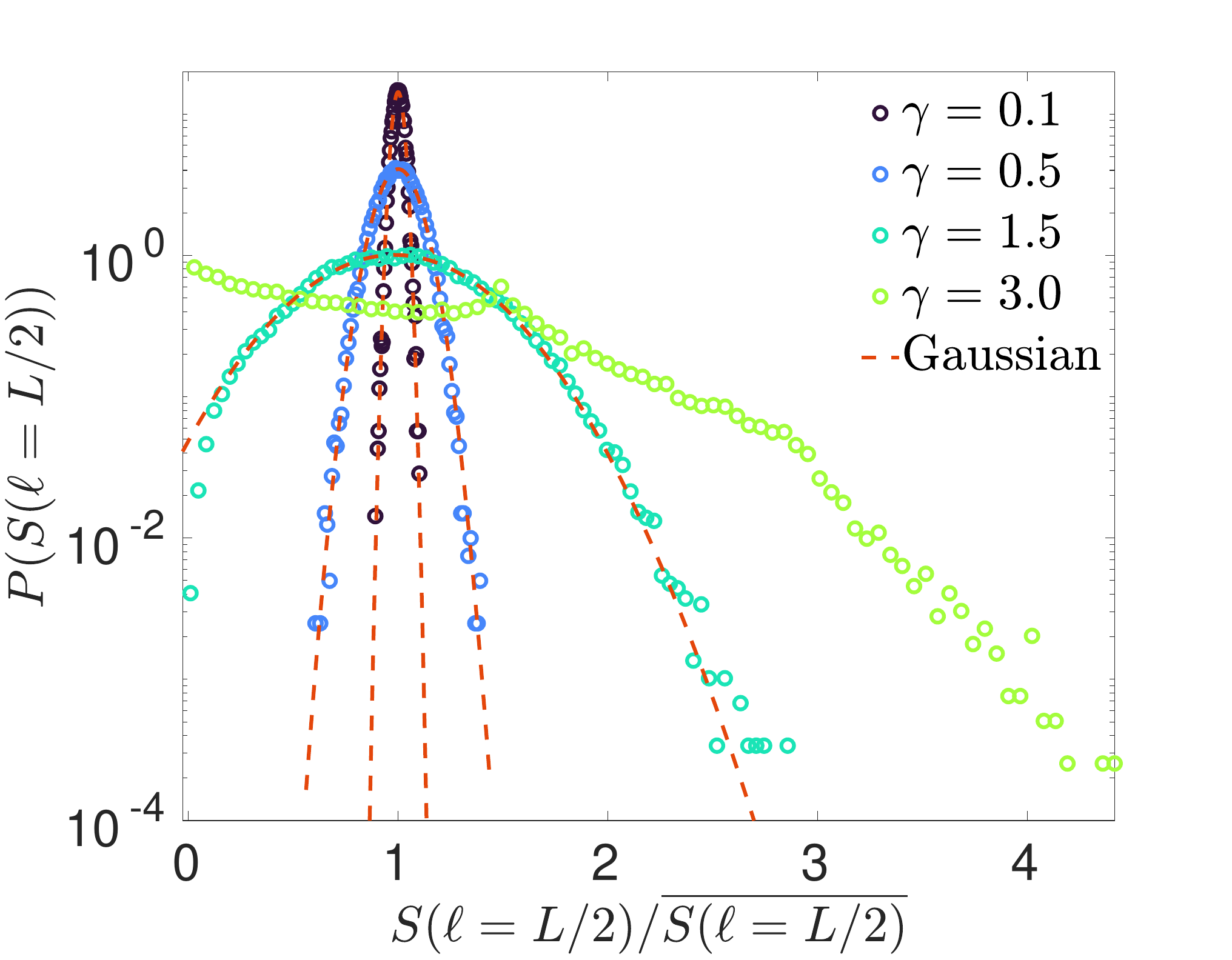} }\hspace{-0.2cm}
		\subfigure[]{ \label{fig.PM_L512_Pro_EE_vs_g_fit}
			\includegraphics[width=5.4cm]{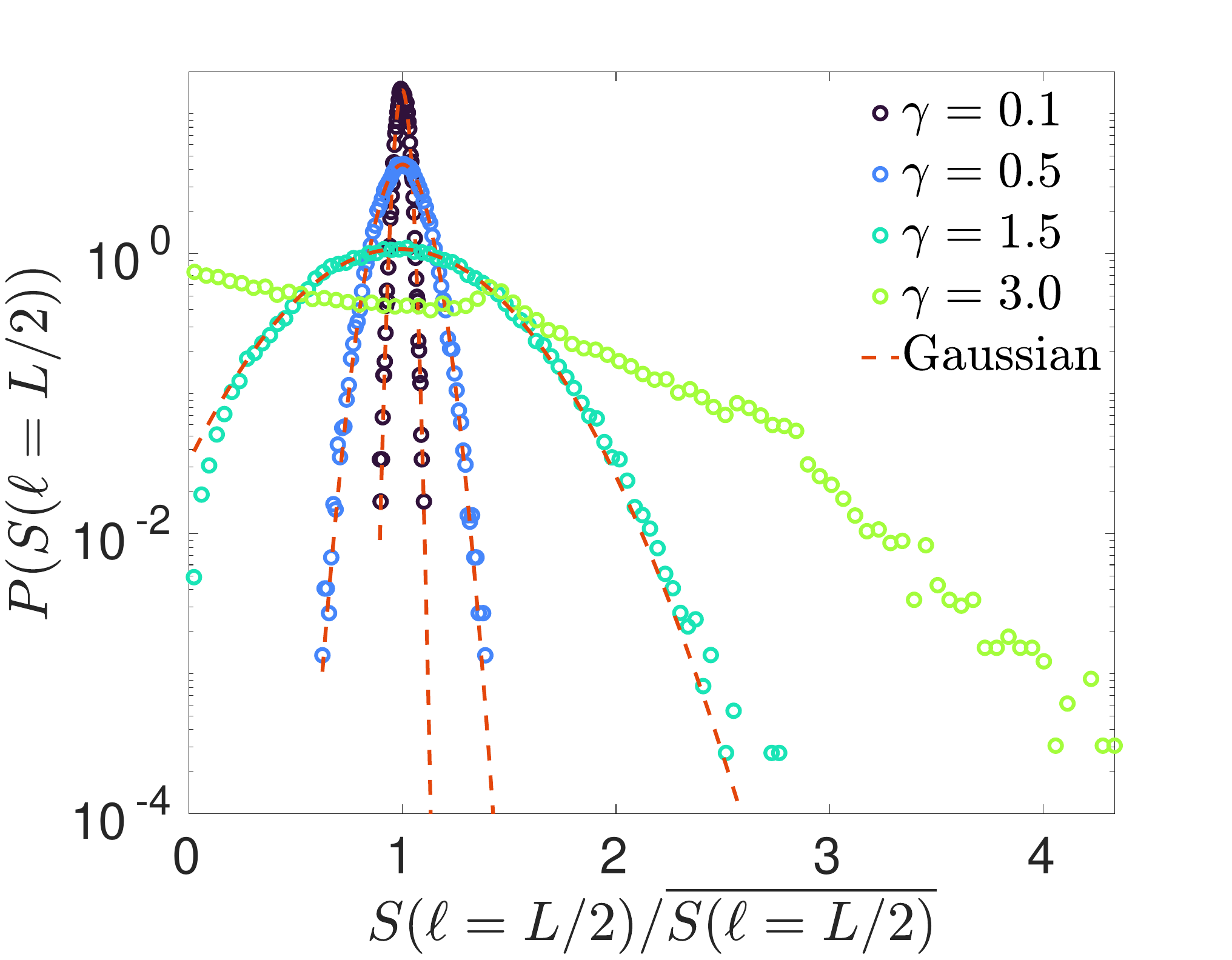} }
		\caption{
			The probability distribution of the global entanglement entropy under different measurement protocols: \subref{fig.QSD_L512_Pro_S_vs_g_fit} QSD, \subref{fig.QJ_L512_Pro_EE_vs_g_fit} QJ and \subref{fig.PM_L512_Pro_EE_vs_g_fit} PM. When the measurement strength is not very strong, the probability distributions agree well with a Gaussian distribution (red dashed line). The system size is $L=512$, and $S(\ell = L/2)$ is rescaled by its average $\overline{S(\ell = L/2)}$.
		}\label{Fig:Pro_EE_vs_g}
	\end{center}
\end{figure}

Strictly speaking, rather than with the distribution of the full EE at those long times, our main interest is the distribution of the EE {\it difference} between the state right before and right after a measurement -- for homodyne detection this corresponds to one time step. However, for the sake of completeness, we depict in Fig.~\ref{Fig:Pro_EE_vs_g} the probability distribution of the full EE in this long time region. As is observed, in the weak monitoring limit, the distribution is close to Gaussian for the three protocols. 
Deviations from Gaussianity, characterized by broad tails and a bimodal distribution are qualitatively similar in all protocols. Such features become evident only when the monitoring rate is sufficiently strong, which we believe corresponds to a region where the quantum Zeno effect \cite{misra1977,degasperis1974,itano1990,raizen2001,biella2021} is dominant.
We comment that similar results for the full EE distribution were obtained earlier for  continuously monitored Ising \cite{turkeshi2021} and free-fermion chains \cite{alberton2021a}. 	
These results suggest that the distribution function of the full EE seems to be capable of capturing a transition in the EE (if present). As we mentioned in the introduction, this model is in the area-law phase for $L \to \infty$, and any $\gamma > 0$, but we still believe that in the range of $L$ that we explore, and for small $\gamma$, our results describe the distribution of a system in a regime where the EE scales with the logarithmic of the system size. 
In any case, it is promising that within the range of sizes we can explore, the full distribution captures sharp differences between the weak and strong monitoring rates.

We now turn to the study of the distribution before and after a measurement which is the main interest of this paper.

\section{The quantum state diffusion measurement protocol}
We initiate our analysis with the study of the distribution function for the change of EE  for the case of continuous measurements characterized by the QSD protocol, see Appendix \ref{app:QSD} for details. We shall show that for this distribution, the distribution is Gaussian for weak monitoring. As monitoring strength increases, the tails becomes broader and the center develops a peak that suggests the growing importance of the Zeno effect.  
 This protocol, related to homodyne measurements, is characterized by weak measurements of the particle number on each site of the lattice at each time step $dt$. Therefore,  by change in entanglement after one measurement, we really mean the change in EE after one time step because, by definition, at a given time, measurements are carried out on all sites at once so the change after a single measurement is not possible. 
It is natural to study the {\it speed} of the entanglement change at each step $dt$ of the dynamical evolution defined as,
\begin{equation}\label{eq:deltaqsd}
	\delta S_{qsd}(t) = \frac{S_A(t+dt) - S_A(t)}{dt}
\end{equation}
where $t$ is sufficiently large so that the EE has already reached its saturation value and $S_A$ is computed using Eq.~(\ref{eq:EE}).
\subsection{Main Results}
\begin{figure}[!htbp]
	\begin{center}
		\subfigure[]{ \label{fig.QSD_g0p1_Pro_dS_vs_L_fit}
			\includegraphics[width=8.cm]{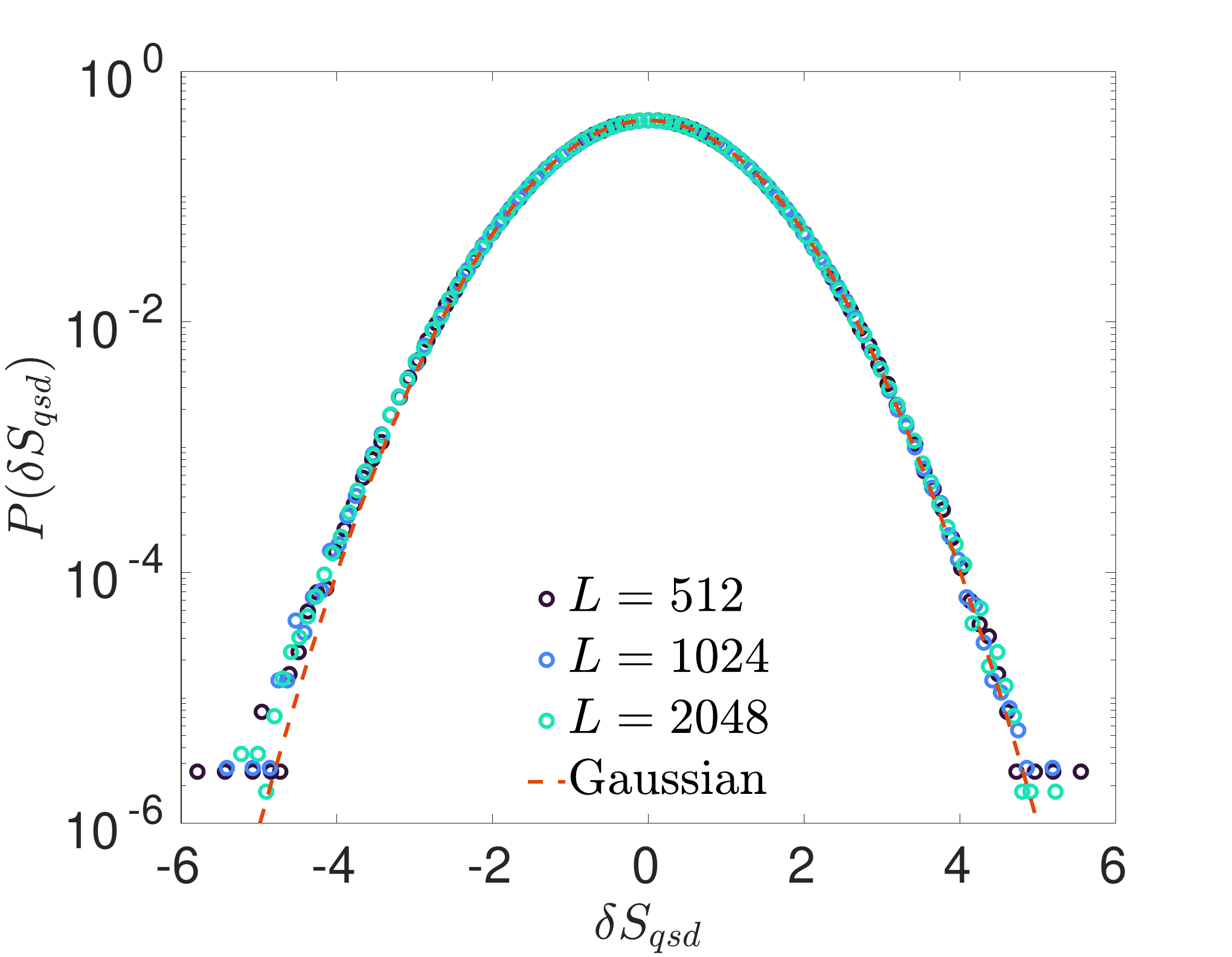} }
		\subfigure[]{ \label{fig.QSD_g0p5_Pro_dS_vs_L_fit}
			\includegraphics[width=8.cm]{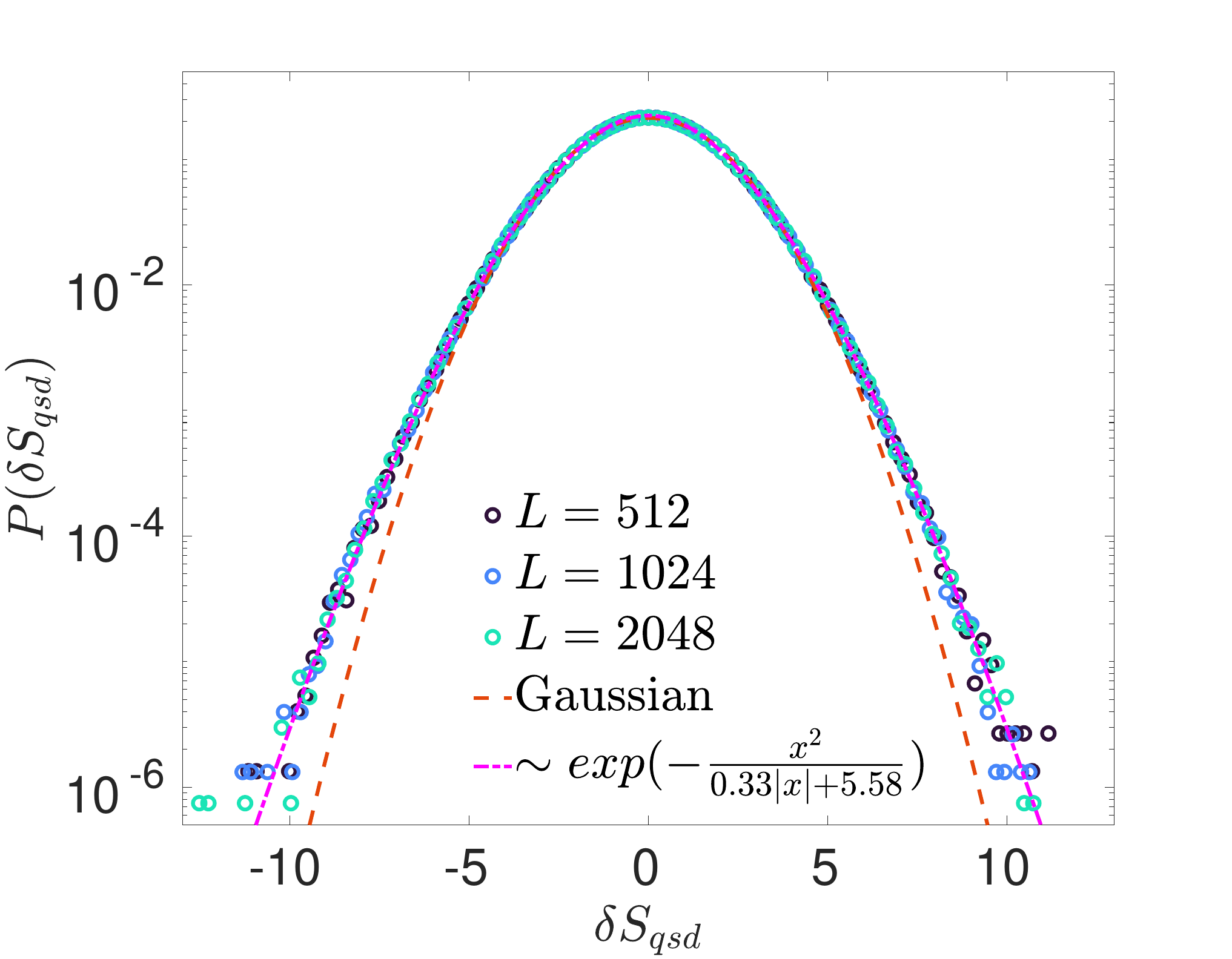} }
		\subfigure[]{ \label{fig.QSD_g1p5_Pro_dS_vs_L_fit}
			\includegraphics[width=8.cm]{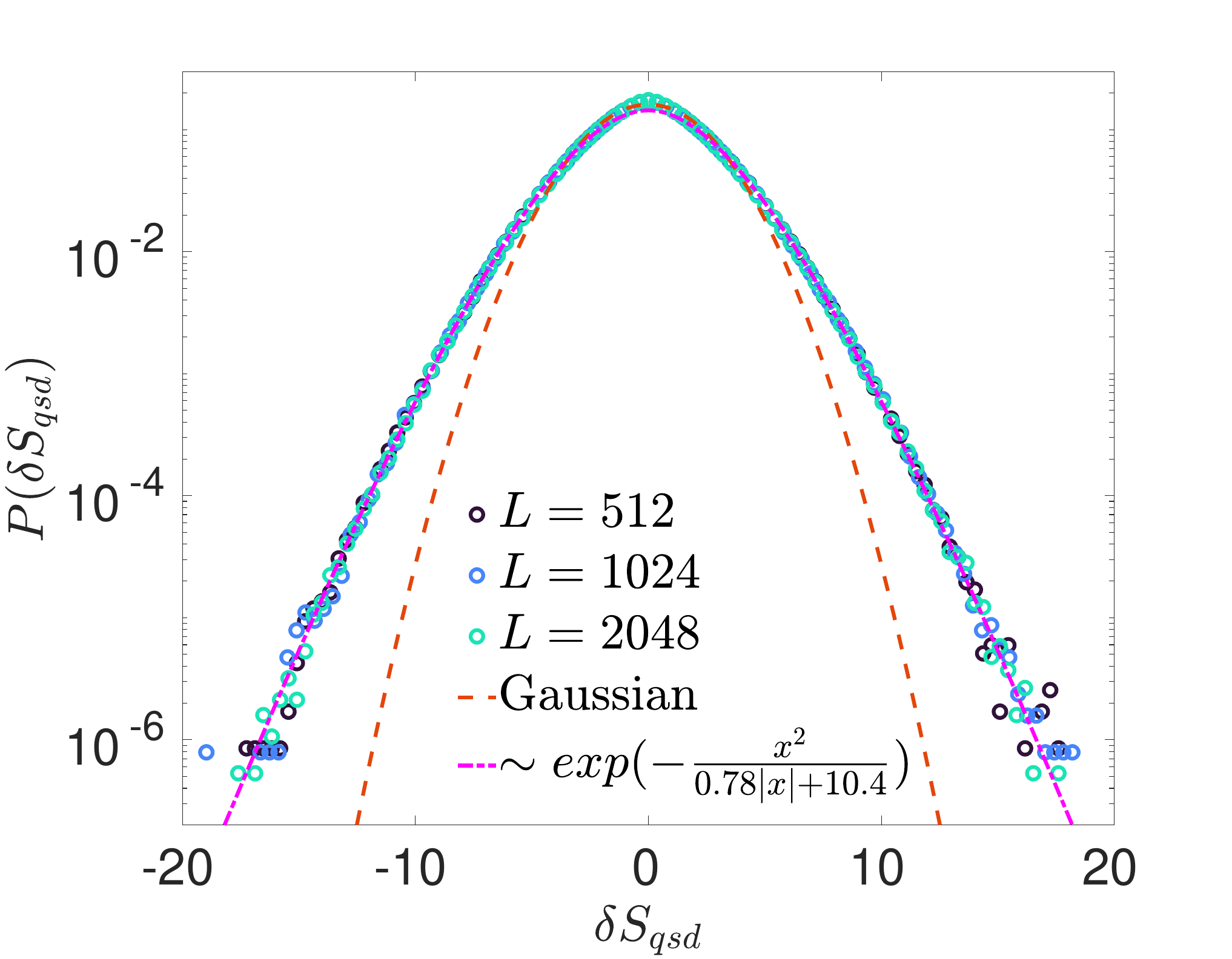} }
		\subfigure[]{ \label{fig.QSD_g3p0_Pro_dS_vs_L_fit}
			\includegraphics[width=8.cm]{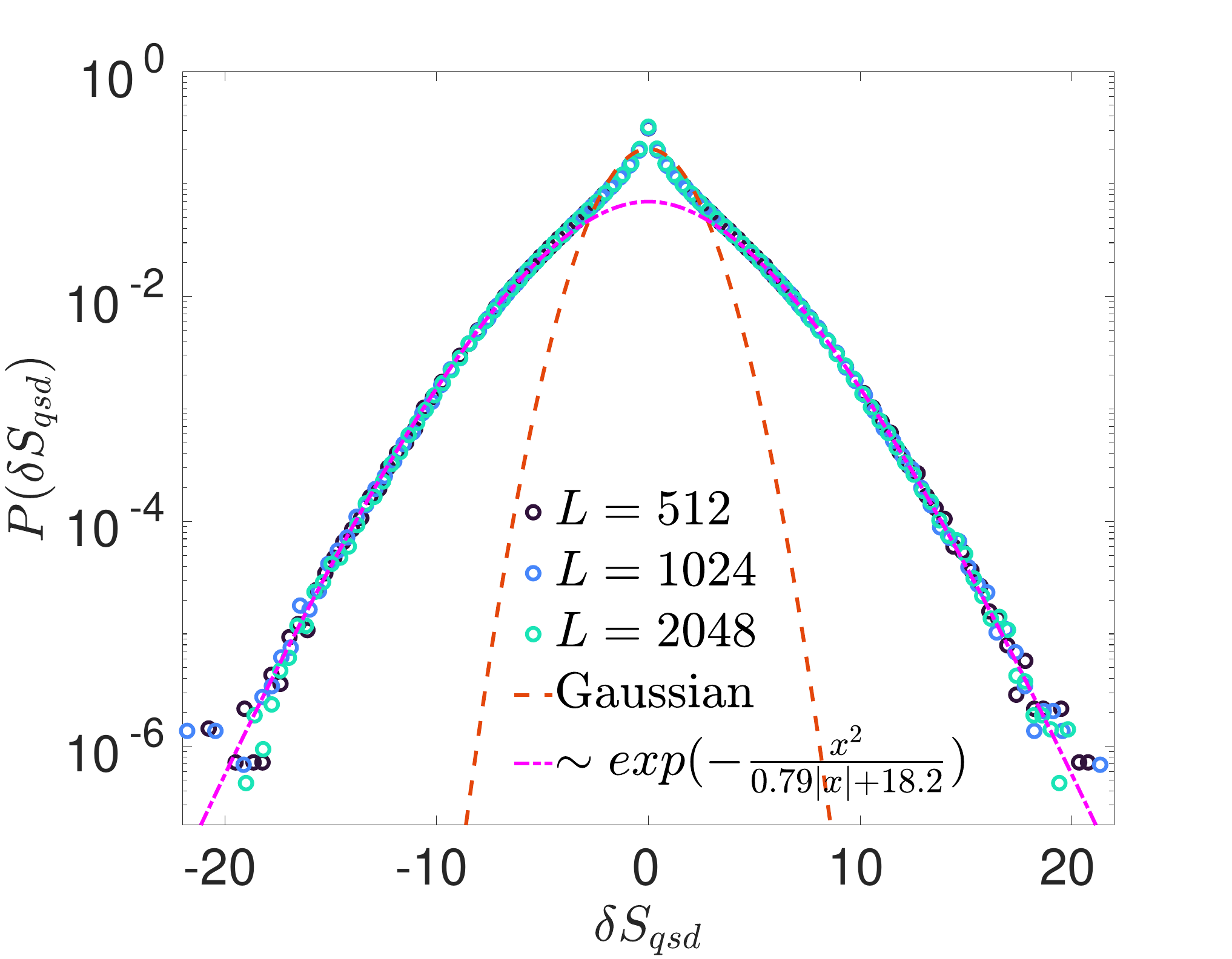} }
		\caption{
			The probability distribution of $\delta S_{qsd}$ for the QSD protocol in Eq.~(\ref{eq:deltaqsd}), employing Eq.~(\ref{eq:EE}) after the saturation time of the average EE. The monitoring rates are $\gamma = 0.1, 0.5, 1.5$ and $3.0$ from \subref{fig.QSD_g0p1_Pro_dS_vs_L_fit} to \subref{fig.QSD_g3p0_Pro_dS_vs_L_fit}. The distribution agrees well with a Gaussian distribution in the weak monitoring limit. For intermediate monitoring strength, the distribution is well described by a simple function interpolating from Gaussian, for small $\delta S_{qsd}$, to exponential, for large $\delta S_{qsd}$. In the strong monitoring limit, the exponential tails are still observed. However, for small $\delta S_{qsd}$, the distribution is strongly peaked around zero which indicates the dominance of quantum Zeno effect. We do not observe any size dependence in our results, see the text for an explanation of this feature.
		}\label{Fig:QSD_Pro_dS_vs_L}
	\end{center}
\end{figure}

The probability distribution of $\delta S_{qsd}$, depicted in Fig.~\ref{Fig:QSD_Pro_dS_vs_L}, show a Gaussian distribution in the weak monitoring limit, consistent with the central limit theorem for small, uncorrelated measurement perturbations. As the monitoring strength increases, deviations from Gaussianity are gradually observed. The full distribution is well described by $\sim \exp(-x^2/(a|x|+b))$ with $a, b$ fitting parameters, which suggests that stronger monitoring induces large fluctuations leading to broader exponential (not Gaussian) tails. The strong monitoring limit is characterized by a peak around zero, consistent (fitting not shown) with a soft power-law singularity at this point which indicates that the entanglement dynamics starts to be dominated by the quantum Zeno effect. The tails are still well described by the same expression that interpolates between a Gaussian and an exponential. Fittings with a stretched exponential (not shown) function also provide a good description of the numerical data so further research would be needed to fully describe the deviations from Gaussianity in this region.

Finally, we address an important feature of the distribution $P(\delta S_{qsd})$: its independence on the system size $L$. We observe this feature for all monitoring strengths $\gamma$ and in the full range of sizes we can explore numerically so it is likely a robust feature of the protocol not related to limitations in the maximum size we can reach numerically. 
Although we do not have a clear explanation of this scale invariance, we believe that the reason is that the change in the EE is governed by the measurement of sites around the boundary between the two subsystems which does not scale with $L$. This is a feature in the QJ and PM protocols, see Fig.~\ref{Fig:QJ_DSqj_vs_r}, which we think it is at work for the QSD protocol as well. Since all sites are measured simultaneously, it is difficult to differentiate the role of each site without changing the protocol.  

We note that the observed size independence and Gaussianity of $\delta S_{qsd}$ is not related to the size dependence of the entanglement entropy itself. Therefore, at least for small $\gamma$, we cannot make any statement about the existence of the area law for $L \to \infty$ by looking at the distribution of $\delta S_{\rm qsd}$. For that purpose, it would be necessary to map the model onto a non-linear sigma model, following the procedure of Ref.~\cite{poboiko2023}, for the PM protocol. Indeed, it has been recently shown \cite{fava2025,chahine2023,starchl2025} that other measurement protocols the EE still verifies an area law at any finite monitoring rate provided that the symmetry of the model does not change.  
\section{The quantum jump measurement protocol}
We now study the monitored dynamics employing the quantum jump protocol where the jump operator ${\hat L}_i = \sqrt{\gamma}\hat{n}_i$, representing the observable being measured, is proportional to the local occupation number $\hat{n}_i$, see Appendix \ref{app:QJ} for definitions and additional technical details. %
We shall see that the main differences with respect to the QSD protocol are that the distribution has a peak at zero even for weak monitoring, and that its tails for positive and negative changes in entropy are qualitatively different. Moreover, the change in entropy is quite sensitive to the site position.  
 In order to study the effects of quantum jumps to the entanglement entropy, we define
\begin{equation}\label{eq:Deltaqj}
	\Delta S_{qj} = S_A(t+\tau^+) - S_A(t+\tau^-),
\end{equation}
where $S_A$ is computed using Eq.~(\ref{eq:EE}) and $\tau^\pm$ the time right before ($-$) and right after ($+$) the quantum jump.

\subsection{Distribution of $\Delta S_{qj}$ after each measurement}

\begin{figure}[!htbp]
	\begin{center}
		\subfigure[]{ \label{fig.QJ_g0p1_Pro_DSqj_vs_L}
			\includegraphics[width=8.cm]{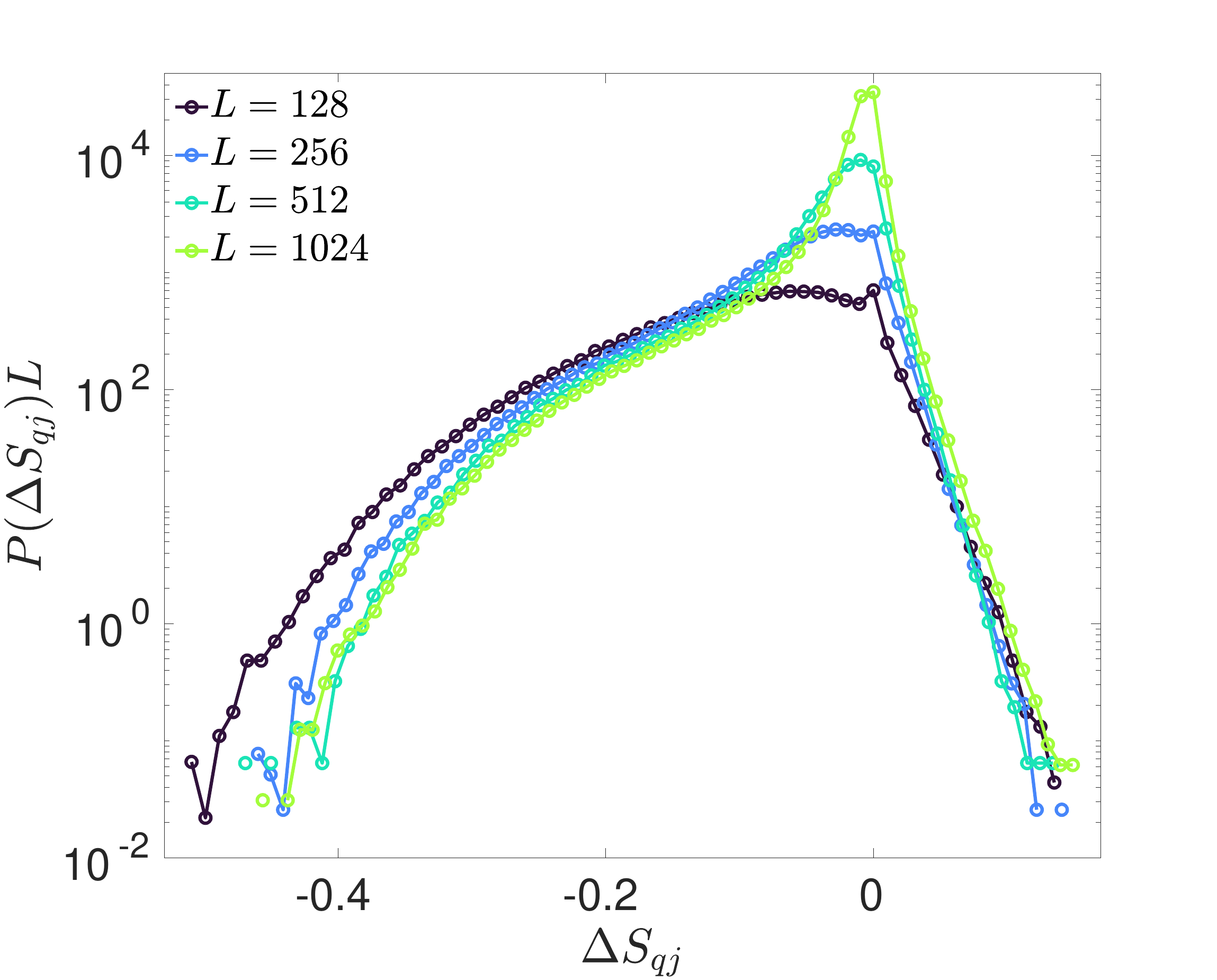} }
		\subfigure[]{ \label{fig.QJ_g0p5_Pro_DSqj_vs_L} 
			\includegraphics[width=8.cm]{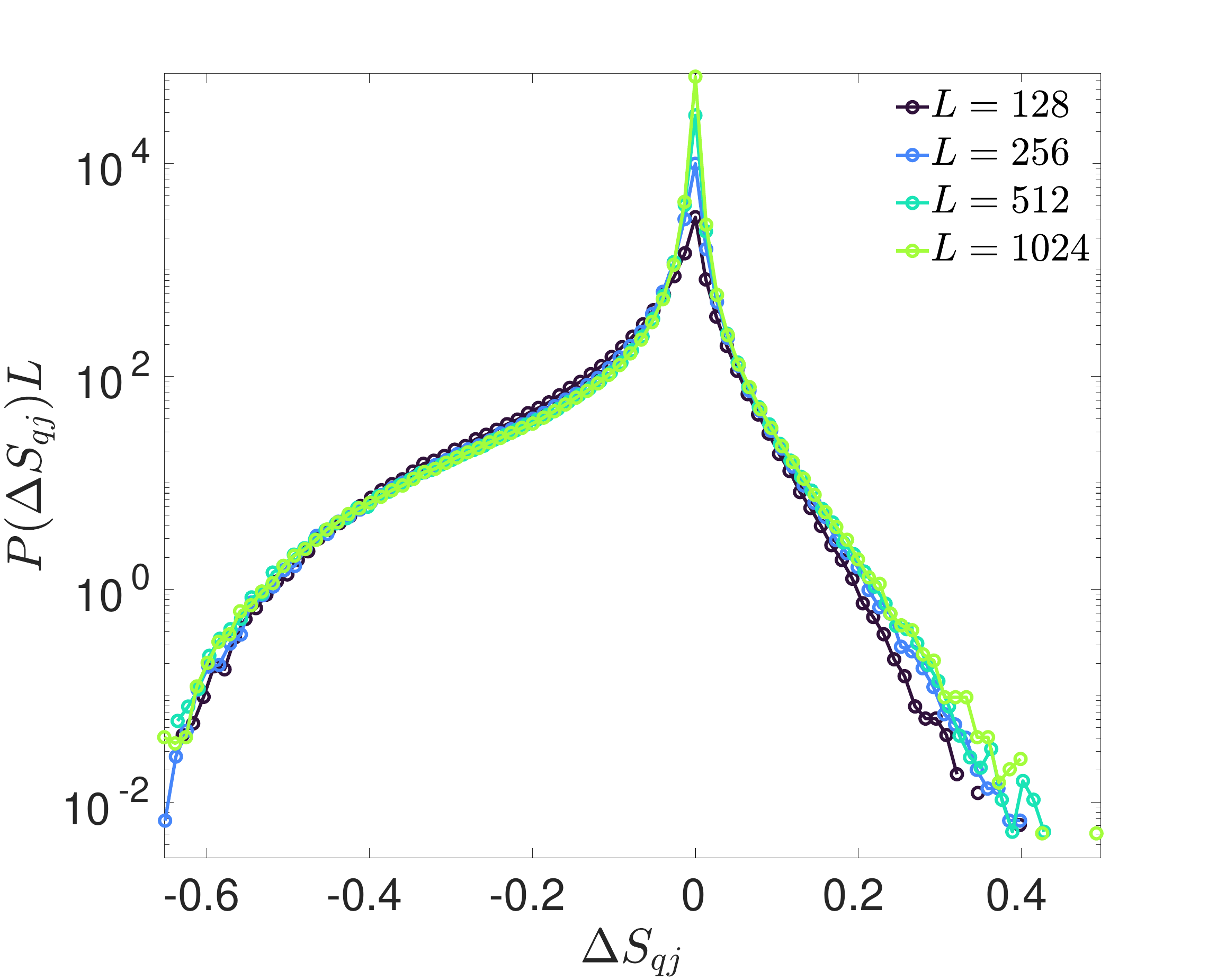} }
		\subfigure[]{ \label{fig.QJ_g1p5_Pro_DSqj_vs_L}
			\includegraphics[width=8.cm]{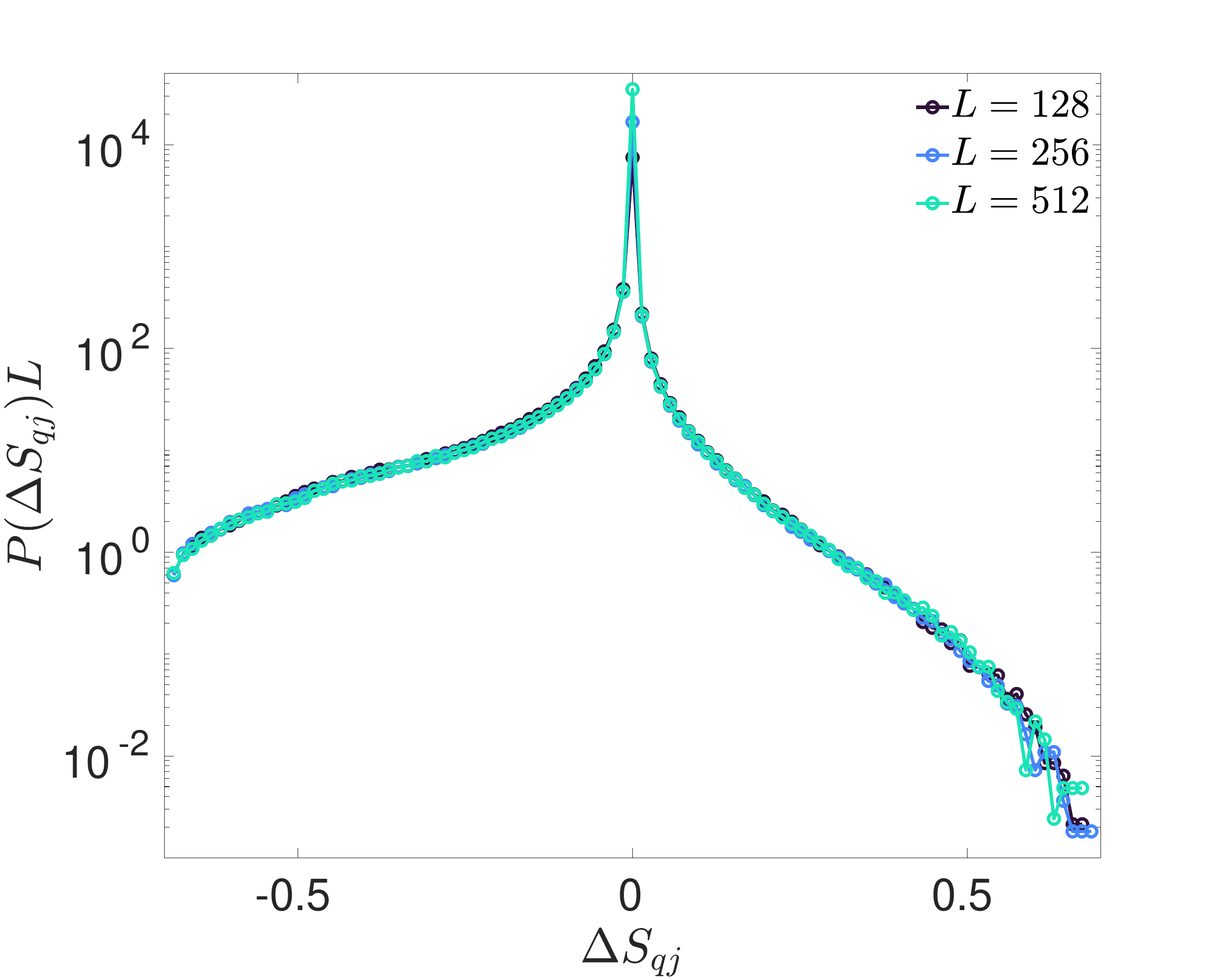} }
		\subfigure[]{ \label{fig.QJ_g3p0_Pro_DSqj_vs_L}
			\includegraphics[width=8.cm]{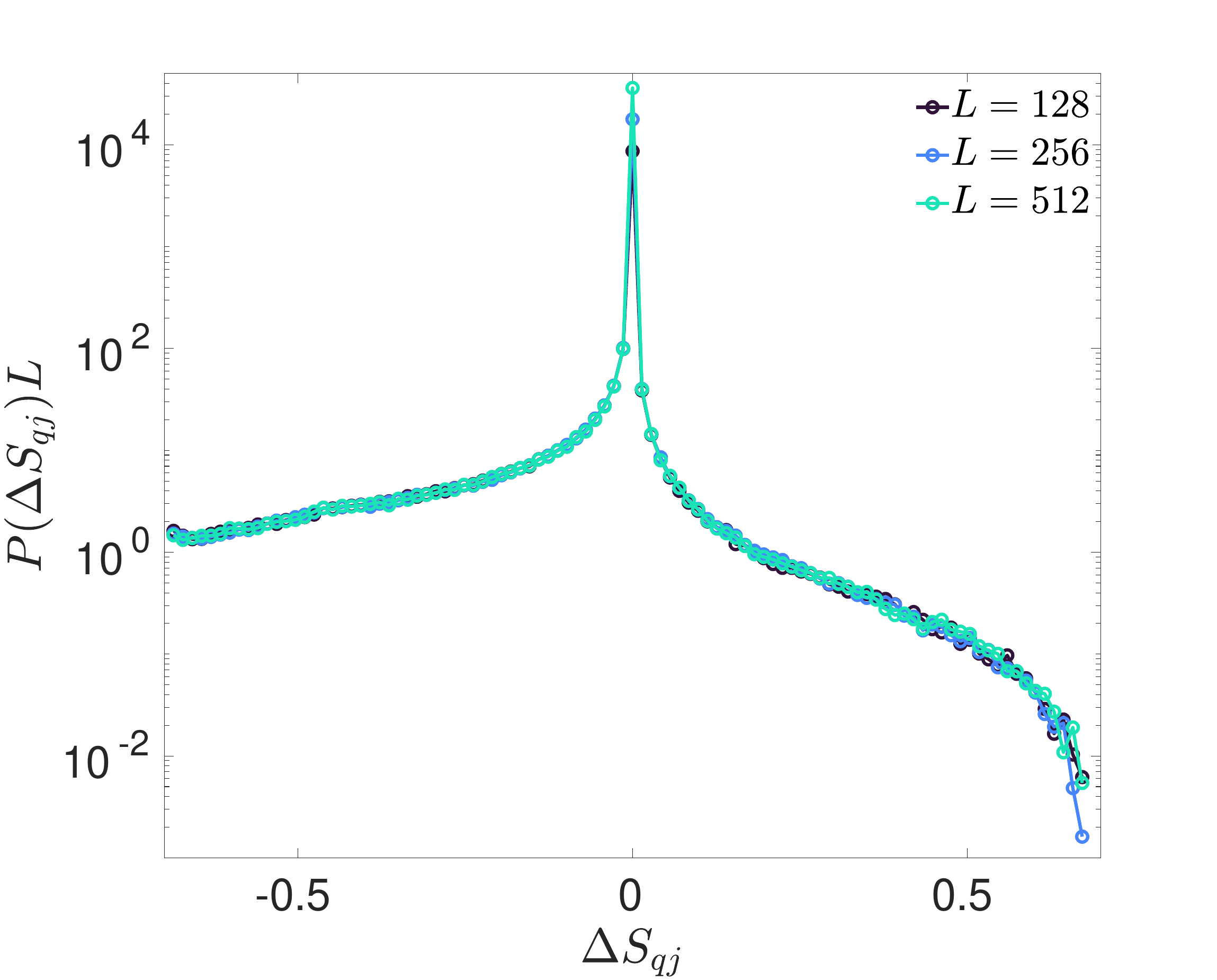} }
		\caption{
			The probability distribution of the change in entanglement entropy $\Delta S_{qj}$ in Eq.~(\ref{eq:Deltaqj}), computed employing Eq.~(\ref{eq:EE}), due to a quantum jump in the QJ protocol, for sufficiently late times so that EE has reached the saturation value. The $P(\Delta S_{qj})$ is rescaled by multiplying the system size $L$. The tail is independent of system size after rescaling, signaling the tail is dominated by the boundary region. The monitoring rates are $\gamma = 0.1, 0.5, 1.5$ and $3.0$ from \subref{fig.QJ_g0p1_Pro_DSqj_vs_L} to \subref{fig.QJ_g3p0_Pro_DSqj_vs_L}.
		}\label{Fig:QJ_Pro_DSqj_vs_L_g}
	\end{center}
\end{figure}

Results for the probability distribution of $\Delta S_{qj}$, presented in Fig.~\ref{Fig:QJ_Pro_DSqj_vs_L_g}, are quite different from those of the QSD protocol, because for the latter we are performing measurements on all sites at each time step.
Even for weak monitoring rates, the distribution has a strong peak at zero that becomes sharper as the monitoring strength, or the system size $L$ increases. This means that many measurements have no effect on the EE and that those 
measurements become more important as the thermodynamic limit $L \to \infty$ is approached. These results are consistent with the analytical prediction \cite{poboiko2023} for the PM protocol that for $L \to \infty$, the EE is in the area law phase for any finite measuring rate.

The rest of features of the distribution are also different from those of the QSD protocol.
The tails of $P(\Delta S_{qj})$ collapse after rescaling them by the system size $L$, but they are still asymmetric for all monitoring strength. For $\Delta S_{qj} > 0$, the tail is exponential for any $\gamma$ if we neglect the effect of the peak at zero and, for strong monitoring, the region $\Delta S_{qj} \geq 0.5$ which decays faster but has a comparatively small weight. The existence of states so that $\Delta S_{qj} > 0$ indicates that, counterintuitively, in some cases measurements characterized by quantum jumps can increase the EE, this feature was earlier observed in Ref.~\cite{schiro2024}.
By contrast, the left tail is more sensitive to the monitoring strength. We could not find a simple analytic expression for general $\gamma$, yet for weak monitoring, putting aside the effect of the peak at zero, the distribution is well described by a Gaussian though the agreement seems to become worse as the system size increases.

\subsection{Spatial dependence of $\Delta S_{qj}$}
\begin{figure}[!htbp]
	\begin{center}
		\subfigure[]{ \label{fig.QJ_Delta_Sqj_vs_site_gm0p05_L512}
			\includegraphics[width=8.cm]{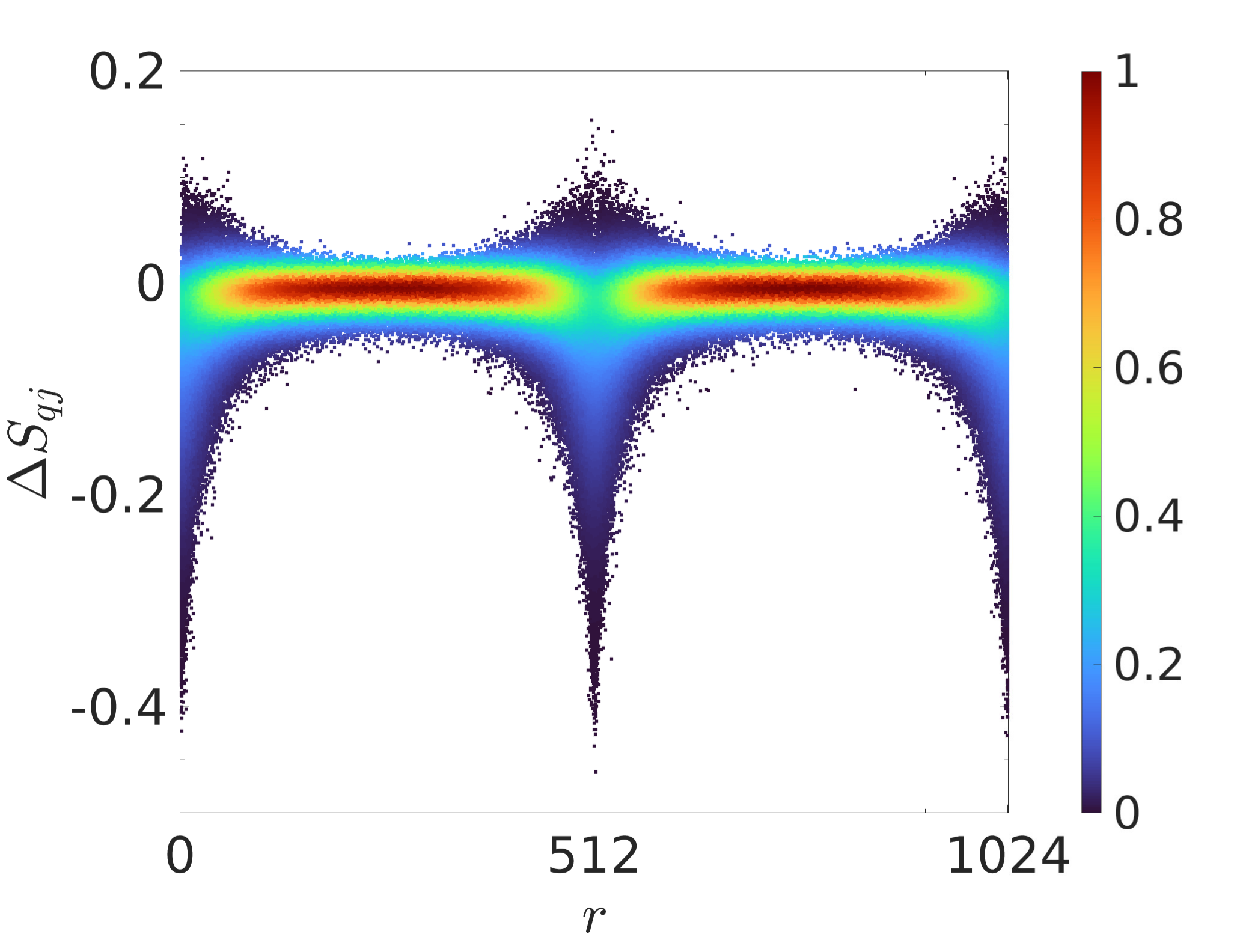} }
		\subfigure[]{ \label{fig.QJ_delta_SnH_vs_site_gm0p05_L512}
			\includegraphics[width=8.cm]{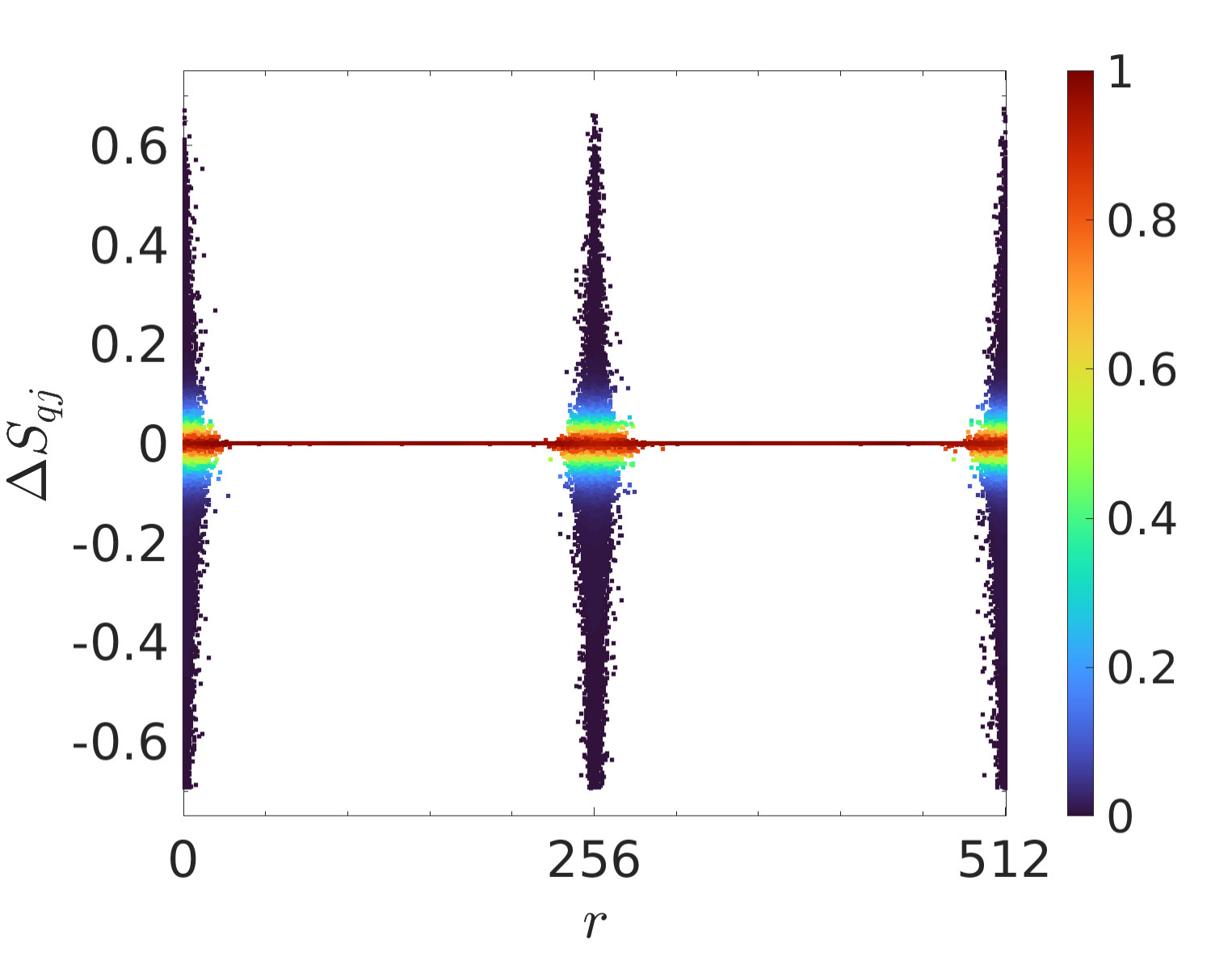} }
		\caption{$\Delta S_{qj}$ in Eq.~(\ref{eq:Deltaqj}) as a function of the measured sites $r = 1, \ldots, L$. The color scale indicates the normalized density of measurement events. The left plot is for the weak monitoring rate $\gamma=0.1$ and system size $L=1024$, while the right plot is for the strong monitoring rate $\gamma=3$ and the system size is $L=512$. Note the stark differences between boundary points, separating the two subsystems defining the EE, and the rest.
		}\label{Fig:QJ_DSqj_vs_r}
	\end{center}
\end{figure}

In order to gain a more quantitative understanding of the distribution, especially this left tail, we investigate the spatial dependence of $\Delta S_{qj}$ defined in Eq.~(\ref{eq:Deltaqj}), see
Fig.~\ref{Fig:QJ_DSqj_vs_r}. Interestingly, we find very different behavior for sites corresponding to the boundary between the two halves in which we split the system to compute the EE. For weak monitoring strength, the distribution of points at, and around, the subsystem boundary is broader, asymmetric, and with a much softer peak at zero. By contrast, for the rest of sites not very close to the boundary, the distribution has a peak at zero with a much narrower distribution. In the strong monitoring limit, and for sites far from the boundary, we observe a complete quantum Zeno effect $\Delta S_{qj} = 0$ while in the region around the boundary, the distribution, though peaked at zero, has a well defined symmetric distribution.
These observations indicate qualitative differences between the region around boundary and the rest of sites. The precise length for this boundary region may depend on the monitoring strength. Far from the boundary, the effect of measurements becomes weak in the sense that the change of EE is small even for relatively weak monitoring strengths.

\begin{figure}[!htbp]
	\begin{center}
		\subfigure[]{ \label{fig.QJ_g0p1_Pro_DSqj_vs_ell}
			\includegraphics[width=8.cm]{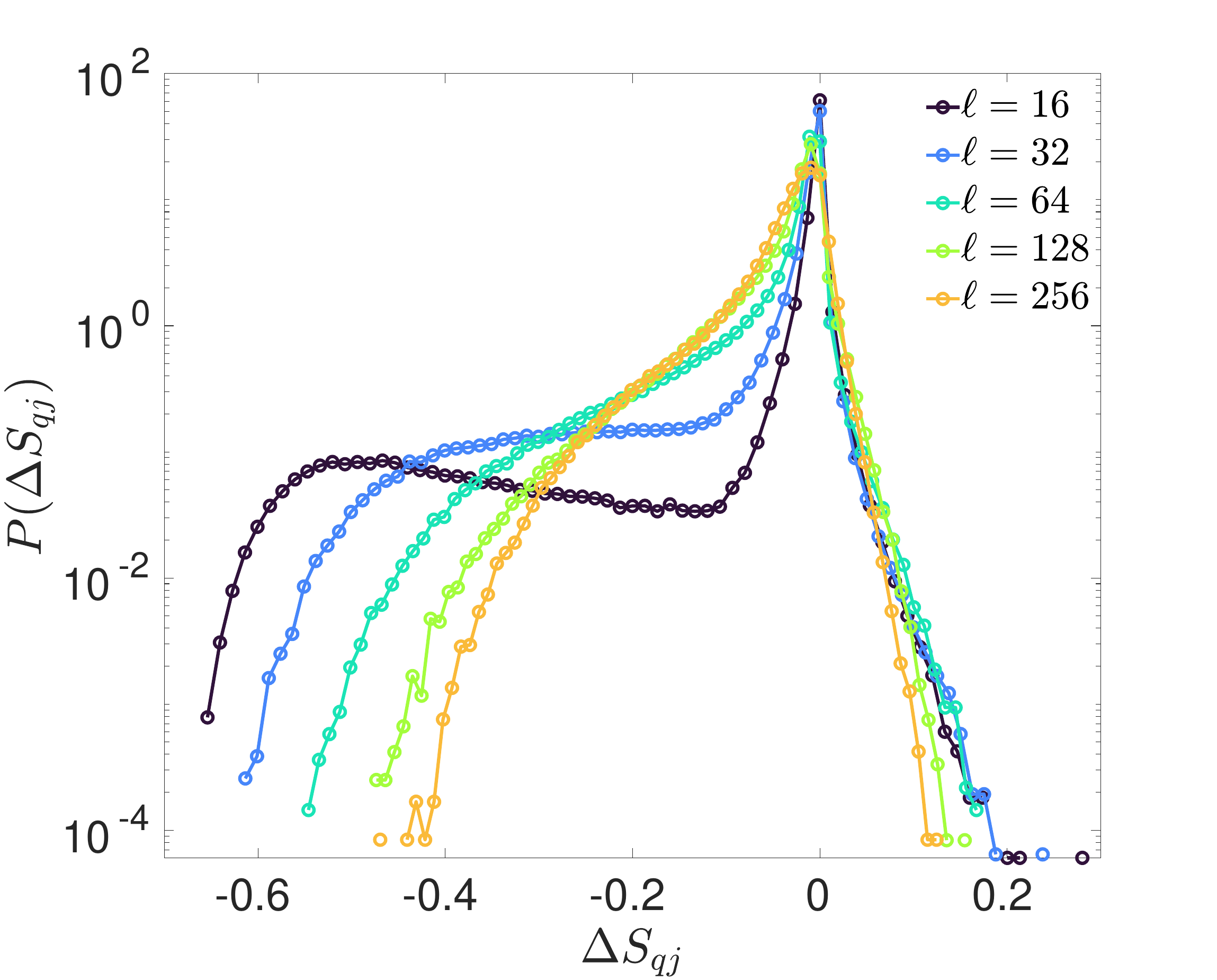} }
		\subfigure[]{ \label{fig.QJ_g0p5_Pro_DSqj_vs_ell}
			\includegraphics[width=8.cm]{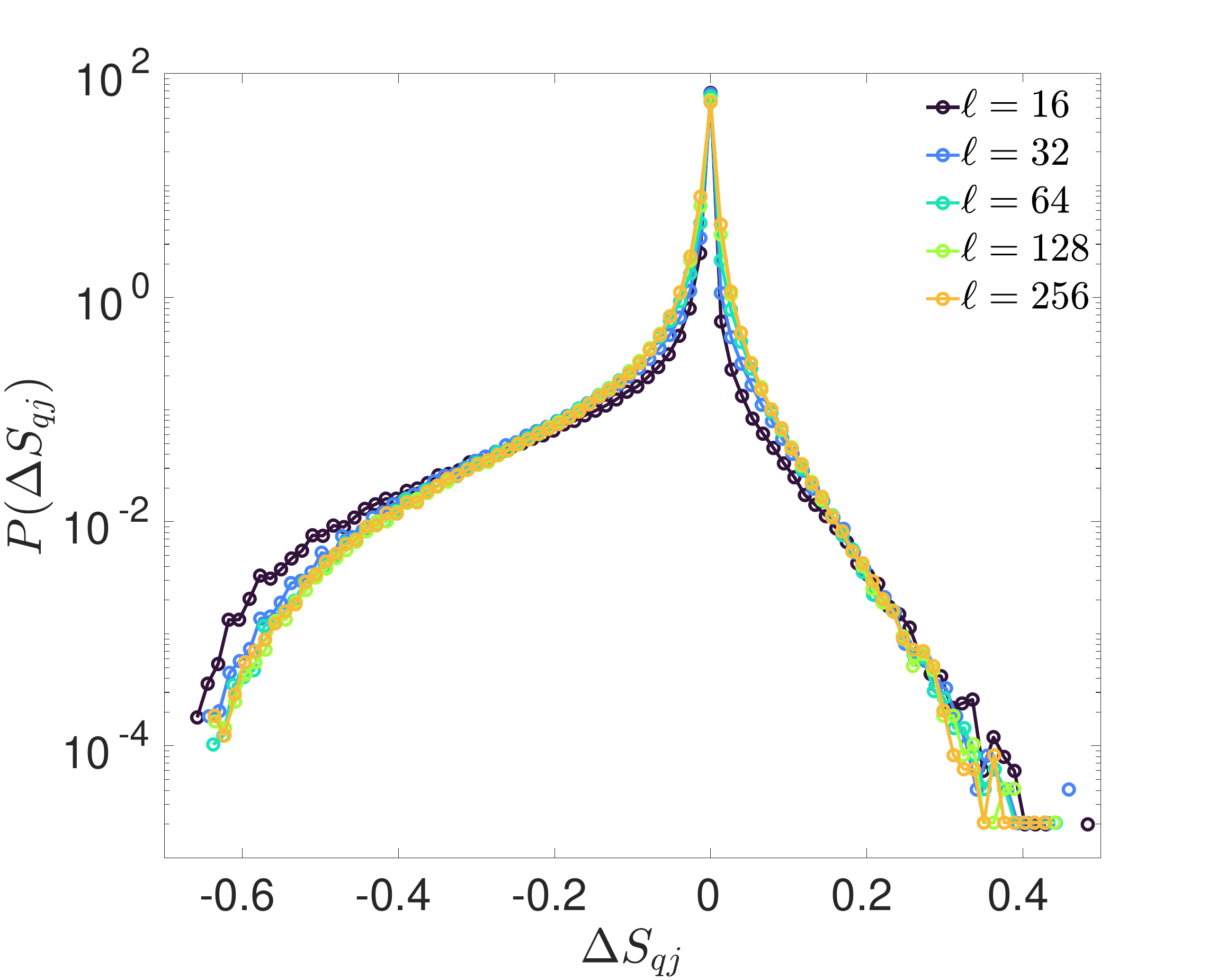} }
		\subfigure[]{ \label{fig.QJ_g1p5_Pro_DSqj_vs_ell}
			\includegraphics[width=8.cm]{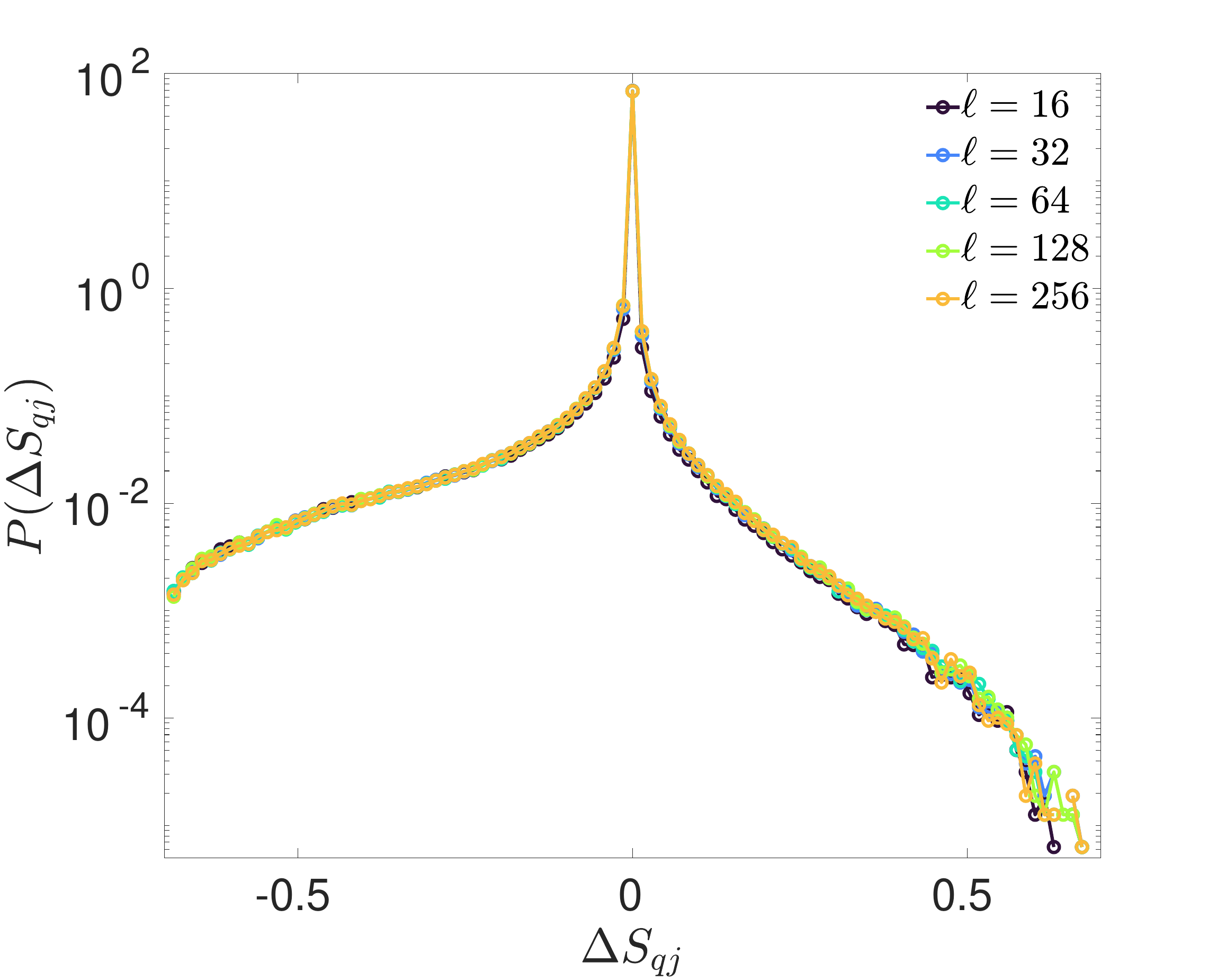} }
		\subfigure[]{ \label{fig.QJ_g3p0_Pro_DSqj_vs_ell}
			\includegraphics[width=8.cm]{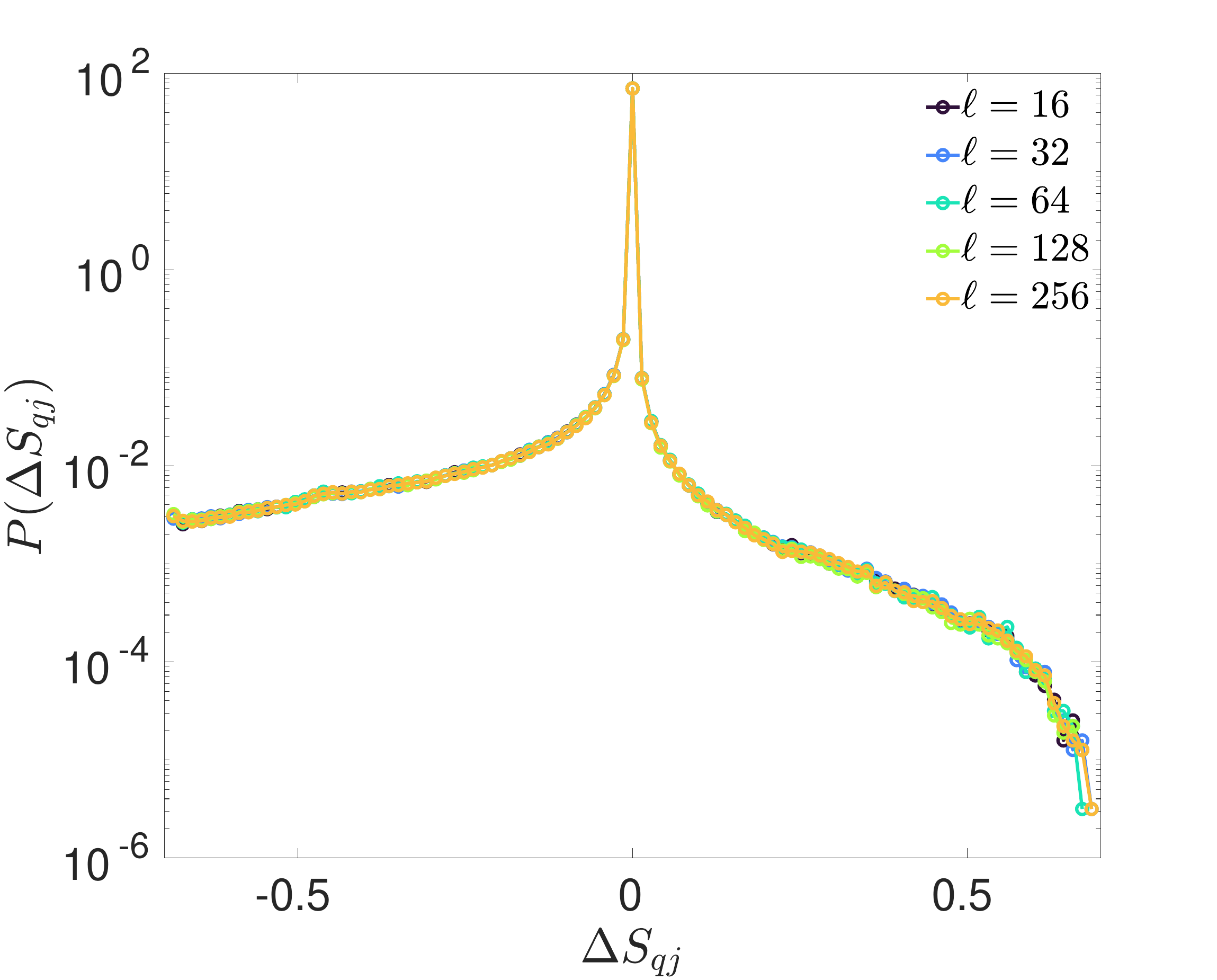} }
		\caption{
		The probability distribution $P(\Delta S_{qj})$ for different subsystem sizes $\ell$. Only in the weak measurement limit $\gamma = 0.1$, $P(\Delta S_{qj})$ depends on $\ell$ due to finite size effects. For stronger monitoring, $P(\Delta S_{qj})$ is insensitive to $\ell$ which may be indicative of an area-law behavior. The monitoring rates are $\gamma = 0.1, 0.5, 1.5$ and $3.0$ from \subref{fig.QJ_g0p1_Pro_DSqj_vs_ell} to \subref{fig.QJ_g3p0_Pro_DSqj_vs_ell}. The system size is $L = 512$. }
		\label{Fig:QJ_Pro_DSqj_vs_ell_g}
	\end{center}
\end{figure}

Therefore, we expect that when the subsystem size is much larger than this boundary region, $P(\Delta S_{qj})$ would be independent of the subsystem size $\ell$, which is illustrated in Fig.~\ref{Fig:QJ_Pro_DSqj_vs_ell_g}. In the weak monitoring limit $\gamma = 0.1$, when the subsystem size is smaller than $128$, $P(\Delta S_{qj})$ changes substantially from a broader distribution to a narrower distribution and finally become insensitive to $\ell$. When the monitoring is stronger, $\gamma \ge 0.5$, the boundary region is much smaller than subsystem size $\ell \ge 16$ that we investigate, so $P(\Delta S_{qj})$ do not depend on the subsystem size. For the projective measurement protocol, we also obtain similar results (not shown). 

\begin{figure}[!htbp]
	\begin{center}
		\subfigure[]{ \label{fig.QJ_L1024_g0p1_Pro_DSqj_sites123}
			\includegraphics[width=8.cm]{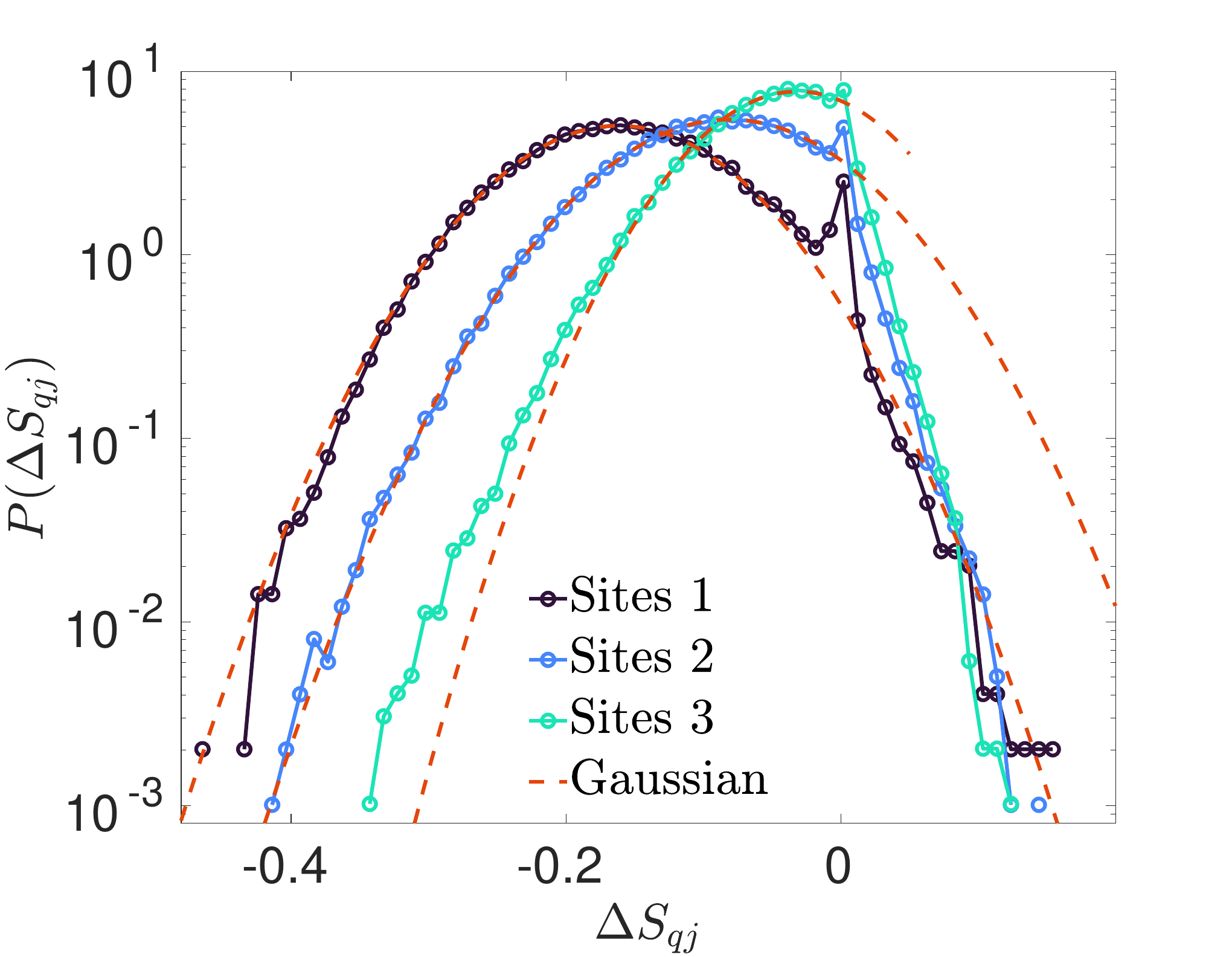} }
		\subfigure[]{ \label{fig.QJ_L1024_g0p1_Pro_DSqj_sites4}
			\includegraphics[width=8.cm]{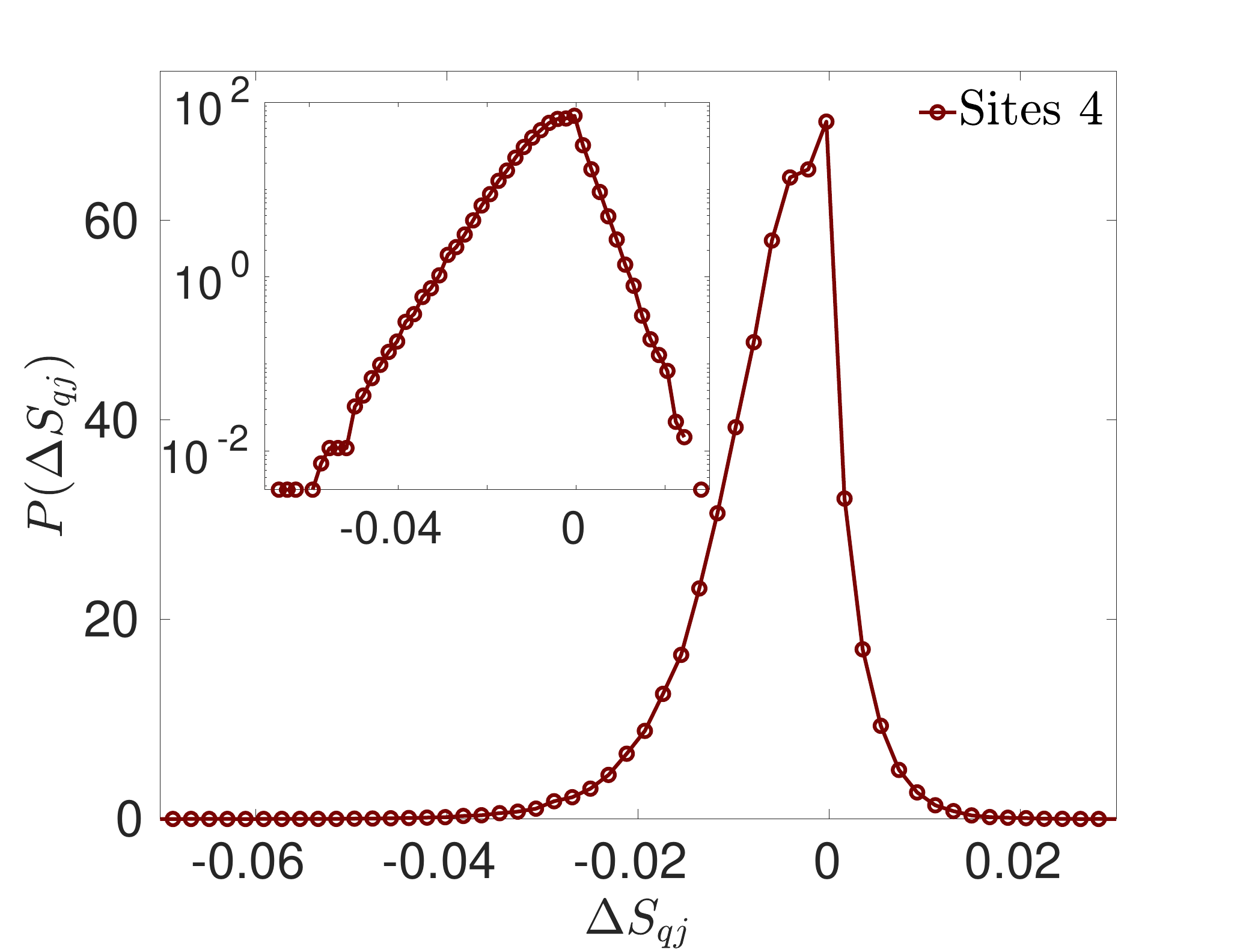} }
		\subfigure[]{ \label{fig.QJ_L512_g3p0_Pro_DSqj_sites1}
			\includegraphics[width=8.cm]{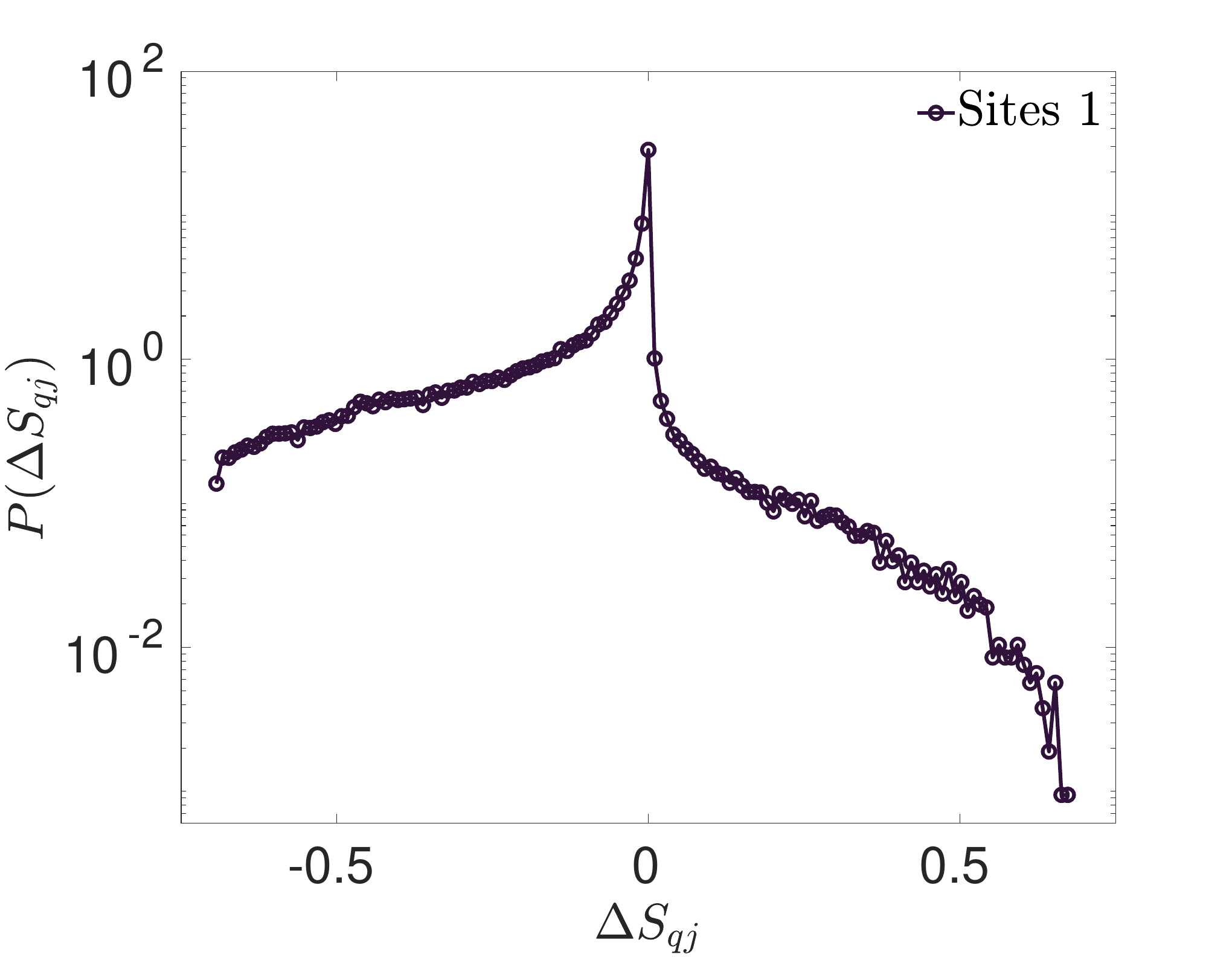} }
		\subfigure[]{ \label{fig.QJ_L512_g3p0_Pro_DSqj_sites4}
			\includegraphics[width=8.cm]{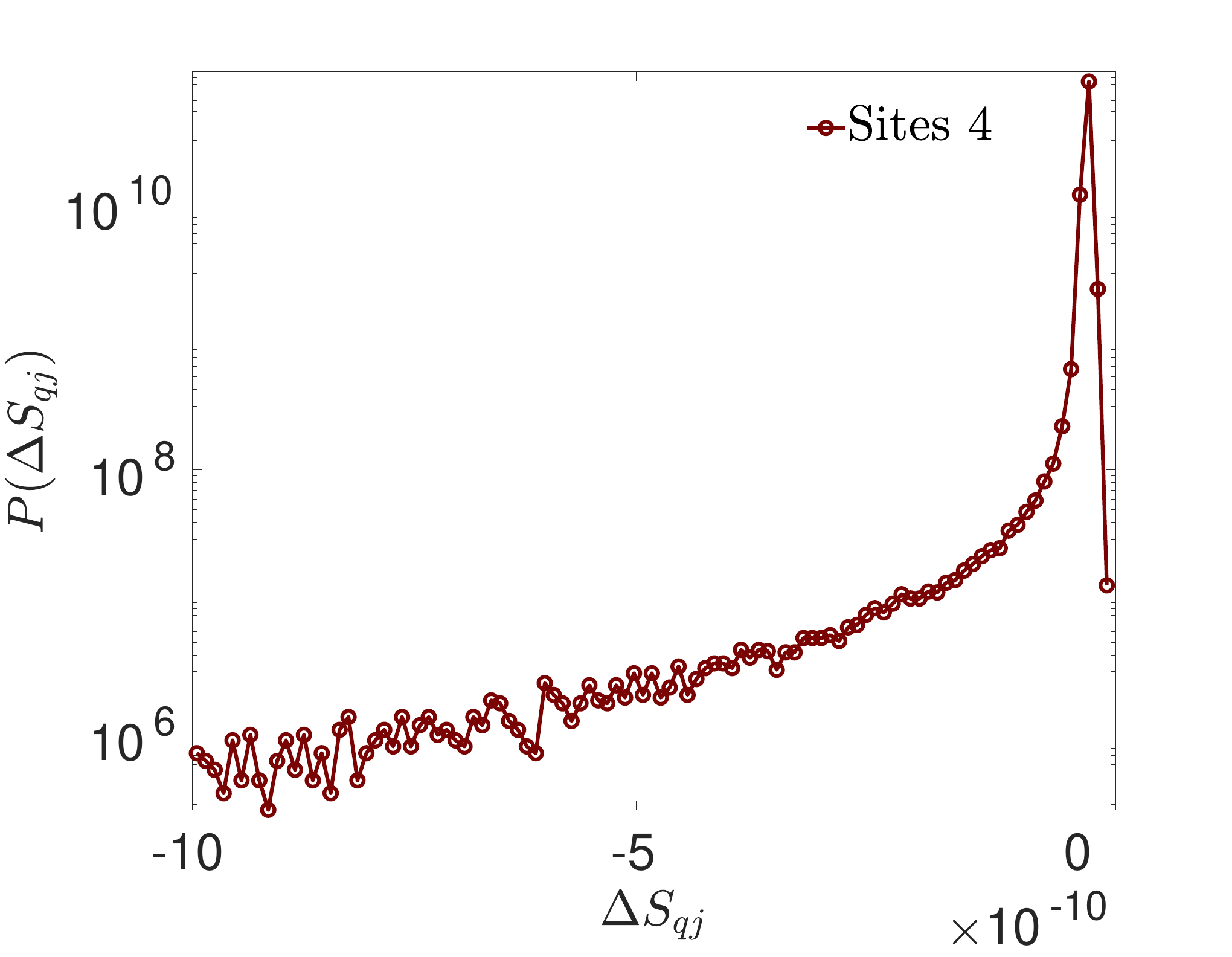} }
		\caption{
			The distribution function of $\Delta S_{qj}$ in Eq.~(\ref{eq:Deltaqj}) for specific sites. The upper panel: the monitoring rates is $\gamma = 0.1$ and the system size is $L=1024$. The bottom panel: the monitoring rates is $\gamma = 3.0$ and the system size is $L=512$. Here sites 1, 2 and 3 stand for sites around the boundary of the subsystem. Site 4 stands for the site at the center of the subsystem. Specifically, sites 1 are sites $[r=1,r=L/2,r=L/2+1,r=L]$, sites 2 are $[r=1+5,r=L/2-5,r=L/2+6,r=L-5]$, sites 3 are $[r=1+15,r=L/2-15,r=L/2+16,r=L-15]$ and sites 4 $[r=L/4,r=3L/4]$. Note that for Fig.~\subref{fig.QJ_L512_g3p0_Pro_DSqj_sites4}, the scale of $\Delta S_{qj}$ is of the order $10^{-10}$ and the maximum is $10^{10}$ so it is effectively a Dirac delta function.
		}\label{Fig:QJ_Pro_DSqj_sites}
	\end{center}
\end{figure}

An explicit calculation of the distribution function separating sites close to the boundary from those in the bulk, see Fig.~\ref{Fig:QJ_Pro_DSqj_sites}, confirms the qualitative picture above.
For weak monitoring, the distribution function for the boundary point is a Gaussian with a small peak at zero. As more sites around the boundary are added, the distribution develops a narrow exponential tail for $\Delta S_{qj} \gtrsim 0$. We note that rare instances of $\Delta S_{qj} \gtrsim 0$ occur due to the fact that the measurement not only disentangles the target site but also rearranges the remaining particles across the rest of the $L-1$ sites, effectively enhancing entanglement in unmeasured regions. This counterintuitive effect highlights the non-locality
inherent in quantum measurement dynamics. For bulk sites, the distribution is asymmetric with exponential tails characterized by different decay rates.
For strong monitoring, the distribution is effectively a delta function at zero for bulk points (note the scale of the axis), while for boundary points, though there is also a peak at zero, we observe an asymmetric distribution with a broad support that covers all possible values of $\Delta S_{qj}$.

Because of the asymmetry of the distribution, though $P(\Delta S_{qj})$ always has a sharp peak at zero, the averaged $\Delta S_{qj}$ is always slightly negative. We shall see this is necessary so that $\Delta S_{qj}$ balances the positive contribution from the non-Hermitian evolution to the dynamics which will be investigated later. This is necessary so that the total change is zero because our study is focused on the long time region where the EE has reached the saturation value.

\begin{figure}[!htbp]
	\begin{center}
		\subfigure[]{ \label{fig.QJ_L1024_g0p1_densitymap_DSqj_vs_ni}
			\includegraphics[width=8.cm]{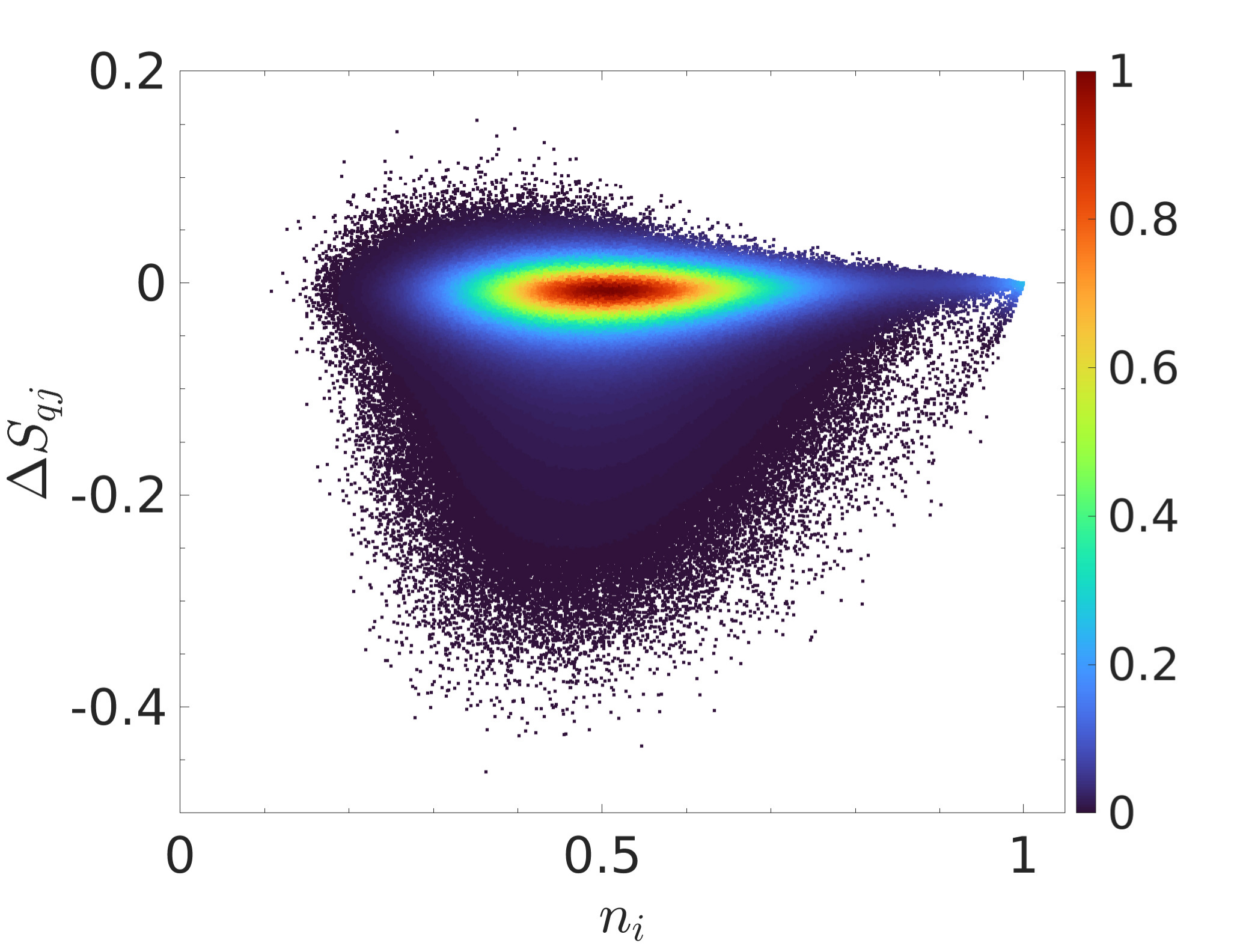} }
		\subfigure[]{ \label{fig.QJ_L1024_g0p5_densitymap_DSqj_vs_ni}
			\includegraphics[width=8.cm]{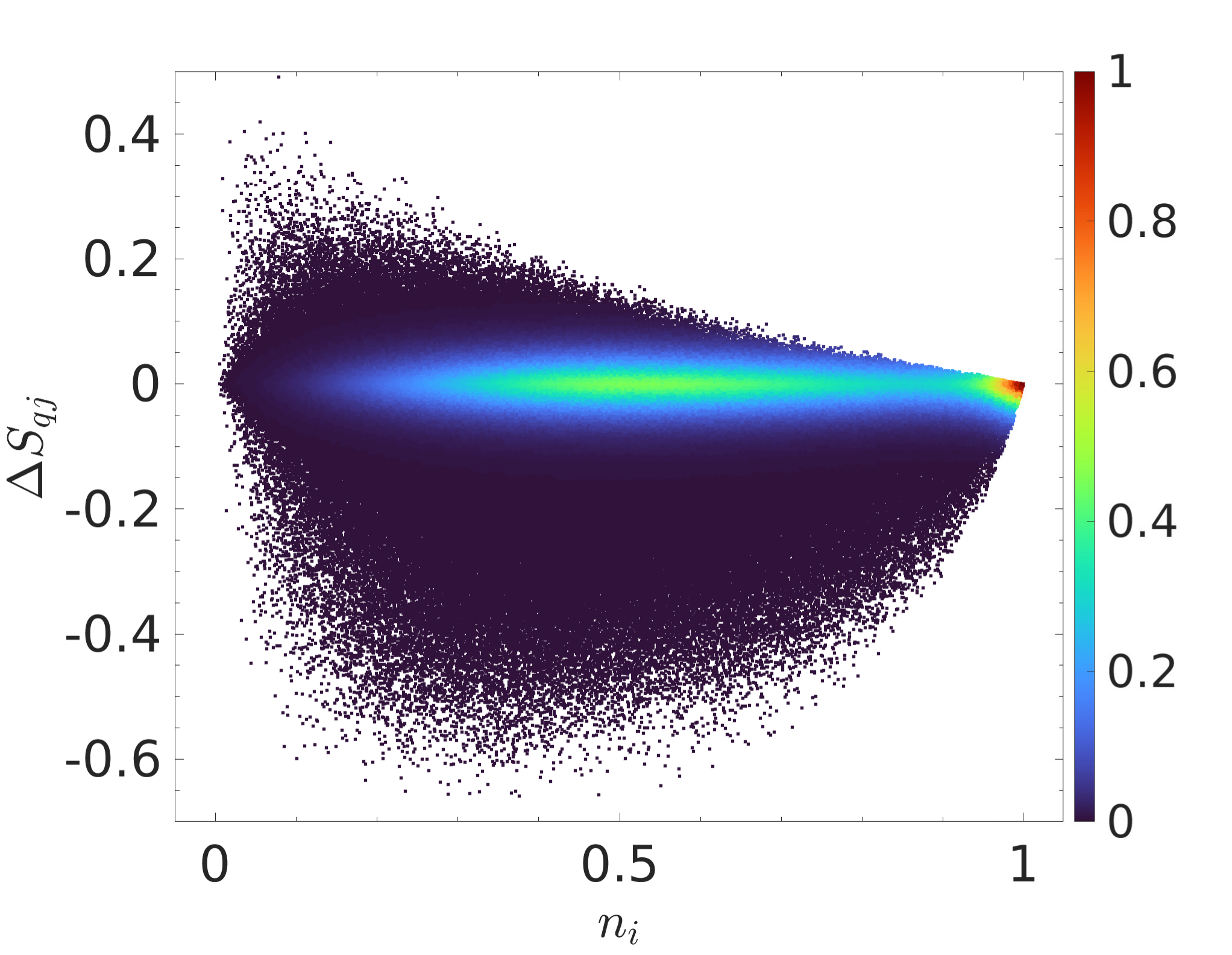} }
		\subfigure[]{ \label{fig.QJ_L512_g1p5_densitymap_DSqj_vs_ni}
			\includegraphics[width=8.cm]{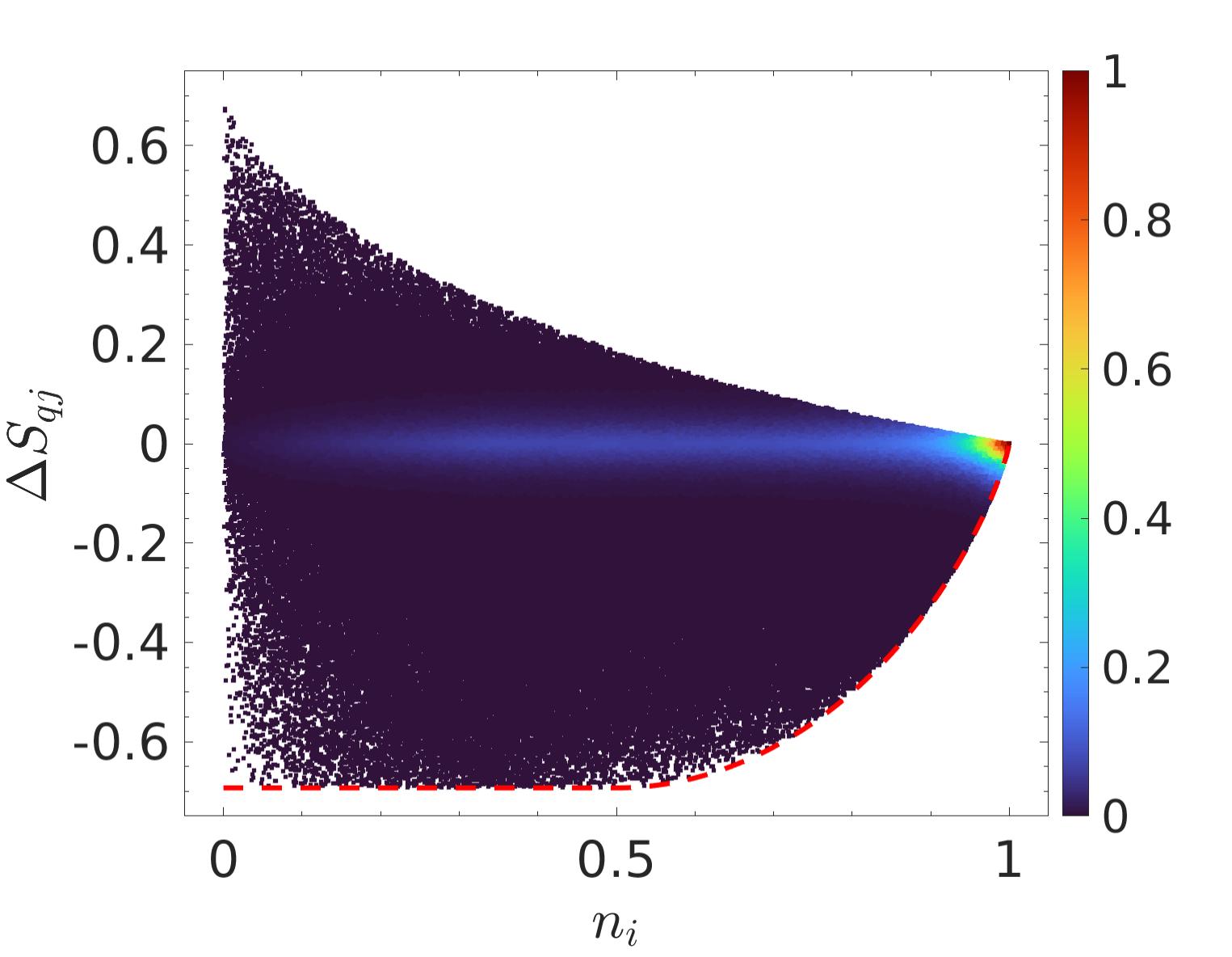} }
		\subfigure[]{ \label{fig.QJ_L512_g3p0_densitymap_DSqj_vs_ni}
			\includegraphics[width=8.cm]{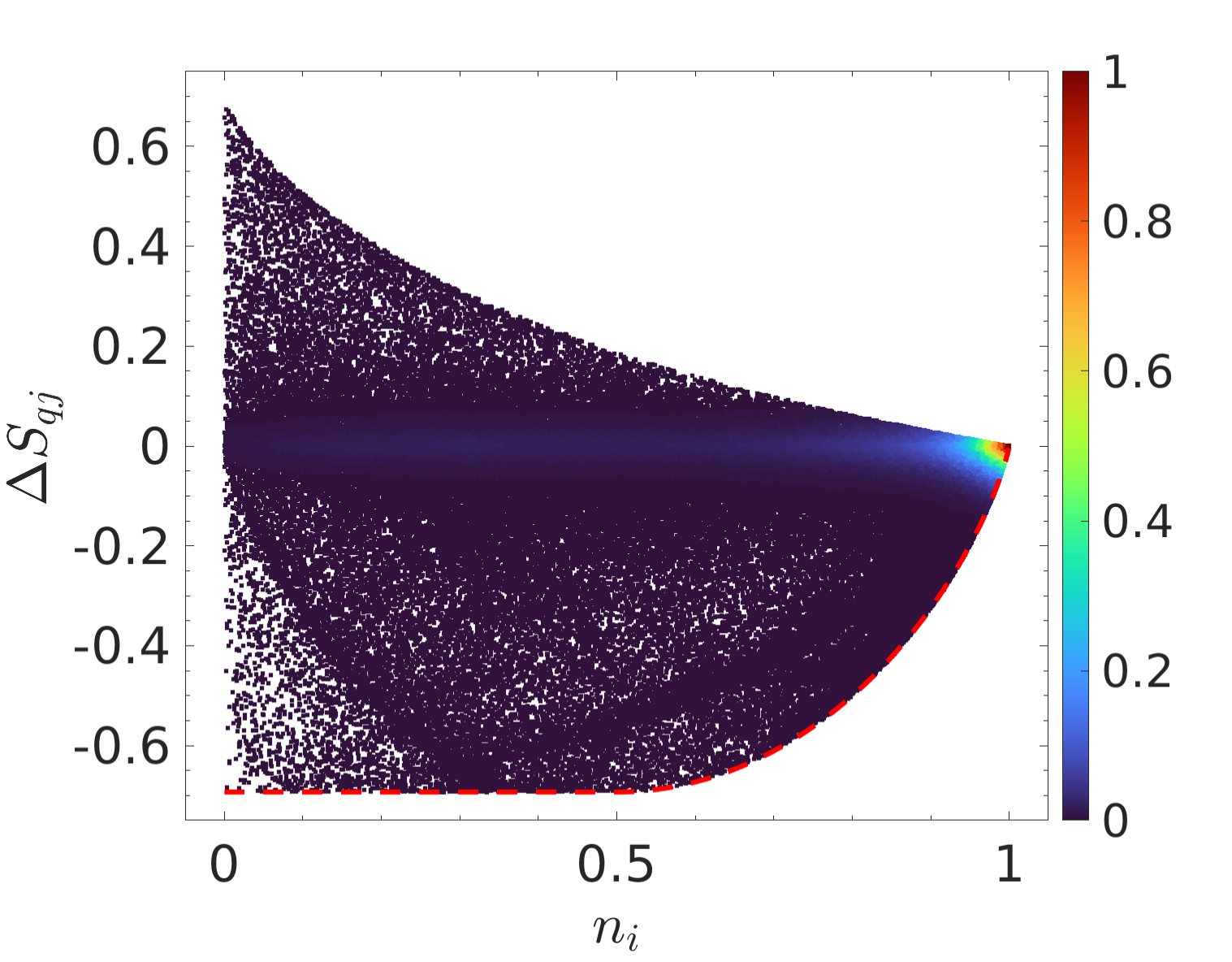} }
		\caption{The density-map of $\Delta S_{qj}$ in Eq.~(\ref{eq:Deltaqj})  as a function of the occupation number of the selected site that are going to be measured.  The color scale indicates the normalized density of measurement events. The monitoring rates are $\gamma = 0.1, 0.5, 1.5$ and $3.0$ from \subref{fig.QJ_L1024_g0p1_densitymap_DSqj_vs_ni} to \subref{fig.QJ_L512_g3p0_densitymap_DSqj_vs_ni}. The system size for the upper panel is $L=1024$, and the lower panel is $L=512$.  The red dashed line is Eq.\eqref{eq.toy}, signals the maximum reduction of the entanglement entropy after a single quantum-jump measurement.}\label{Fig:QJ_densitymap_DSqj_vs_ni}
	\end{center}
\end{figure}

\subsection{Effects of the occupation number $n_i$ on $\Delta S_{qj}$ }

In order to gain a more comprehensive understanding of the effect of the quantum jumps on the EE, we study its dependence on the occupation number $n_i$ at site $i$ right before the measurement occurs. We show in Fig.~\ref{Fig:QJ_densitymap_DSqj_vs_ni}, the density-map of $\Delta S_{qj}$ as a function of $n_i$.
We note that in the QJ protocol only the outcome $n_i = 1$ is measured.
It is expected \cite{schiro2024} that the smaller the occupation number, the larger the change in the EE after a measurement. From the results in Fig.~\ref{Fig:QJ_DSqj_vs_r}, sufficiently large changes are restricted to the boundary of the subsystem.
As a result, a change in $n_i = 0 \rightarrow 1$ near the subsystem boundary, induced by a measurement, is expected to generate a more pronounced change to the EE.
In the weak monitoring limit, most sites will have a filling rate near $1/2$ with virtually no sites with a filling rate near $0$, so that we cannot observe the expected large entanglement change for sites with $n_i\sim 0$. Instead, we observe the expected behavior of a maximum at half filling and, consistent with previous results for the distribution of $\Delta S_{qj}$, that a relatively strong peak occurs at $\Delta S_{qj} = 0$. It is also worth noting that even in this weak monitoring limit, there is already a small peak at $n_i = 1$ as a consequence of the measurements. Indeed, even for intermediate monitoring strength, the peak rapidly moves to $n_i = 1$ which illustrate the strong impact of measurement on the entanglement dynamics.

In the strong monitoring limit, even though $n_i$ can take any value within its domain of definition $[0,1]$, the change in EE has a maximum value of $\ln{2} \approx 0.69$, reflecting the maximal entanglement contribution from a single particle.
As mentioned earlier, this maximum value can only occur at the subsystem boundary sites because, for the rest of sites, measurements induce little change in $\Delta S_{qj}$. Therefore, a measurement would only affect particles around the boundary. In this strong coupling limit, it is possible to compute analytically the maximum decrease of the entanglement entropy after each quantum jump measurement. 
For that purpose, we assume that after each measurement, only the measured particle is affected, while the rest of the system remains unchanged. Therefore, the measurement only affects the entropy associated with site $i$, with no rearrangement effects that could increase the total EE.
 We have to distinguish two cases, 
if the occupation number of the site $i$ before the measurement is $n_i > 0.5$, the maximum bipartite entanglement entropy is dominated by the entanglement between this site $i$ and the other half subsystem, which is given by $S = -n_i\ln{n_i} - (1-n_i)\ln{(1-n_i)}$. 
If $n_i < 0.5$, the maximum entanglement is no longer dominated by the entanglement between the site $i$ and the other half system because the particle wave function can be extended over several sites across the subsystem boundary, namely, effectively the particle is fully entangled only in the subsystem boundary, leading to a maximum single-particle entanglement entropy $S = \ln{2}$. After measurement, the entanglement entropy of this particle drops to $0$.
Therefore, the maximum drop in entanglement entropy induced by a single quantum-jump measurement is then given by
\begin{equation}
\Delta S_{qj} = 
	\begin{cases}
		-\ln{2} &  n_i < 0.5 \\
		n_i\ln{n_i} + (1-n_i)\ln{(1-n_i)} & n_i \ge 0.5 
	\end{cases} 
\label{eq.toy}
\end{equation}
This assumption shows excellent agreement with our numerical results, as illustrated in Fig.~\ref{Fig:QJ_densitymap_DSqj_vs_ni}.
In contrast, the maximum increase of entanglement entropy after a measurement, which also shows a sharp bound, remains less understood, because it requires a coordinated rearrangement of many fermions near the subsystem boundary so it cannot be expressed solely in terms of $n_i$. A detailed characterization of this upper bound is beyond the scope of this paper.

\subsection{The recovery of entanglement due to non-Hermitian evolution}

\begin{figure}[!htbp]
	\begin{center}
		\subfigure[]{ \label{fig.QJ_L512_dSnH_DSqj_vs_gm}
			\includegraphics[width=8.cm]{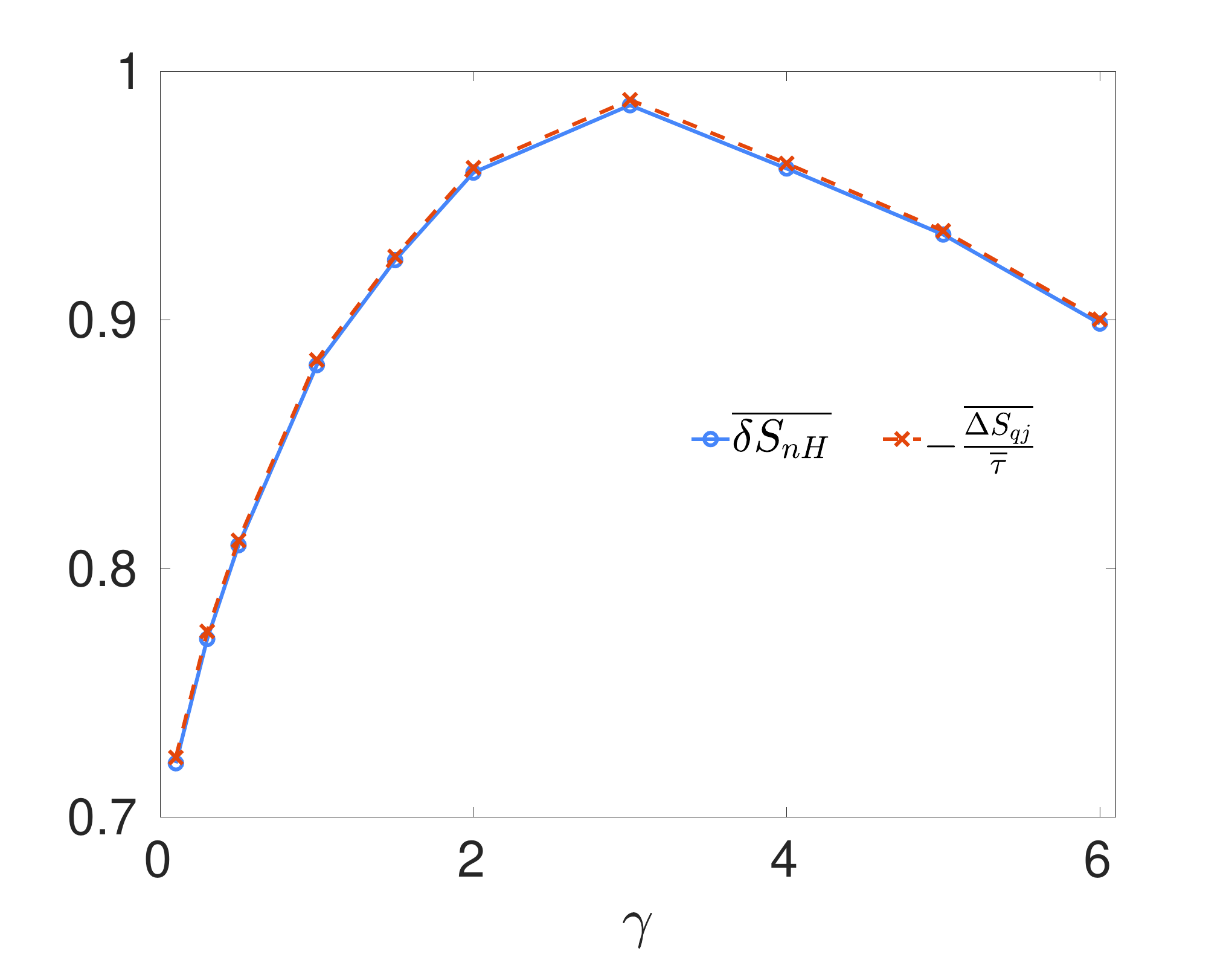} }
		\subfigure[]{ \label{fig.PM_L512_dSun_DSpm_vs_gm}
			\includegraphics[width=8.cm]{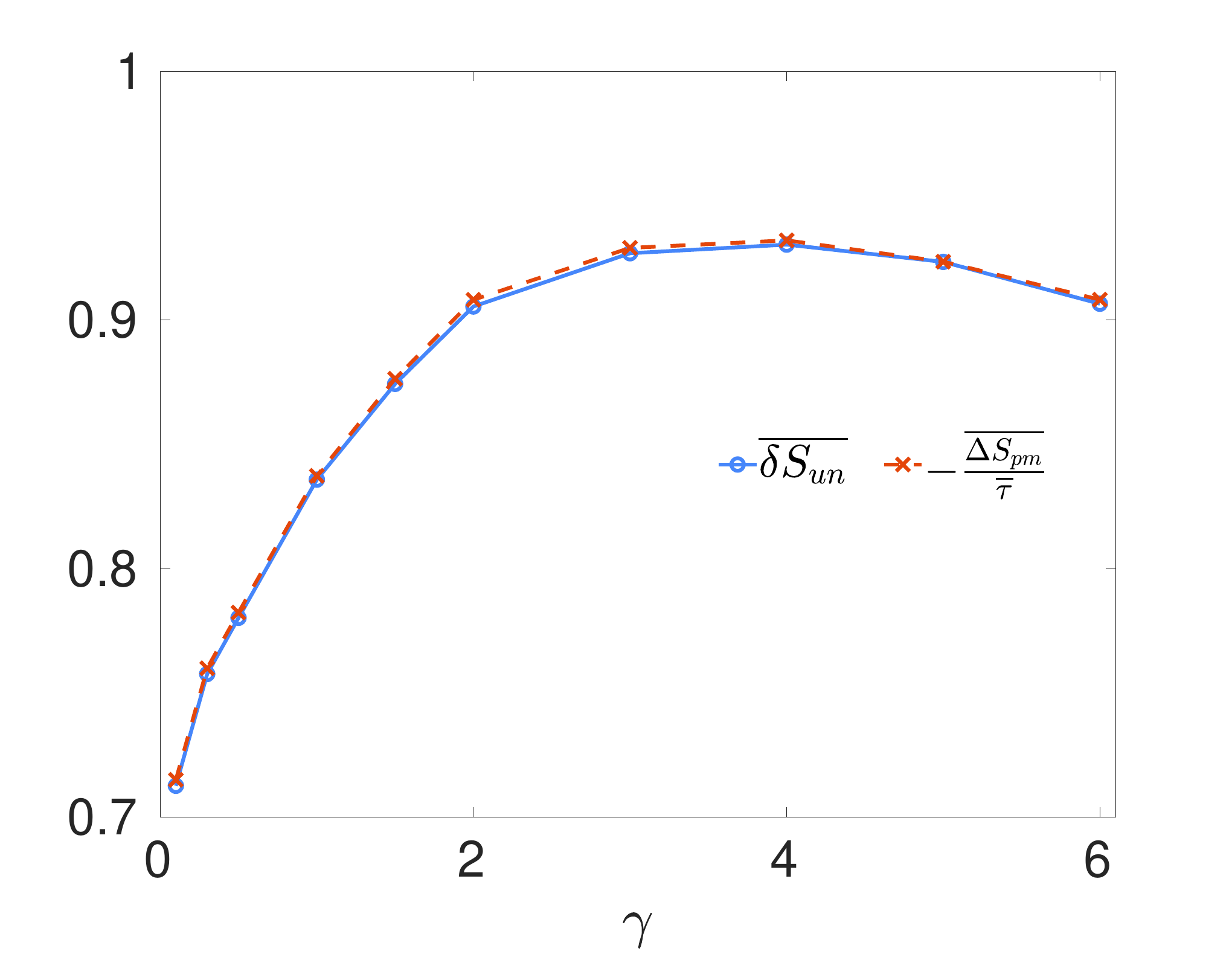} }
		\caption{The average change in EE after a measurement as a function of the monitoring rate $\gamma$ using the  \subref{fig.QJ_L512_dSnH_DSqj_vs_gm} QJ protocol (left) and the \subref{fig.PM_L512_dSun_DSpm_vs_gm} PM protocol (right). This is compared with the average speed of entanglement gain between measurements in the QJ protocol ($\delta S_{nH}$) (left) and the PM protocol ($\delta S_{un}$) (right). As is expected, the creation of entanglement is compensated by the loss due to the measurement. The system size is $L=512$.
		}\label{Fig:L512_dSnH_DSqj_vs_gm}
	\end{center}
\end{figure}

Although our interest is mostly focused on the effect of quantum jumps on the EE, the change in EE due to the non-Hermitian evolution namely, the evolution between quantum jumps, is also important in order to have a complete understanding of the entanglement dynamics. For that purpose, we define the speed of the entanglement entropy change
\begin{equation}\label{eq:deltanH}
	\delta S_{nH} = \Delta S_{nH}/\tau 	
\end{equation}
where $\tau$ is the time between quantum jumps and $\Delta S_{nH} = S_A(t+\tau) - S_A(t)$ computed from Eq.~(\ref{eq:EE}). We use $\delta$ in $\delta S_{nH}$ to distinguish the time dimension from $\Delta$ in $\Delta S_{nH}$, consistent with $\delta S_{qsd}$. Therefore, $\delta S_{nH}$ captures the system intrinsic capacity to regenerate entanglement between measurements.
Since we are interested in the late time dynamics of entanglement for which the average EE has reached the saturation value and therefore does not have a net growth in time, we expect \cite{schiro2024} that, on average, the generation of entanglement due to the non-Hermitian evolution will compensate its destruction due to measurements  $\overline{\delta S_{nH}} \approx \overline{\Delta S_{qj}}/\bar{\tau}$, where $\overline{~\bullet~}$ stands for the average over quantum trajectories.
Results depicted in the left plot of Fig.~\ref{Fig:L512_dSnH_DSqj_vs_gm} confirm that this is the case for all considered monitoring strengths. We stress that this cancellation is a property of the average only. We shall see that the distribution of $\delta S_{nH}$ and other properties are qualitatively different from those due to quantum jumps $\Delta S_{qj}$.

\begin{figure}[!htbp]
	\begin{center}
		\subfigure[]{ \label{fig.QJ_g0p1_Pro_dSnH_vs_L}
			\includegraphics[width=8.cm]{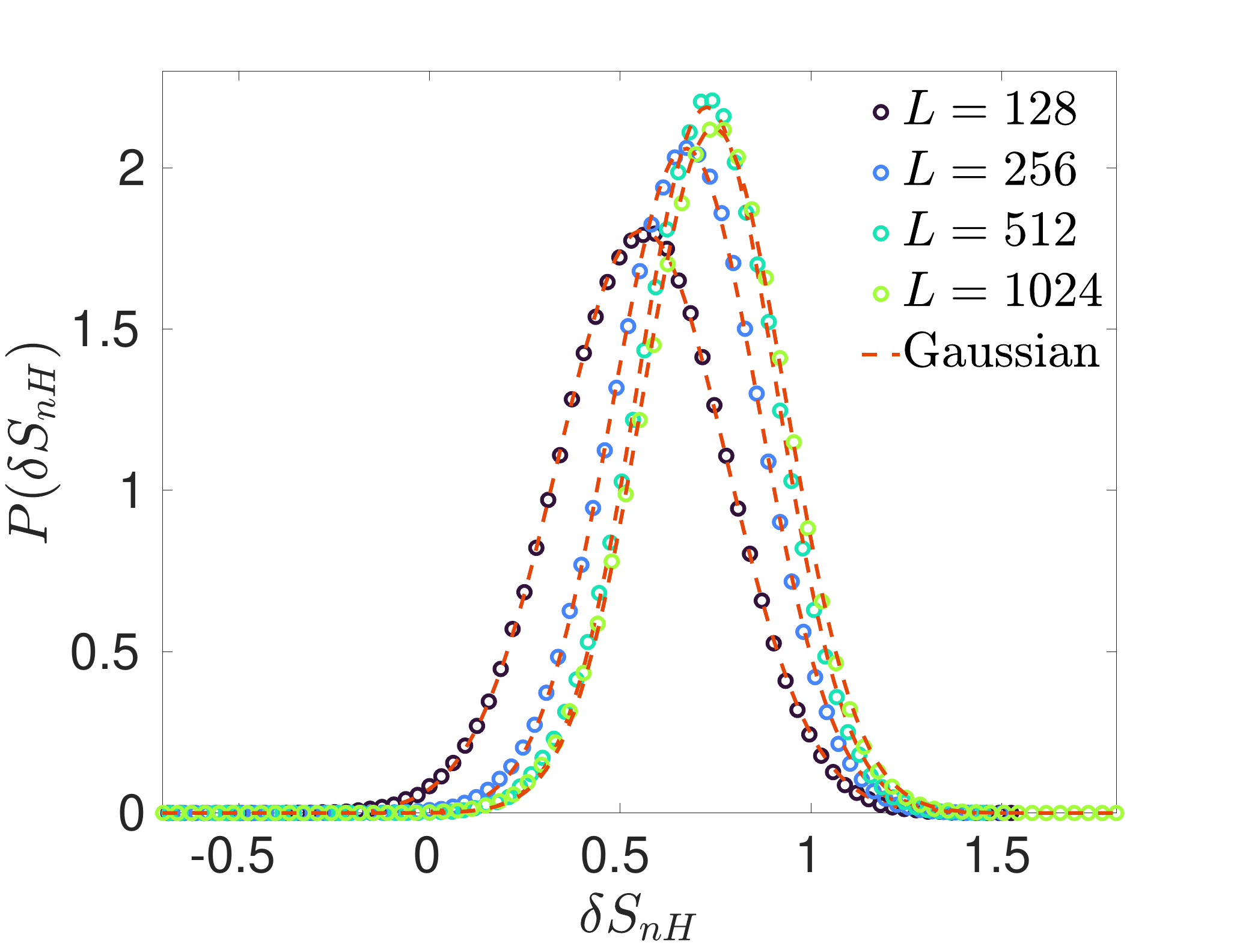} }
		\subfigure[]{ \label{fig.QJ_g0p5_Pro_dSnH_vs_L}
			\includegraphics[width=8.cm]{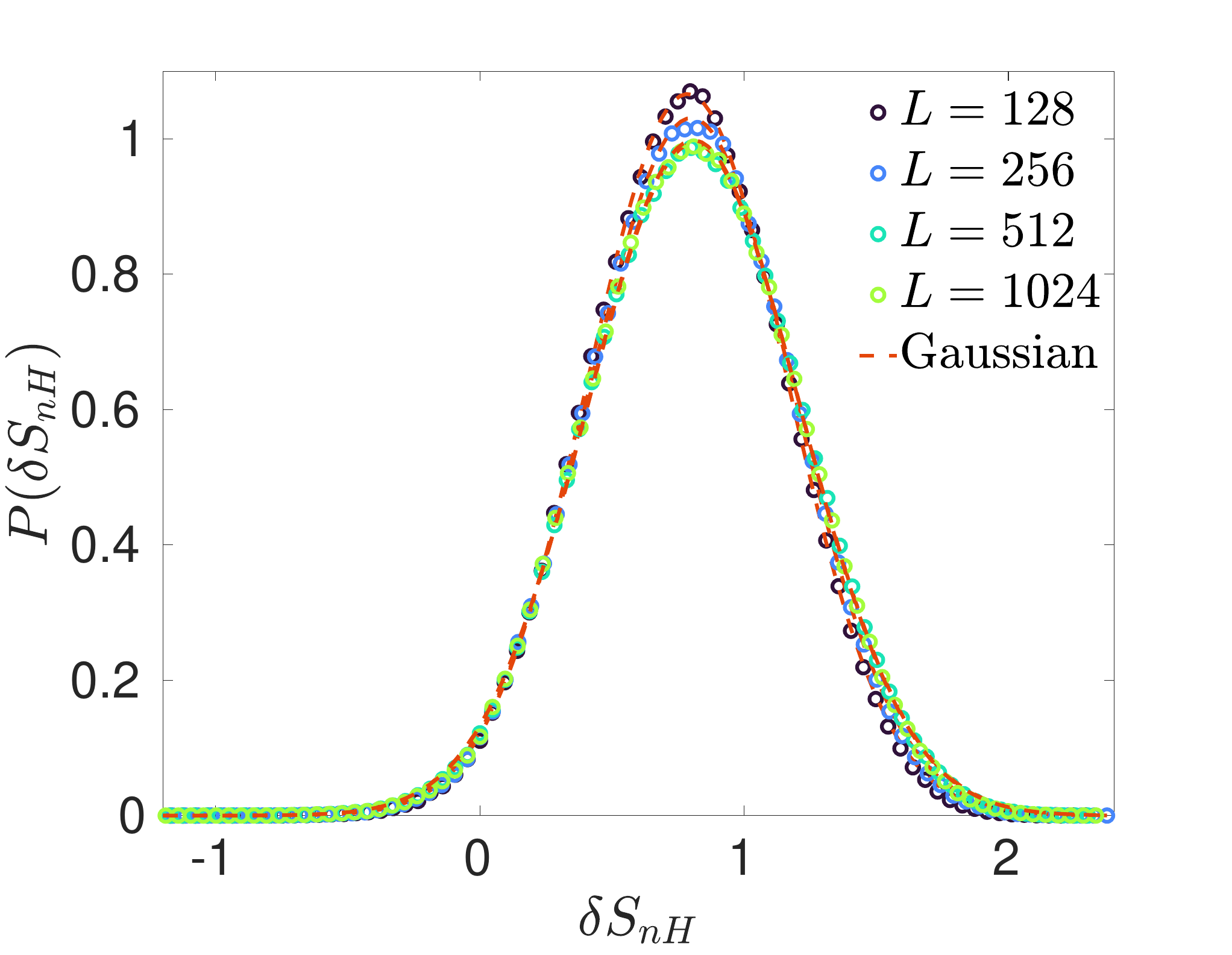} }
		\subfigure[]{ \label{fig.QJ_g1p5_Pro_dSnH_vs_L}
			\includegraphics[width=8.cm]{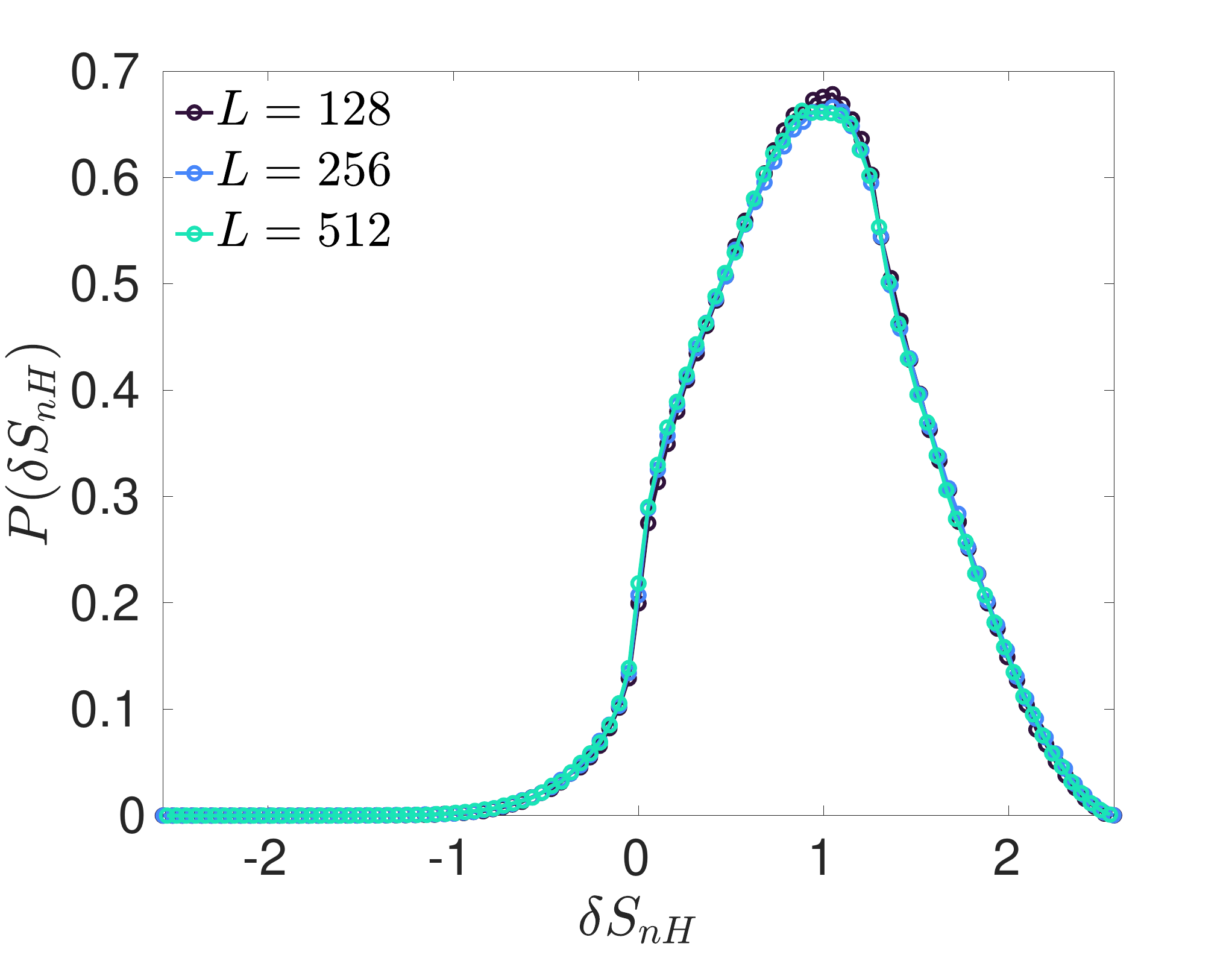} }
		\subfigure[]{ \label{fig.QJ_g3p0_Pro_dSnH_vs_L}
			\includegraphics[width=8.cm]{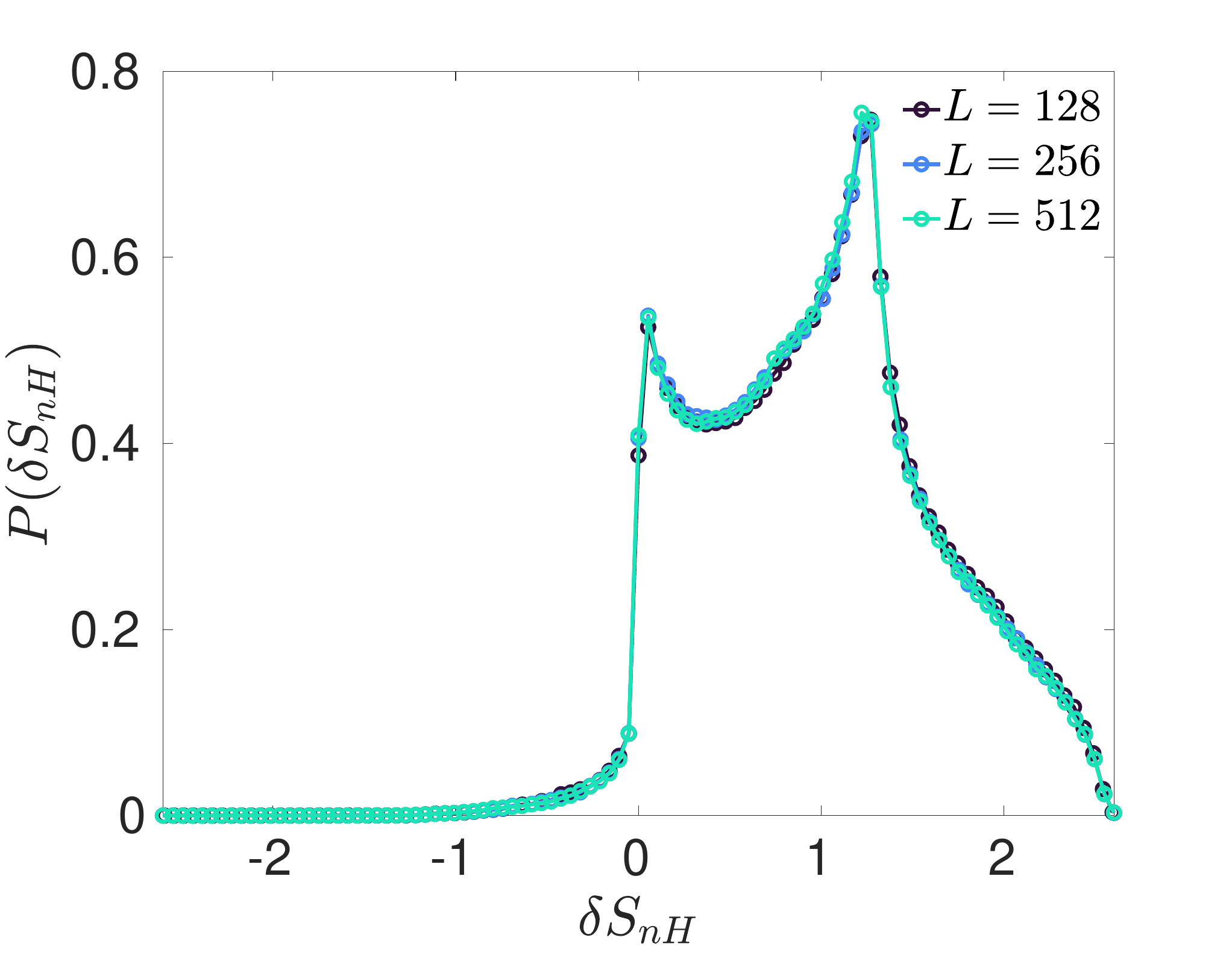} }
		\caption{The probability distribution of the change of EE $\delta S_{nH}$ in Eq.~(\ref{eq:deltanH}), computed using Eq.~(\ref{eq:EE}), between measurements due to the non-hermitian evolution in the QJ protocol for different system sizes and times longer than the saturation time.
			The monitoring rates are $\gamma = 0.1, 0.5, 1.5$ and $3.0$ from \subref{fig.QJ_g0p1_Pro_dSnH_vs_L} to \subref{fig.QJ_g3p0_Pro_dSnH_vs_L}.	The monitoring rate is the same as in Fig.~\ref{Fig:QJ_Pro_DSqj_vs_L_g}.
		}\label{Fig:QJ_Pro_dSnH_vs_L}
	\end{center}
\end{figure}

Interestingly, the dependence on the average is non-monotonic with the monitoring strength $\gamma$. This behavior stems from a competition between two effects.
For monitored systems, $\overline{\delta S_{nH}}$ regenerates entanglement within a region around the subsystem boundary with characteristic length $\xi$ \cite{poboiko2023}, which decreases with increasing $\gamma$.
This typical length $\xi$ grows \cite{poboiko2023} exponentially with the monitoring strength $\gamma$ and therefore it will be larger than the system size $L$ we can reach numerically for sufficiently small $\gamma$. Consequently, when $\gamma = 0.1$, the distribution $P(\delta S_{nH})$, as is shown in Fig.~\ref{Fig:QJ_Pro_dSnH_vs_L}, depends on the system size since $\xi$ exceeds the maximum system size $L$. In contrast, in the stronger monitoring limit, when $\gamma > 0.5$, $\xi$ is smaller than the system size $L$ and, as expected, $P(\delta S_{nH})$ exhibits no size dependence.

In that region, it is expected that the regeneration rate of EE dominates over its measurement-induced destruction rate governed by the quantum jumps.
More specifically, in the weak monitoring limit, the change of EE due to each measurement $\overline{\Delta S_{qj}}$ shows a weak dependence on $\gamma$, while $\bar{\tau} \propto 1/\gamma$, and therefore $\overline{\Delta S_{qj}}/\bar{\tau} \propto \gamma$. This destruction of entanglement is compensated  by $\overline{\delta S_{nH}} \propto \gamma$, which is consistent with the results of Fig.~\ref{Fig:L512_dSnH_DSqj_vs_gm}.
When $\gamma$ is sufficiently large, the typical length $\xi$ becomes on the order of or smaller than the lattice constant. In this regime, the spatial extent for entanglement regeneration is suppressed, and the destruction of entanglement by frequent measurements outpaces regeneration. Consequently, $\overline{\delta S_{nH}} $ decreases as $\gamma \gtrsim 1$ increases while the effects of single local measurements $\overline{\Delta S_{qj}} \rightarrow 0$. The observation of a local maximum at $\gamma \sim 3$ in Fig.~\ref{Fig:L512_dSnH_DSqj_vs_gm} confirms this picture.

\subsection{Probability distribution of $\delta S_{nH}$}
Since $\delta S_{nH}$ provides additional insight into the system's ability to recover entanglement, a study of $P(\delta S_{nH})$, the probability distribution of $\delta S_{nH}$, directly complements our understanding of the impact of measurements in the dynamics. As depicted in Fig.~\ref{Fig:QJ_Pro_dSnH_vs_L}, the distribution of probability $P(\delta S_{nH})$ in the weak monitoring limit is well described by a Gaussian distribution which is not strongly affected by the increase in the monitoring strength. However, in the strong monitoring limit, we observe a drastic change in the distribution, the central part of $P(\delta S_{nH})$ becomes closer to an inverted Gaussian with two non-analytic finite peaks at zero and a positive value. The peak at zero is a consequence of complete Zeno effect characterized by a state that has zero EE before the action of the non-Hermitian Hamiltonian and that therefore has a larger probability to remain in this state.
We believe that the second peak at a larger positive value of $\delta S_{nH}$ is related to the boundary points at which the system can still regenerates entanglement fast.

\begin{figure}[!htbp]
	\begin{center}
		\subfigure[]{ \label{fig.QJ_L1024_g0p1_densitymap_dSnH_vs_r}
			\includegraphics[width=8.cm]{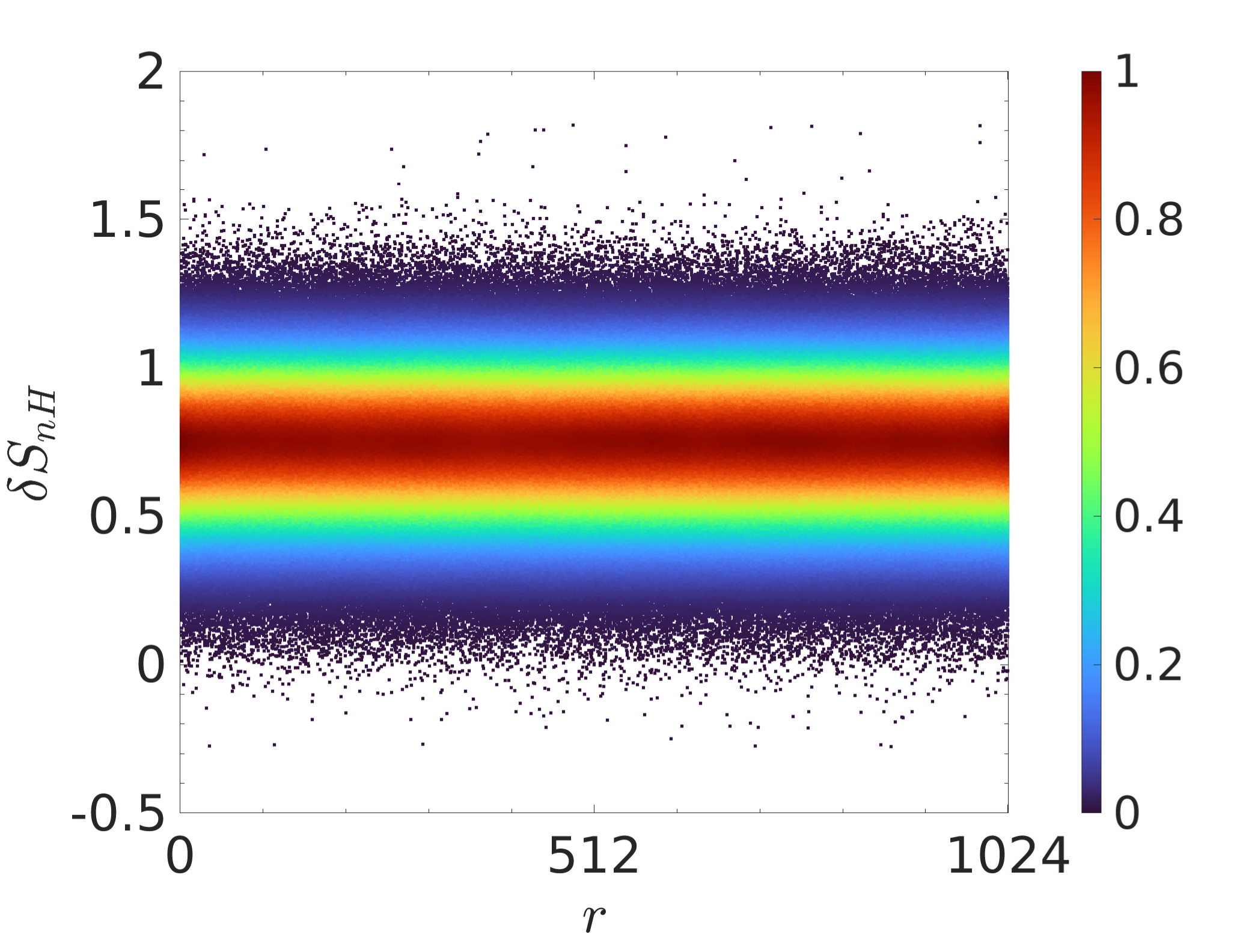} }
		\subfigure[]{ \label{fig.QJ_L512_g3p0_densitymap_dSnH_vs_r}
			\includegraphics[width=8.cm]{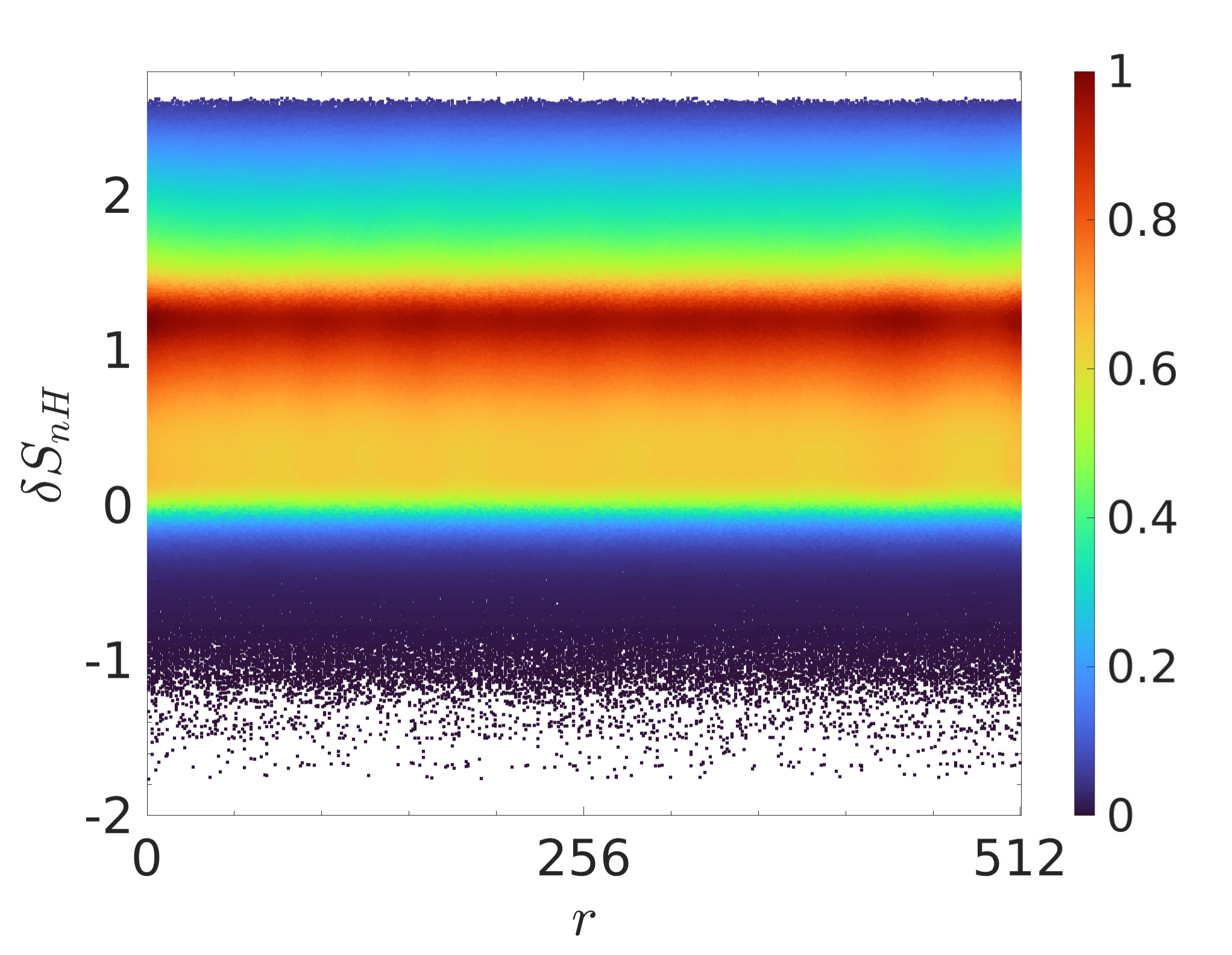} }
		\caption{The entanglement change $\delta S_{nH}$ in Eq.~(\ref{eq:deltanH}) due to the non-Hermitian evolution in the QJ protocol after measuring site $r$. The color scale indicates the normalized density of measurement events. The left plot is for the weak monitoring rate $\gamma=0.1$ and system size $L=1024$, while the right plot is for the strong monitoring rate $\gamma=3$. The system size is $L=512$. Different from the effect of quantum jumps $\Delta S_{qj}$, see Fig.~\ref{Fig:QJ_DSqj_vs_r}, $\delta S_{nH}$ due to the non-Hermitian evolution does not show any dependence on the measuring site.}\label{Fig:QJ_dSnH_vs_r}
	\end{center}
\end{figure}

We now address the site dependence of $\delta S_{nH}$, which reveals the regeneration of entanglement after measuring the selected sites. An important difference with respect to the quantum jump contribution $\Delta S_{qj}$ is that, see Fig.~\ref{Fig:QJ_dSnH_vs_r}, $\delta S_{nH}$ is identical for all sites. The stark difference between boundary and bulk sites observed for the change in EE induced by quantum jumps is not present at all in the non-Hermitian contribution to the EE. The absence of any site dependence in $\delta S_{nH}$ indicates that earlier measurements $\Delta S_{qj}$ do not affect the later non-Hermitian evolution $\delta S_{nH}$. 

$\delta S_{nH}$ is an intrinsic property of the system that depends solely on a typical length scale governing the spread of entanglement between two consecutive measurements near the boundary of the two subsystems.
This length scale captures the short-time entanglement dynamics and is different from the finite correlation length \cite{poboiko2023} characterizing the area-law phase of the dynamical equilibrium state.

This qualitative difference between the different contributions to the EE change is a further illustration of the importance, and relevance, of studying local distribution functions for a full characterization of the entanglement dynamics in a many-body quantum system.

\begin{figure}[!htbp]
	\begin{center}
		\subfigure[]{ \label{fig.QJ_L1024_g0p1_densitymap_dSnH_vs_ni}
			\includegraphics[width=8.cm]{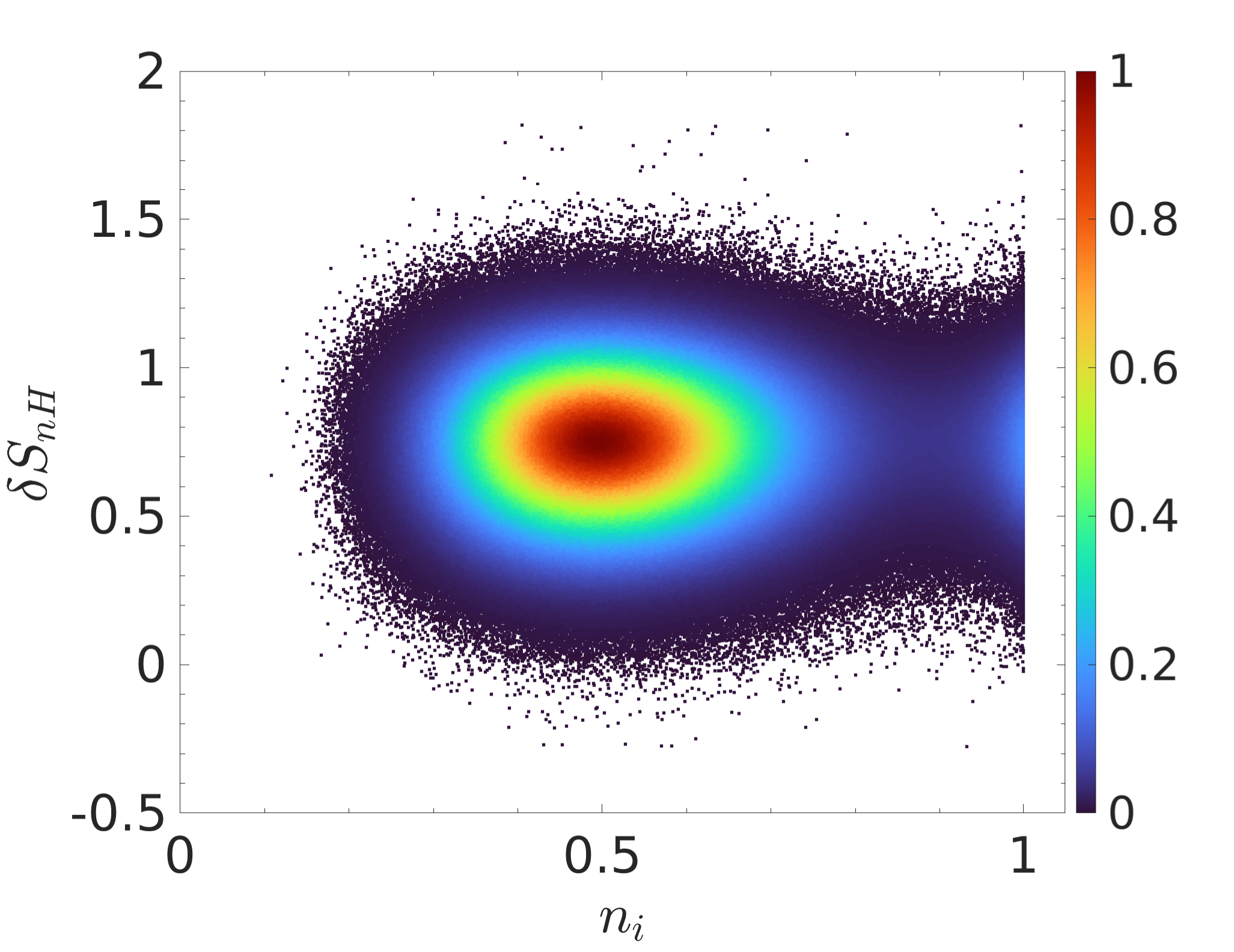} }
		\subfigure[]{ \label{fig.QJ_L512_g3p0_densitymap_DSnH_vs_ni}
			\includegraphics[width=8.cm]{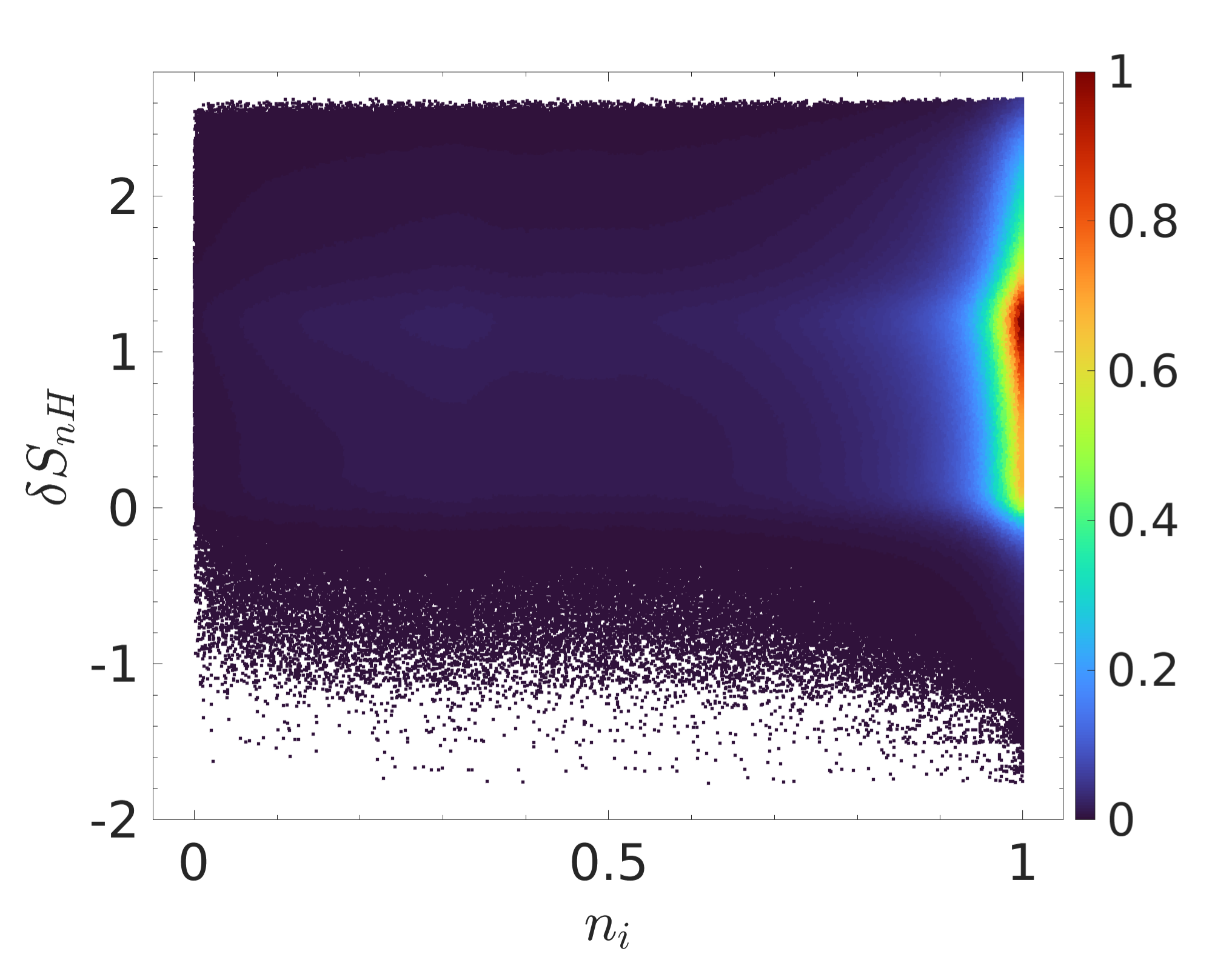} }
		\caption{The density-map of $\delta S_{nH}$ in Eq.~(\ref{eq:deltanH}) for the QJ protocol as a function of the occupation number before the measurements. The color scale indicates the normalized density of measurement events. The monitoring rate are $\gamma = 0.1$ (left), and $\gamma = 3.0$ (right).
		}\label{Fig:QJ_densitymap_dSnH_vs_ni}
	\end{center}
\end{figure}
\subsection{Site dependence of $\delta S_{nH}$}
We now investigate the dependence of $\delta S_{nH}$ on the occupation number $n_i$ by a density map depicted in Fig.~\ref{Fig:QJ_densitymap_dSnH_vs_ni}. For weak monitoring $\gamma \ll 1$, we observe that the density-map is concentrated, and largely symmetric, around half-filling. However, we note that, as in the density map of $\Delta S_{qj}$, there is also a larger density around $n_i = 1$ corresponding to an incipient quantum Zeno effect caused by measurements in the bulk sites.

In the strong monitoring limit $\gamma=3.0$, the density becomes concentrated around $n_i = 1$, in line with the analogue results for the quantum jump Fig.~\ref{Fig:QJ_densitymap_DSqj_vs_ni}, which signals the dominance of Zeno effect, a signature of the area-law phase. The long tails of $\delta S_{qj}$ at small values of $n_i$ are expected, because they are needed to compensate for the entanglement destroyed, or in some cases created, by measurements at the boundary points, see Fig.~\ref{Fig:QJ_densitymap_DSqj_vs_ni}.

\subsection{The scaling behavior of the mutual information  $I(d_r)$ }
\begin{figure}[!htbp]
	\begin{center}
		\subfigure[]{ \label{fig.QJ_L512_g0p1_0p5_I_vs_dr_rescaled_powerfit}
			\includegraphics[width=8.cm]{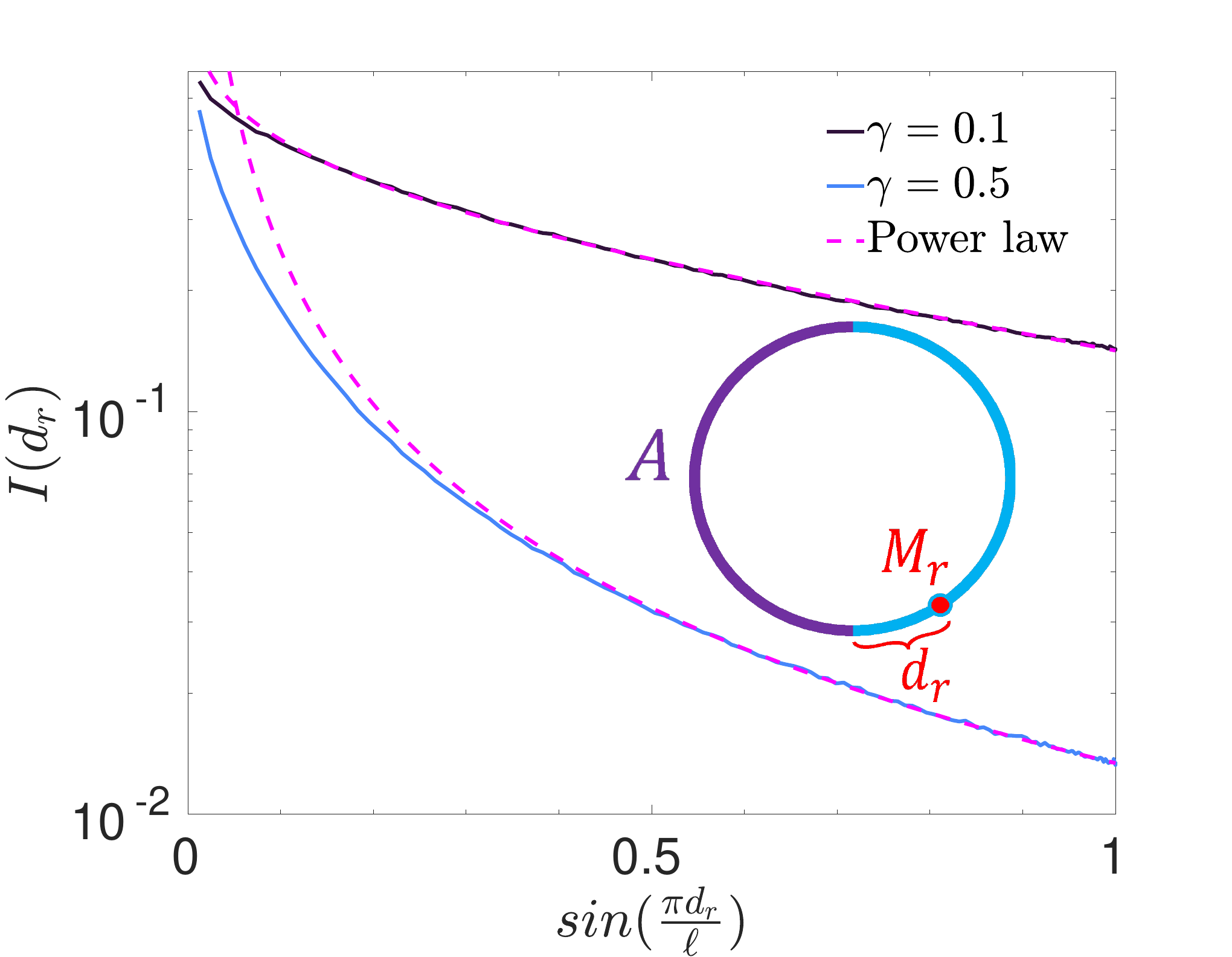} }
		\subfigure[]{ \label{fig.QJ_L512_g1p5_3p0_I_vs_dr_rescaled_expfit}
			\includegraphics[width=8.cm]{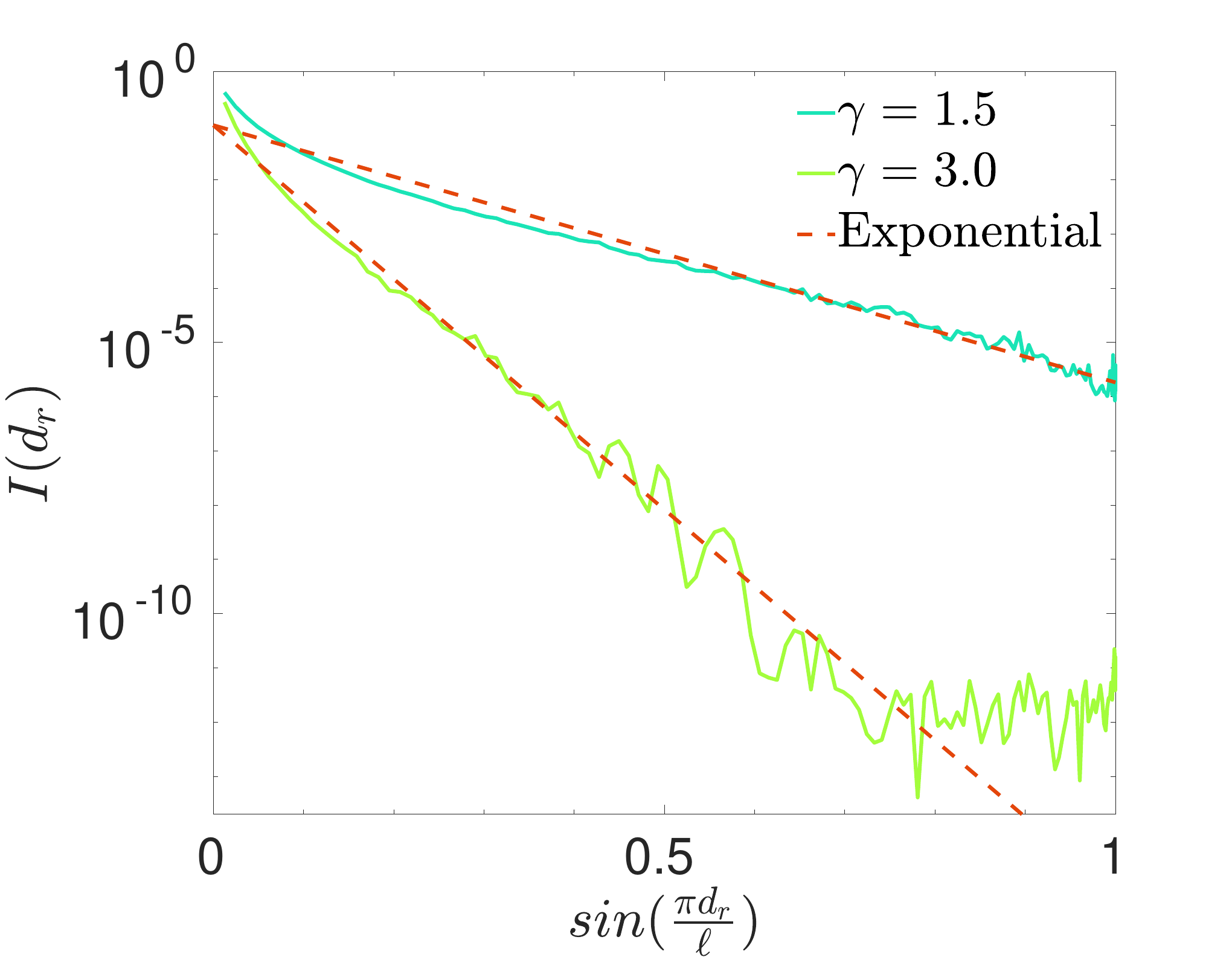} }
		\caption{The mutual information $I(d_r)$ in Eq.~(\ref{eq:mutual}) between the subsystem $A$ and a site $M_r$ in the other subsystem as a function of the distance $d_r$ for the QJ protocol. The pink dashed line is a power-law fitting and the red dashed line is an exponential fitting. The system size is $L = 512$.
		}\label{Fig:QJ_L512_I_vs_dr_rescaled}
	\end{center}
\end{figure}

Finally, we investigate to what extent the results from the distribution function are consistent with the average usually studied in the literature. Since we have already computed the EE, we will focus on the mutual information $I(d_r)$ that provides a local probe of the impact of a measurement in a selected site on sites located at a certain distance. The mutual information is expressed in terms of the EE,
\begin{equation}\label{eq:mutual}
	I(d_r) = S_A + S_{M_r} - S_{A\cup M_r}
\end{equation}
where $M_r$ denotes the measured site at distance $d_r$ from the subsystem $A$. See Fig.~\ref{fig.QJ_L512_g0p1_0p5_I_vs_dr_rescaled_powerfit} for a pictorial representation.
Therefore, $I(d_r)$ could provide qualitative insights into the change in EE, $\Delta S_{qj}$, after measuring different sites in one of the subsystems.
In the weak monitoring limit, $I(d_r)$ exhibits a power law decay with distance, consistent with the long-range entanglement correlations expected in finite size systems, but not in the $L \to \infty$ limit. While in the strong monitoring limit, $I(d_r)$ decays exponentially, indicating the measurement-induced destruction of entanglement.

In the strong monitoring limit, these results are fully consistent with those obtained from the distribution functions. For instance,
the sharp peak at zero in the distribution arises because most measurements, except those near the subsystem boundaries, can only affect few sites around the one being measured so the global EE cannot change much.

For weak monitoring, the power-law decay in the average mutual information is a signature of long-range entanglement and therefore of a EE that scales with the logarithmic of the system size. This is expected because \cite{poboiko2023} for small $\gamma$ the correlation length is too large to observe the area law phase in the range of sizes we can reach numerically.
 However, previous results in the paper considering the full distribution function, instead of the average, point to a more nuanced picture. Even for weak monitoring, the distribution of $\Delta S_{qj}$ has a peak at zero which increases with the system size. The change in EE due to a quantum jump is specially pronounced in the boundary points, for the rest of sites, it has a relatively narrow distribution around zero. The decay of the mutual information $I(d_r)$ further distinguishes the two regimes \cite{poboiko2023}. In the logarithmic growth regime, where $I(d_r)$ decays as a power law due to finite size effects, $P(\Delta S_{qj})$ exhibits a power-law divergent peak around $\Delta S_{qj} = 0$. This explains that, increasing with system size, a soft peak occurs in the weak monitoring limit. In contrast, in the strong monitoring limit, where $I(d_r)$ decays exponentially, $P(\Delta S_{qj})$ collapses into a $\delta$-function peak at $\Delta S_{qj} = 0$.   Therefore, even a weak monitoring strength has a profound impact on the entanglement dynamics that is not captured by the average of the mutual information,

Indeed, although our results do not provide direct support of the analytical prediction \cite{poboiko2023} for the similar PM protocol that for $L \to \infty$, the system is in the area law phase for any $\gamma > 0$, they are certainly consistent with this prediction and moreover show that the calculation of distribution functions, and its site dependence, provides a much more detailed picture of the mechanism of creation and destruction of the EE in quantum many-body systems.

\section{The projective measurement protocol}	
Finally, we study the PM protocol corresponding to the standard picture in quantum mechanics where, after a measurement, the wavefunction {\it collapses} to one of the eigenstates of the Hermitian operator representing the observable being measured with a probability given by the Born's rule.
Details of the protocol are given in Appendix \ref{app:PM}.
 This protocol have similarities with the quantum jump one 
but with three differences. First, in the projective measurement protocol, at each time step, the probability of measuring each site is the same while in the quantum jump protocol it depends on the result of the previous measurement. This is a consequence of the fact that only the quantum jump is a continuous measurement protocol. In the projective measurement protocol, the evolution between measurements is unitary.
Second, the outcome of the quantum jump protocol is only $n_i = 1$, while the outcome of the projective measurement is either $n_i = 0$ or $n_i = 1$. As a consequence, in the strong monitoring limit, we expect quantum Zeno effect around these two values.
Third, the waiting time $\tau$ between measurements can be chosen freely in the projective measurement protocol while in the quantum jump protocol is dictated by the non-Hermitian decay. 
Despite these differences, we show next that the results is very similar to that of the quantum jump protocol. 
\begin{figure}[!htbp]
	\begin{center}
		\subfigure[]{ \label{fig.PM_g0p1_Pro_DSpm_vs_L}
			\includegraphics[width=8.cm]{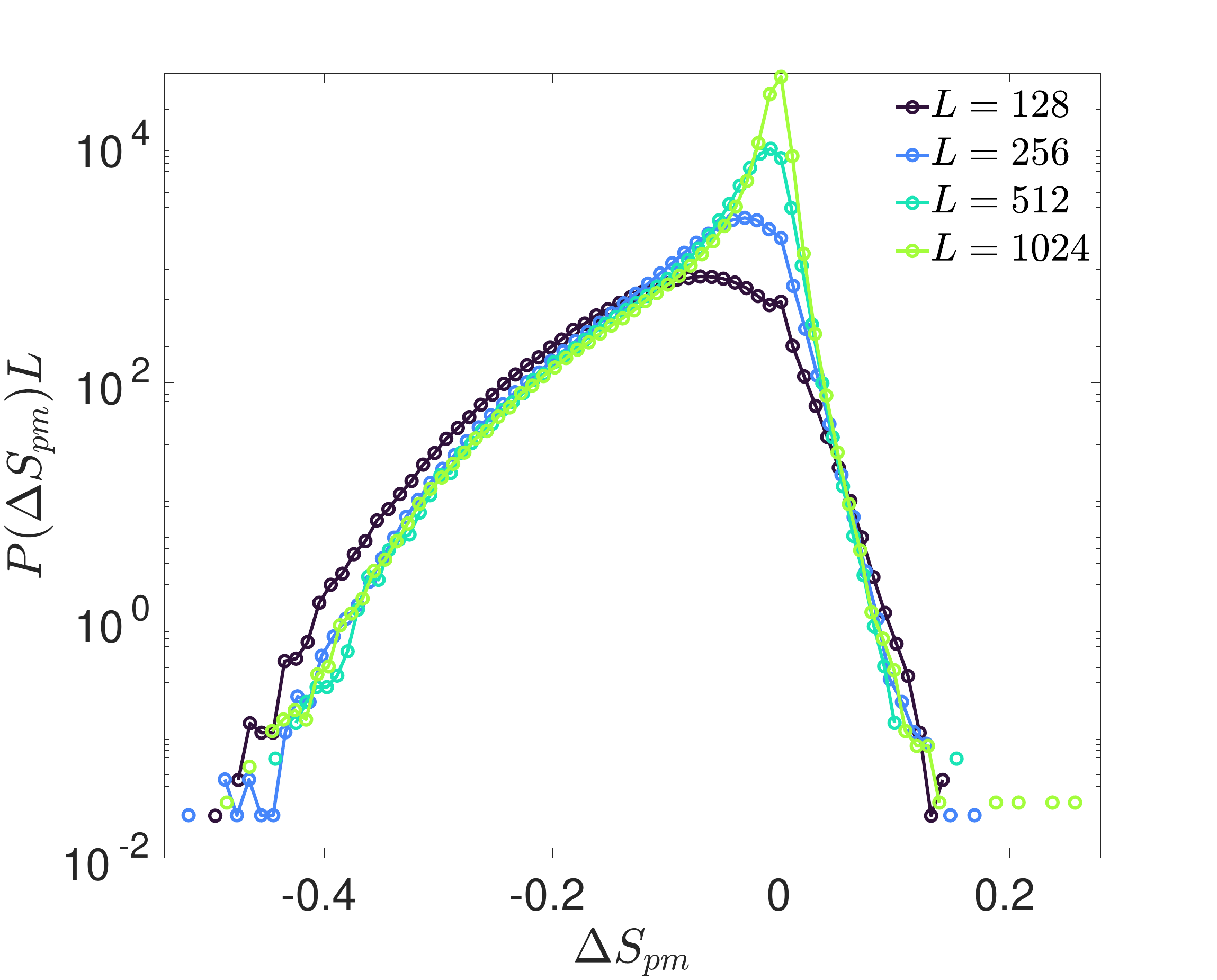} }
		\subfigure[]{ \label{fig.PM_g0p5_Pro_DSpm_vs_L}
			\includegraphics[width=8.cm]{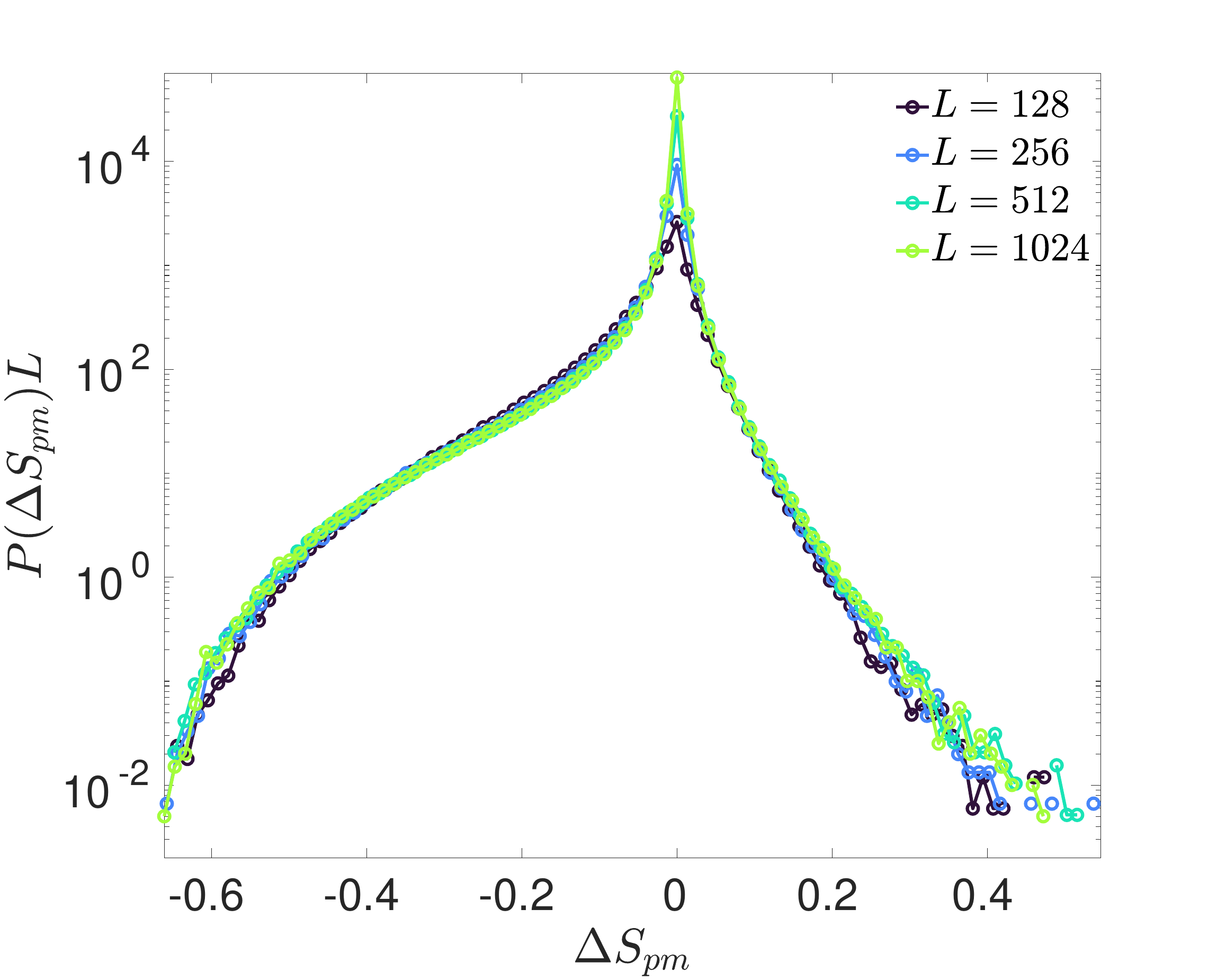} }
		\subfigure[]{ \label{fig.PM_g1p5_Pro_DSpm_vs_L}
			\includegraphics[width=8.cm]{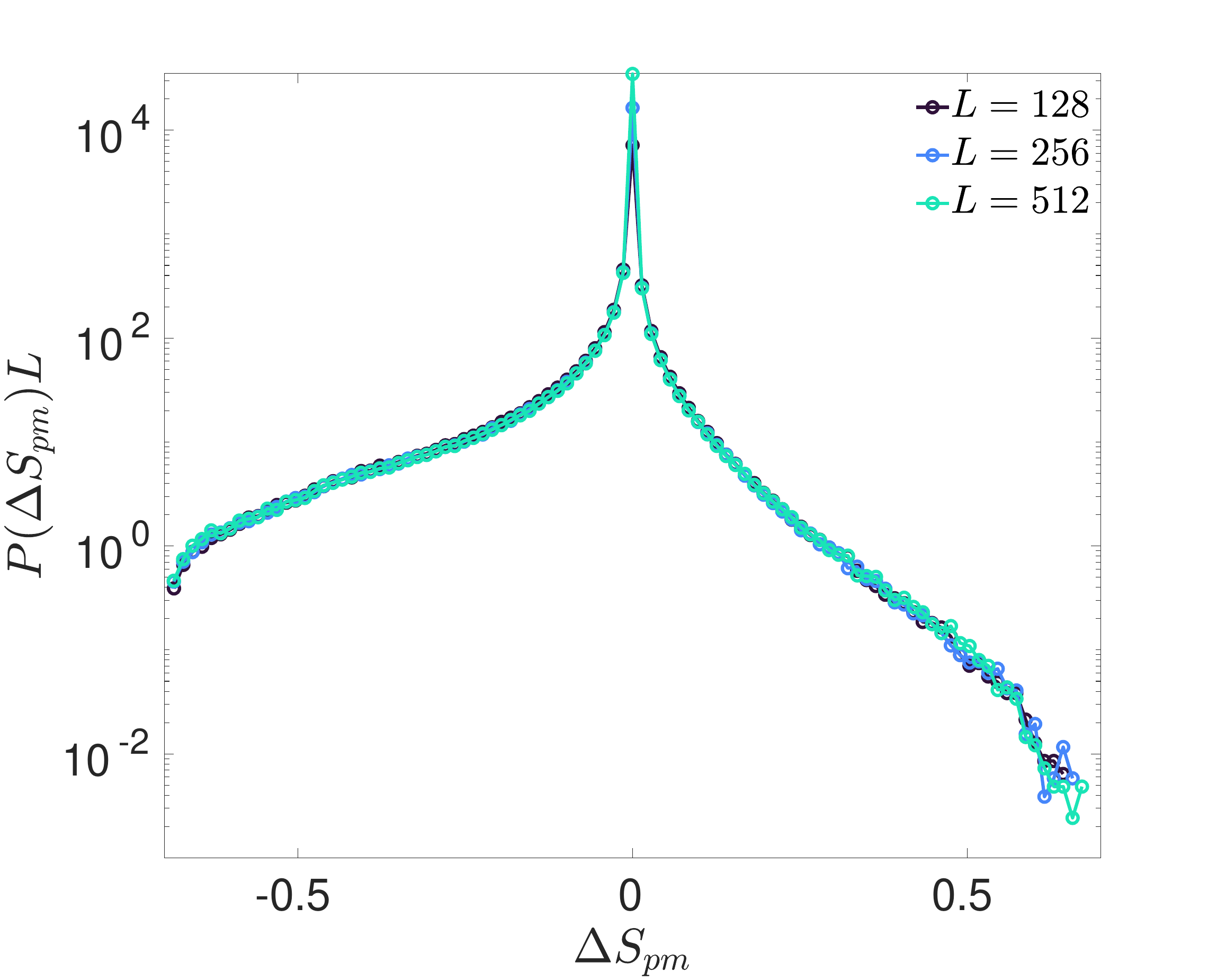} }
		\subfigure[]{ \label{fig.PM_g3p0_Pro_DSpm_vs_L}
			\includegraphics[width=8.cm]{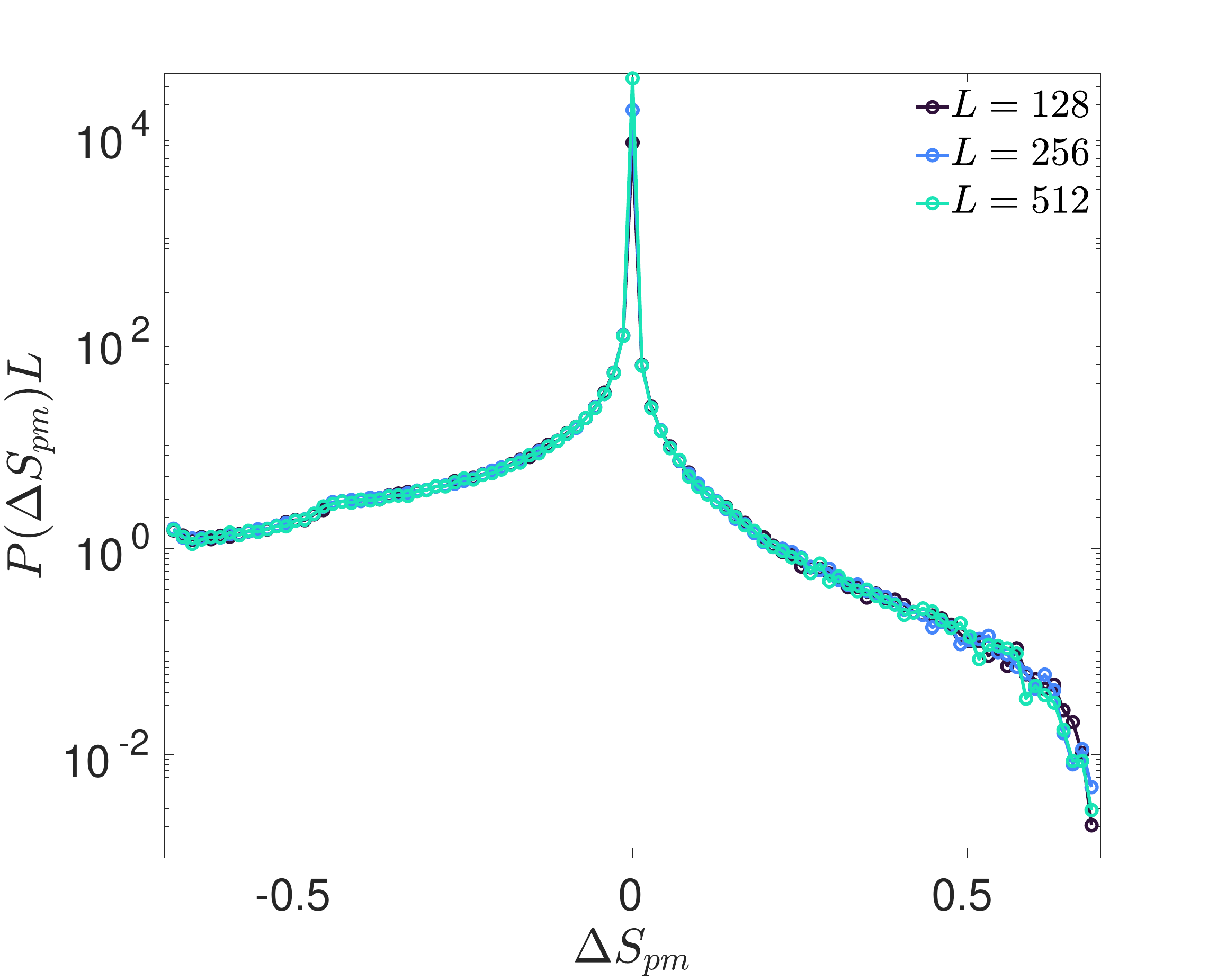} }
		\caption{
			The probability distribution $P(\Delta S_{pm})$ of the entanglement entropy change $\Delta S_{pm}$ in Eq.~(\ref{eq:Deltapm}), computed from Eq.~(\ref{eq:EE}) after saturation of the EE, due to a projective measurement of the occupation number (PM protocol). For convenience, the distribution $P(\Delta S_{pm})$ is multiplied by the system size $L$. The monitoring rates are $\gamma = 0.1, 0.5, 1.5$ and $3.0$ from \subref{fig.PM_g0p1_Pro_DSpm_vs_L} to \subref{fig.PM_g3p0_Pro_DSpm_vs_L}.
		}\label{Fig:PM_Pro_DSqj_vs_L_g}
	\end{center}
\end{figure}

\begin{figure}[!htbp]
	\begin{center}
		\subfigure[]{ \label{fig.PM_L1024_g0p1_densitymap_DSpm_vs_ni}
			\includegraphics[width=8.cm]{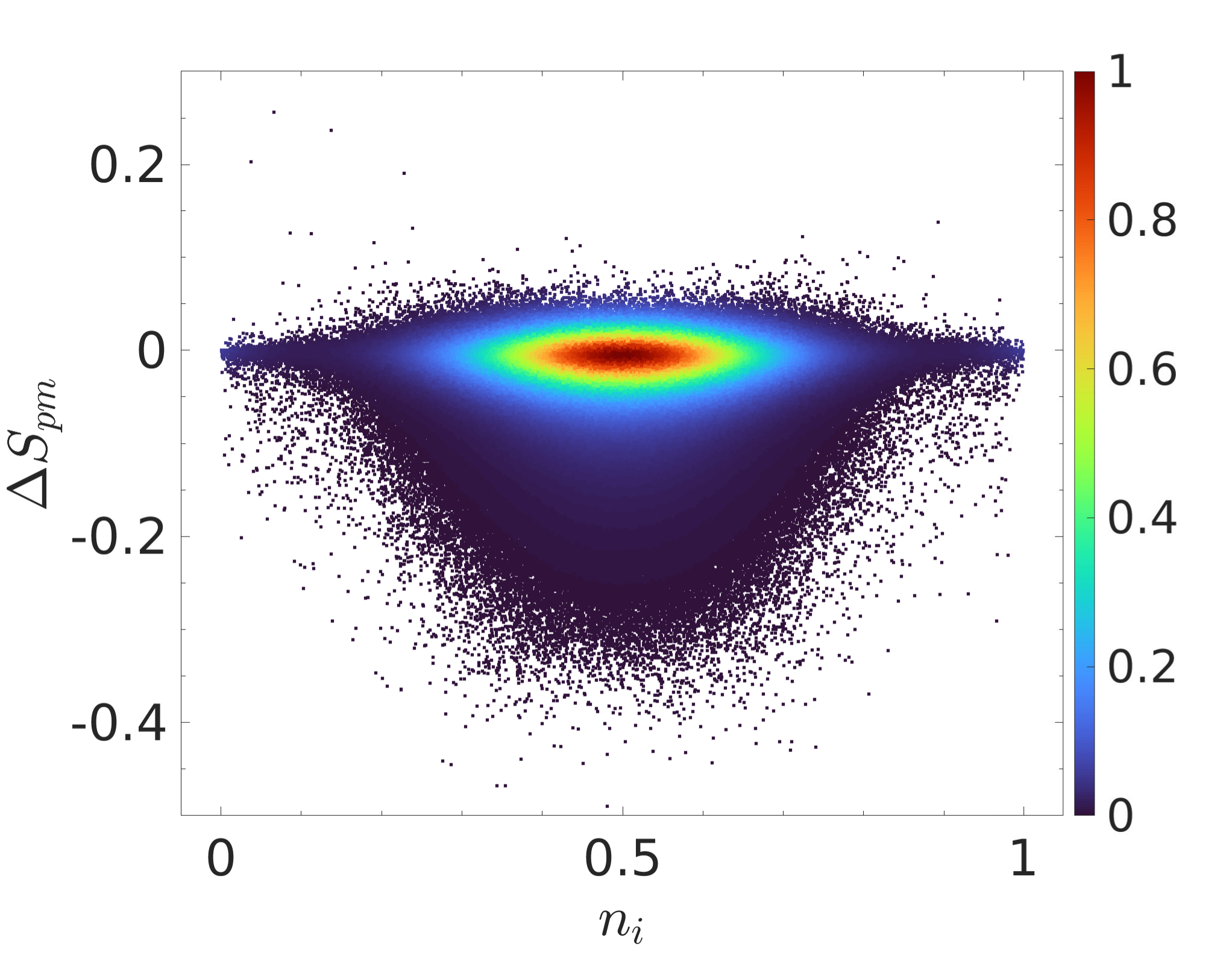} }
		\subfigure[]{ \label{fig.PM_L1024_g0p5_densitymap_DSpm_vs_ni}
			\includegraphics[width=8.cm]{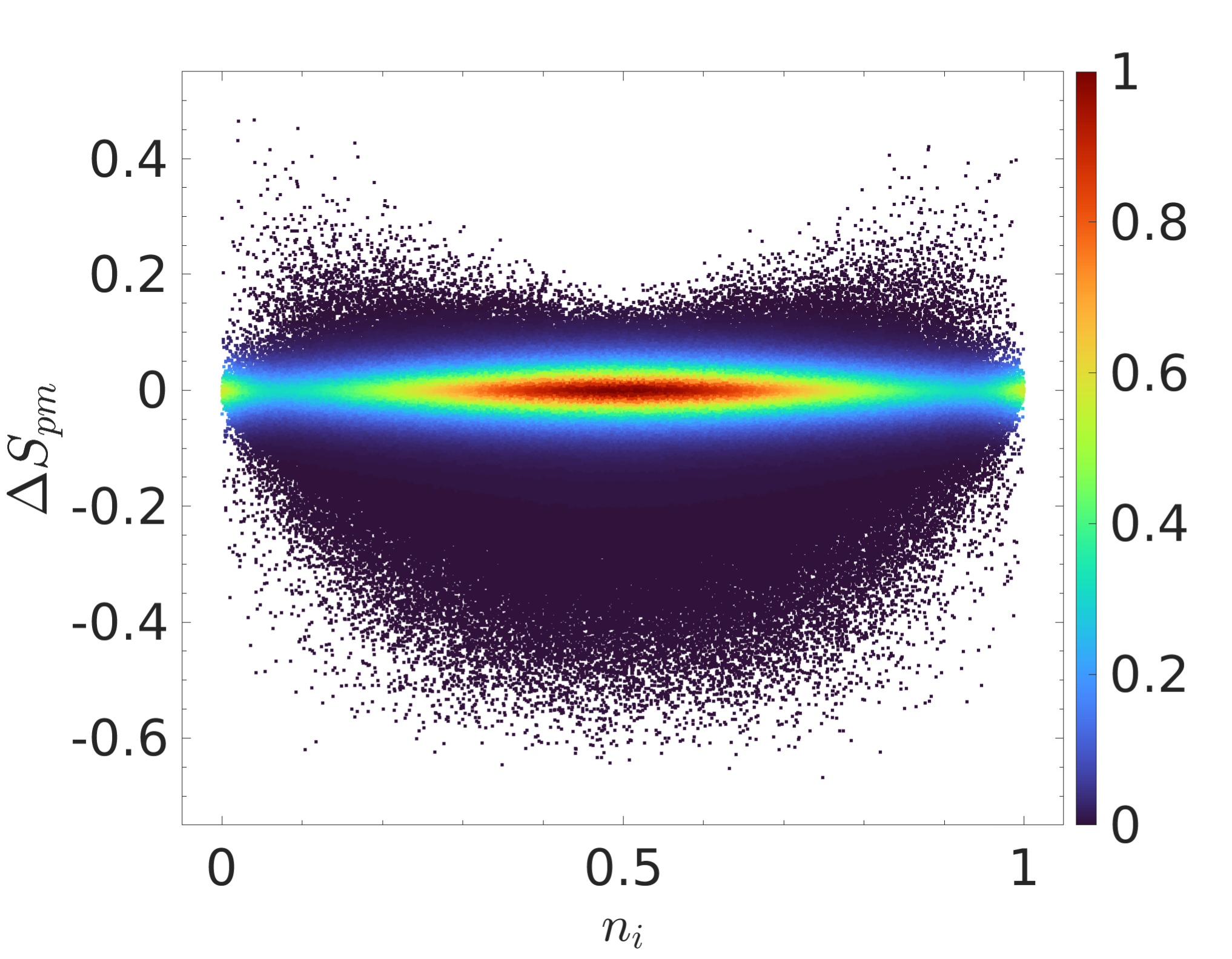} }
		\subfigure[]{ \label{fig.PM_L512_g1p5_densitymap_DSpm_vs_ni}
			\includegraphics[width=8.cm]{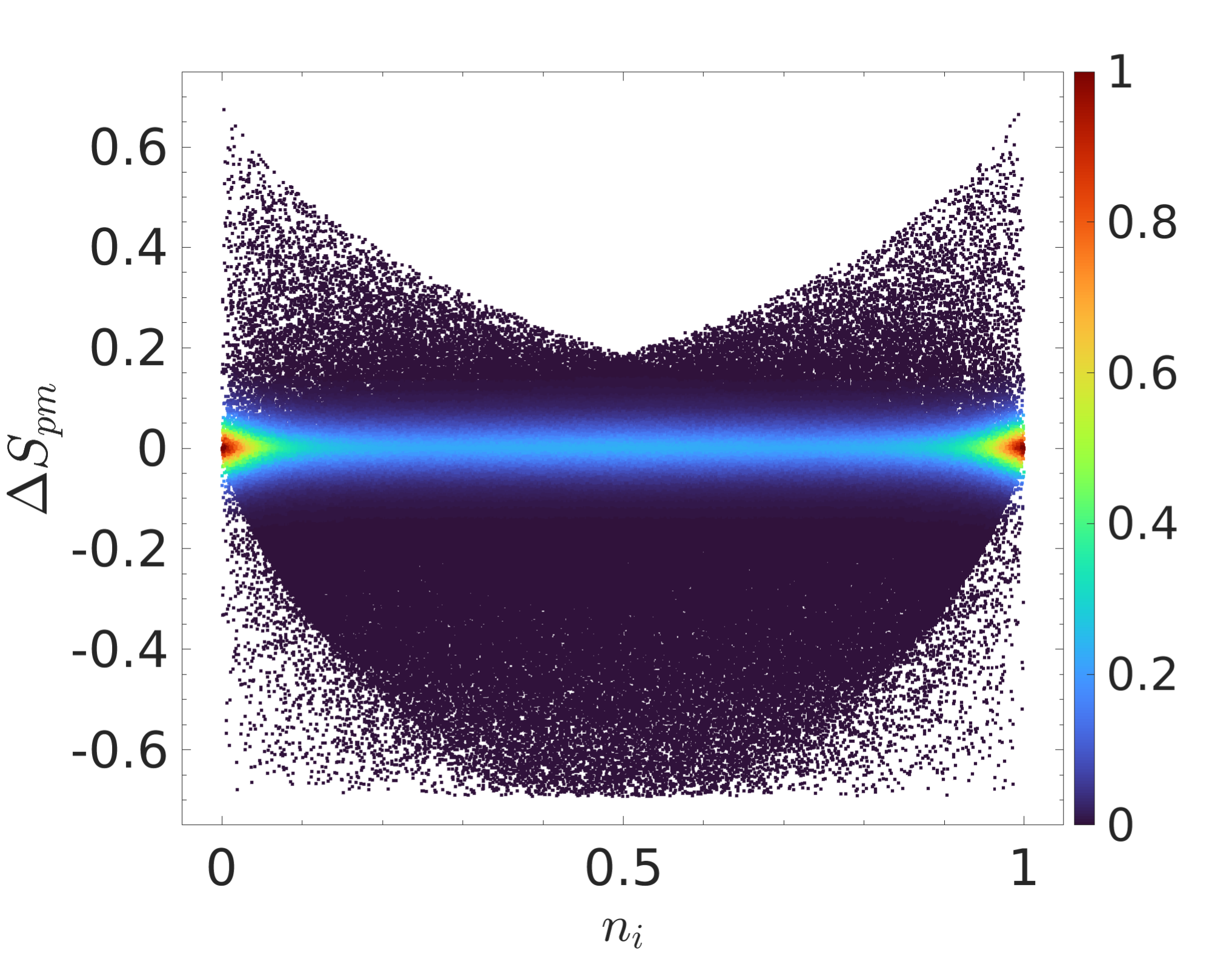} }
		\subfigure[]{ \label{fig.PM_L512_g3p0_densitymap_DSpm_vs_ni}
			\includegraphics[width=8.cm]{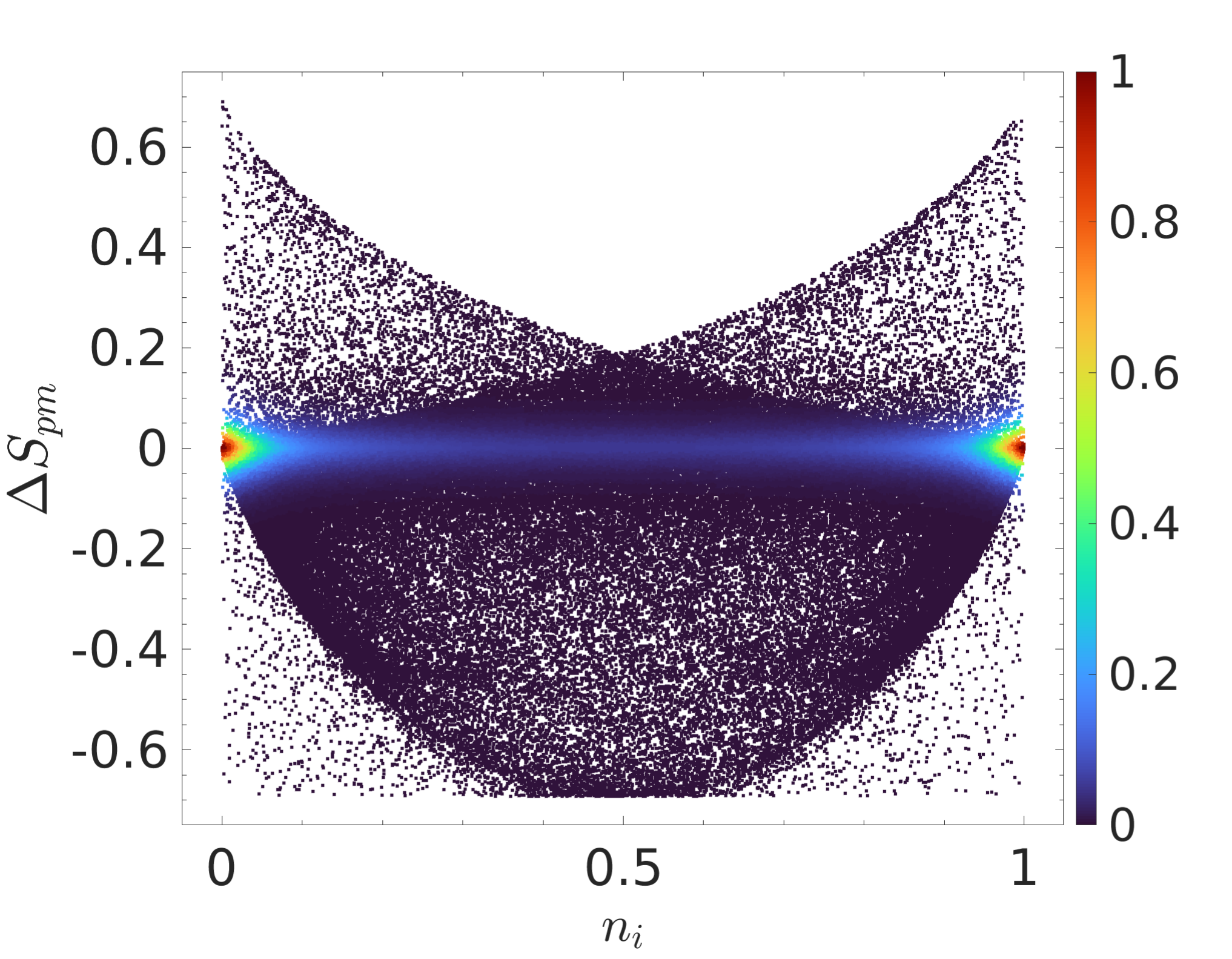} }
		\caption{
			Density map of $\Delta S_{pm}$ in Eq.~(\ref{eq:Deltapm}) with the same parameters as in Fig.~\ref{Fig:QJ_densitymap_DSqj_vs_ni} (QJ protocol) but for the PM protocol.  The color scale indicates the normalized density of measurement events. Results are similar for both protocols after taking into account that the potential outcomes in the PM protocol are $n_i = 0, 1$ while in the QJ protocol is only $n_i =1$.
		}\label{Fig:PM_delta_S_Iac_vs_site_gm0p5_L256}
	\end{center}
\end{figure}
\begin{figure}[!htbp]
	\begin{center}
		\subfigure[]{ \label{fig.PM_L1024_g0p1_densitymap_dSun_vs_ni}
			\includegraphics[width=8.cm]{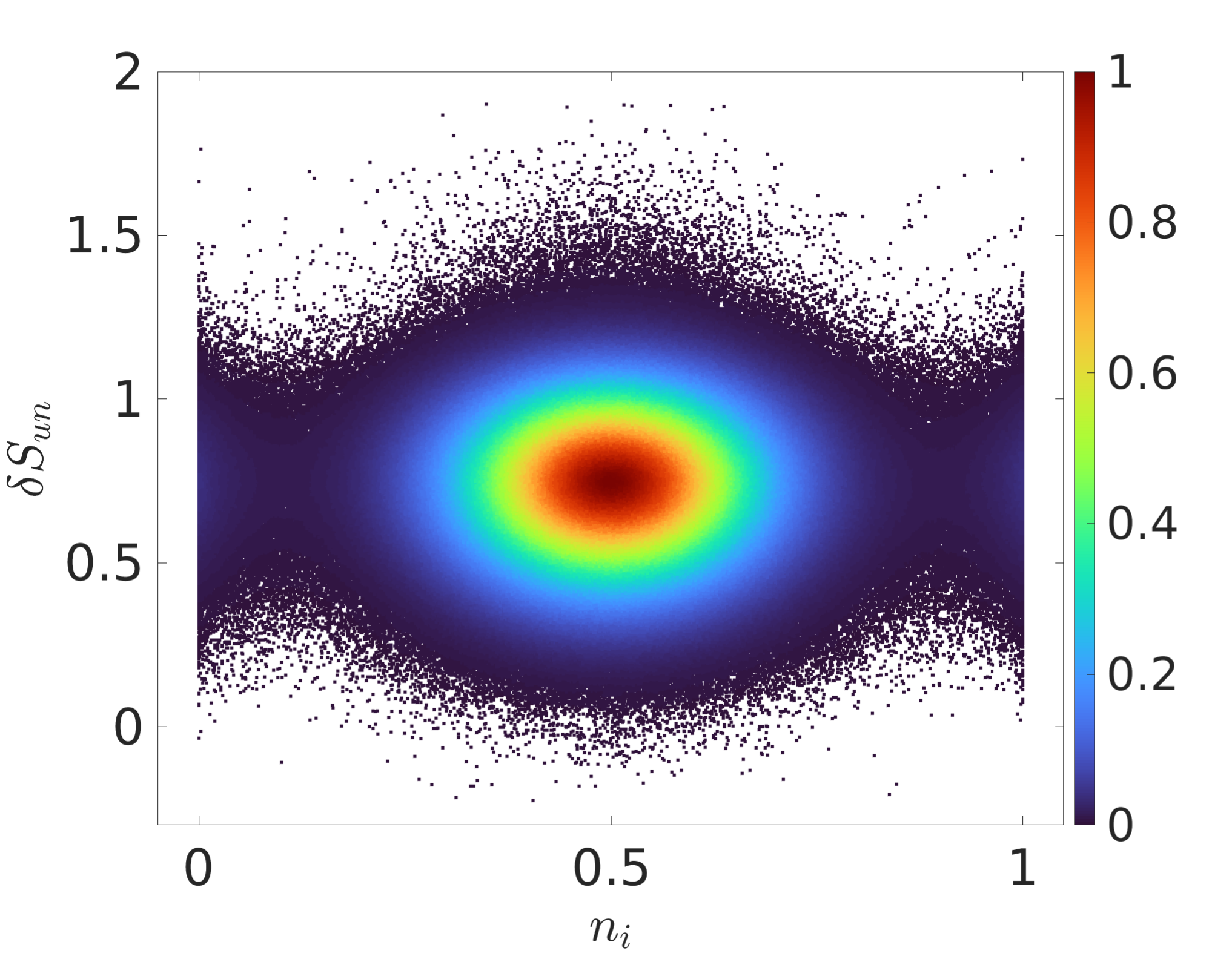} }
		\subfigure[]{ \label{fig.PM_L512_g3p0_densitymap_DSqj_vs_ni}
			\includegraphics[width=8.cm]{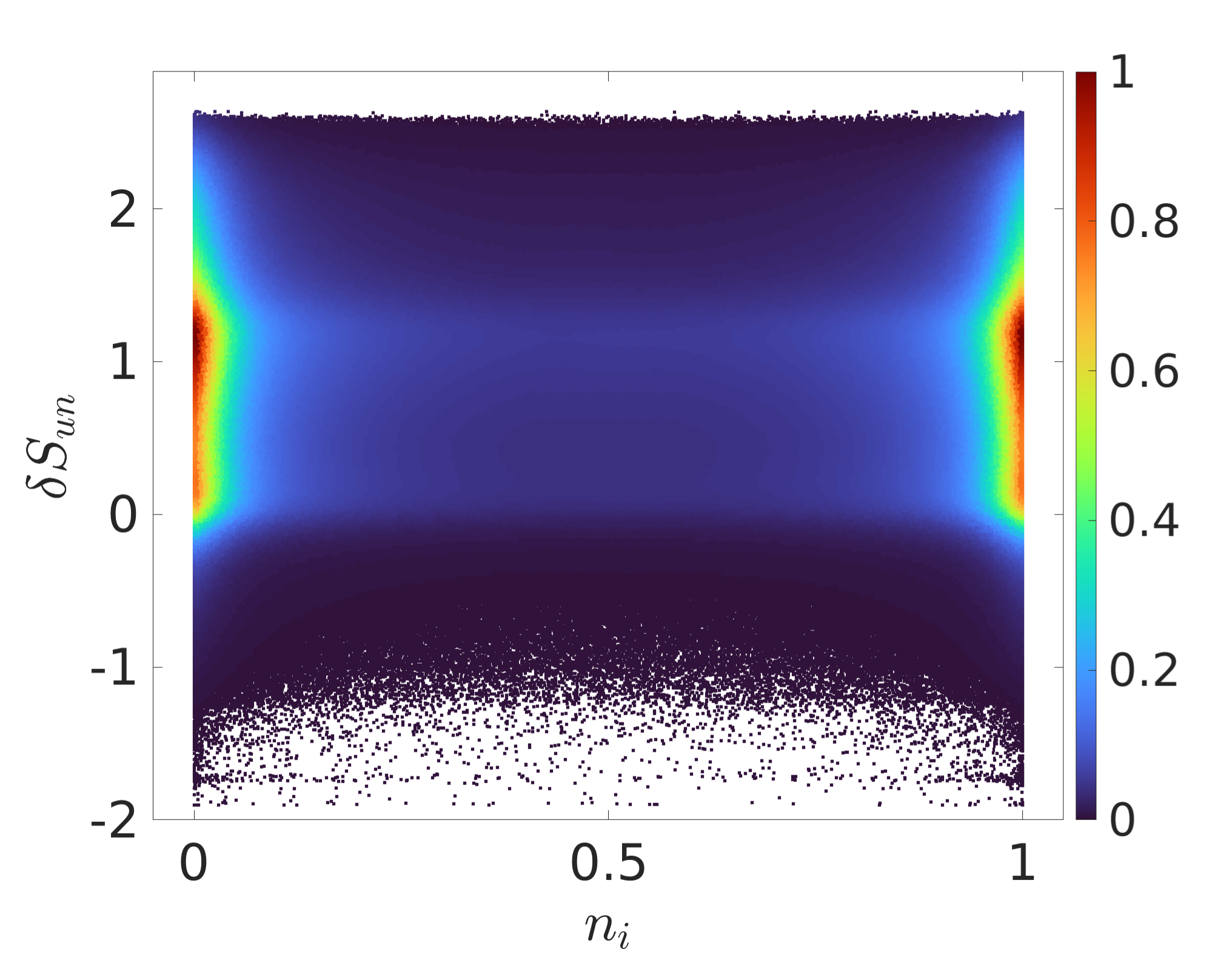} }
		\caption{The density-map of $\delta S_{un}$ in Eq.~(\ref{eq:Deltapm}) for the PM protocol as a function of the occupation number right before the measurement.  The color scale indicates the normalized density of measurement events. The monitoring rates are $\gamma = 0.1$ (left), and $\gamma = 3.0$ (right).
		}\label{Fig:PM_densitymap_dSun_vs_ni}
	\end{center}
\end{figure}

As in the previous cases, we also define the change in EE
\begin{equation}\label{eq:Deltapm}
	\Delta S_{pm} = S_A(t+\tau^+) - S_A(t+\tau^-)
\end{equation}
due to the projective measurement, $\Delta S_{un} = S_A(t+\tau) - S_A(t)$ and
\begin{equation}\label{eq:deltapmu}
	\delta S_{un} = \Delta S_{un}/\tau
\end{equation}
where $\Delta S_{\rm un}$ stands for the change in EE due to the unitary evolution and $\tau$ the time between measurements. Results for the distribution of probability $P(\Delta S_{pm})$, depicted in Fig.~\ref{Fig:PM_Pro_DSqj_vs_L_g}, including the size dependence, are very similar to those for the quantum jump protocol so the same conclusions apply. In Fig.~\ref{Fig:PM_delta_S_Iac_vs_site_gm0p5_L256}, we present density plots of $\Delta S_{pm}$ as a function of the occupation number $n_i$. As in the quantum jump protocol, for weak monitoring, there is a maximum around half-filling that shifts to a bi-modal distribution centered around $n_i = 0,1$ in the strong monitoring limit. This is also in line with the quantum jump case though, by definition, the maximum for the quantum jump is only at $n_i = 1$. The unitary contribution $\Delta S_{un}$, see Fig.~\ref{Fig:PM_densitymap_dSun_vs_ni}, has qualitatively similar features, with the same crossover from a maximum at half filling, for weak monitoring, to maxima at $n_i=0,1$, in the strong monitoring limit.

\section{Discussion and Conclusions}
We have studied the distribution function of the change in EE after a measurement of the occupation number in a chain of free fermions using the QSD, QJ and PM protocols.
The distribution for the QSD protocol is Gaussian and size independent in the weak monitoring limit. As the monitoring strength increases, the distribution is still symmetric but deviations from Gaussianity are clearly observed which are well characterized by a distribution which is Gaussian around zero but with exponential tails. In the strong monitoring limit, the tail are still exponential but the Gaussian around zero turns into a strong peak suggesting a singularity at zero which is a signature of the Zeno effect.

QJ and PM protocols lead to asymmetric non-Gaussian distributions with a peak at zero that becomes stronger as the monitoring strength or the system size increases. More interestingly, for these two protocols, the distribution function is spatially inhomogeneous, namely, it is not the same for all sites. For boundary sites, separating the two sub-systems in the definition of the EE, the distribution in the weak monitoring limit is close to a Gaussian with a small peak at zero and a broad support. However, for the rest of sites, the support is one order of magnitude smaller and the distribution has asymmetric exponential tails. On average, the change in EE is negative, however, we have found that the distribution has an exponential tail for positive values so there is a finite probability that the EE increases as a result of a single measurement. It would be interesting a detailed characterization of the conditions for the existence of such measurements resulting in a positive change of the EE.

For strong monitoring, the distribution of bulk points is well described by a delta function at zero indicating the dominance of the quantum Zeno effect. For boundary sites, the distribution is peaked around zero but, at least for the sizes investigated, still deviates from a delta function. Indeed, it has a broad support covering all possible values of the change in EE with a right tail showing a Gaussian-like decay and a left tail with a much slower decay.

An interesting question to ask is whether these distributions could provide early signals of area-law phase. As is mentioned in the introduction, at least in the PM \cite{poboiko2023}, and likely in the rest of protocols as well, the system is in the area law phase in the limit of $L \to \infty$ for any $\gamma > 0$. 
For a fixed $\gamma$, there exists a crossover between a phase where the scaling of the EE is logarithm with the system size, usually termed critical, and a area-law phase as $L$ is increased. The typical length that controls this crossover \cite{poboiko2023} grows exponentially with $1/\gamma$ so deviations from the phase characterized by a logarithmic growth of the EE are small for $\gamma \ll 1$. However, our results for the density-maps of the change of the EE before and after a measurement show that even for relatively small $\gamma$ there is a crossover between a peak at half filling associated with the logarithmic growth with system size to a peak at $n_i = 1$ (QJ), or $n_i = 0, 1$ (PM), signaling the building up of the quantum Zeno effect. A demonstration, based on a finite size scaling analysis, that the observed Zeno-like features in the distribution function develop into an area law for the EE for any $\gamma > 0$ is numerically costly and beyond the scope of the paper. A finite size scaling analysis of just the average \cite{poboiko2023}, not the full distribution, is enough to confirm the absence of a MIPT.  However, our findings provide evidence that, unlike the average, the distribution function shows changes, even for small values of $\gamma$ and not very large system sizes, which suggests that monitoring is already strongly influencing the entanglement dynamics. 
By contrast, for the QSD protocol, we do not find any sign of the area law phase for weak monitoring. However, we have provided arguments that, for this protocol, the distribution of changes is not relevant for the identification of the transition at least in the weak monitoring limit.

Two general features of our results: the existence of strong peaks at zero in the distribution function, corresponding to no changes in the EE after a measurement, whose strength increases with the monitoring rate, together with the general presence of broader non-Gaussian tails call for further research. As mentioned earlier,
a longstanding problem for the experimental observation of MIPT is the post-selection problem that requires to probe all possible experimental outcomes to observe the transition. However, a detailed understanding of the distribution may at least reduce the problem in some cases by identifying special measurements that carry a large weight, like those corresponding to the peak at zero or to rare measurements leading to the broad tails of the distribution. It would be interesting a more quantitative understanding of the specific features in the distribution function that could make possible to reduce the severity of the post-selection problem.
We plan to address some of these puzzles in the near future.

\vspace{1.0cm}
\centerline{\bf Acknowledgments}
A. M. G. thanks Marco Schiro for illuminating discussions and for bringing to my attention the relevance of distribution functions which was instrumental to start this project.
We acknowledge support from the National Natural Science Foundation of China (NSFC): Individual Grant No. 12374138, Research Fund for International Senior Scientists No. 12350710180. B.F. acknowledges support from the China Postdoctoral Science Foundation (Grant numbers: 2023M732256, 2023T160409, GZB20230420).

\appendix
\section{Definition of the three measurement protocols}

\subsection{Quantum state diffusion (QSD)} \label{app:QSD}

The dynamic evolution of a quantum system under the QSD protocol is mathematically described by the Stochastic Schr\"{o}dinger equation (SSE) \cite{cao2019a,carisch2023,szyniszewski2023,ladewig2022}:
\begin{equation}
	d\left|\psi(t)\right\rangle = -i \hat{H} dt |\psi(t)\rangle +\sum_{i=1}^L\left(\sqrt{\gamma}\left[\hat{n}_i-\left\langle \hat{n}_i\right\rangle_t\right] d \tilde{W}_t^i-\frac{\gamma}{2}\left[\hat{n}_i-\left\langle \hat{n}_i\right\rangle_t\right]^2 dt\right)|\psi(t)\rangle\equiv \hat{K}_t\left|\psi(t)\right\rangle
	\label{eq:Qsd_def}
\end{equation}
where $\hat{H}$ is the Hamiltonian defined in Eq.~\eqref{eq:Ham}, $ \langle \hat{n}_i \rangle_t \equiv \langle \psi(t)|\hat{n}_i|\psi(t)\rangle$ denotes the expectation value of the particle density at site $i$. $\tilde{W}_t^i$ is the Wiener process associated with Gaussian stochastic noise satisfying $d \tilde{W}_t^i d \tilde{W}_t^j =\delta_{ij} dt$. $\hat{K}_t$ denotes the infinitesimal update operator collecting all $O(dt)$ and $O(\sqrt{dt})$ contributions in Eq.~\ref{eq:Qsd_def}. The second term on the right describes measurement effects where the monitoring strength is $\gamma$. It can be derived from the unraveling of the corresponding Lindbladian \cite{lindblad1975,gorini1976} master equation \cite{daley2014}. We stress that in the QSD protocol measurements induce independent Gaussian noise at all sites and continuously in time. The time discretization $dt$ value is an important parameter in our numerical analysis. After numerous tests, we have found that $dt=0.05$ is sufficient given that the state evolution is governed by a first order equation.

The state evolution equation also includes the stochastic term $d\tilde{W}_t^i$ which requires Ito calculus. To leading order $O(\gamma dt)$, we must keep $\hat{K}_t^2$ contribution, since $(\sqrt{\gamma}d\tilde{W}_t^i)^2 =\gamma dt$, 
\begin{equation}
\left|\psi(t+dt)\right\rangle=\exp(d\hat{X}_t)\left|\psi(t)\right\rangle \qquad d\hat{X}_t=\log(\mathbb{I}+\hat{K}_t)=\hat{K}_t-\frac{1}{2}\hat{K}_t^2+O(dt^{\frac{3}{2}})
\end{equation}
where $\mathbb{I}$ is the identity operator. As a result,
\begin{equation}
	|\psi(t+dt)\rangle =\exp \left( -i\hat{H} dt + \sum_{i=1}^{L}\left[\left(\hat{n}_i-\langle \hat{n}_i\rangle_t\right) d W_t^i-\gamma \left(\hat{n}_i-\langle \hat{n}_i\rangle_t\right)^2dt\right]\right)  |\psi(t)\rangle
	\label{eq:QSD_psievo}
\end{equation}
where we redefine $d W_t^i=\sqrt{\gamma} d \tilde{W}_t^i$ as the Gaussian noise of zero average and standard deviation $\sqrt{\gamma dt}$. The state $|\psi(t)\rangle $ can also be expressed in terms of the coefficient matrix $U(t)$, 
$|\psi(t)\rangle = \prod_{k=1}^{N} \left[\sum_{j=1}^L U_{jk}(t) \hat{c}_j^\dagger\right] | \text{vac} \rangle $,  Eq.~\eqref{eq:def_state} in the main text. For any quadratic operator in the second quantization form $\hat{A}=\sum_{ij}\tilde{A}_{ij}\hat{c}_i^\dag \hat{c}_j$, it is possible to show by using the standard anti-commutation relations that 
$A|\psi\rangle=\prod_{k=1}^{N} \left[\sum_{j=1}^L \tilde{A}_{jl}U_{lk}(t) \hat{c}_j^\dagger\right] | \text{vac} \rangle$, which is effectively the multiplication of the coefficients matrices $\tilde{A}U$. As a result, we can reduce the state evolution Eq.~\eqref{eq:QSD_psievo} to the updating of the coefficient matrix $U(t)$ to $U(t+ dt)$ at each time step \cite{cao2019a,carisch2023,szyniszewski2023},
\begin{equation}
	U(t+dt)\propto \exp\left(-i\tilde{H} dt+ dW_t+(2\langle \hat{n}\rangle_t-\mathbf{1})\gamma dt\right) U(t),
	\label{eq:QSD_Uevo}
\end{equation}
where $\tilde{H}_{ij}=J\delta_{i\pm 1,j}$ is the $L\times L$ coefficient matrix of the Hamiltonian $\hat{H}$, namely, $\hat{H}=\sum_{ij}\tilde{H}_{ij}\hat{c}_i^\dag \hat{c}_j$. $d W_t$ and $\langle \hat{n}\rangle_t$ are the $L\times L$ diagonal matrix whose $i$-th diagonal element equals to $d W^i_t$ and $\langle \hat{n}_i\rangle_t$ respectively and $\mathbf{1}$ is the $L\times L$ identity matrix.

It's worthwhile to note that, we can interpret the stochastic state evolution governed by the Ito process Eq.~\eqref{eq:Qsd_def} as the one resulting from the state evolution with the following non-Hermitian Hamiltonian $H_d$,
\begin{equation}
	\hat{H}_d = \hat{H} +i \sum_{i=1}^{L} \hat{n}_i \left(\frac{dW_t^i}{dt}+(2\langle \hat{n}_i\rangle-1)\gamma \right)
	\label{eq:Qsd_Nh}
\end{equation}
The low-energy dynamics of such quadratic disordered Hamilton Eq.~\eqref{eq:Qsd_Nh} has been addressed by using the replica trick and the path-integral formalism \cite{poboiko2023,Jian2023,fava2023} resulting in a Non-Linear Sigma model with symmetries that depend on both the system and the measurement protocol.

Numerically, we evolve the state by updating the coefficient matrix $U(t)$ at each time step. We now explicitly write down the numerical steps to evolve $U(t)$:

1. \textbf{Initialization:} We initialize $U$ in the half-filling N\'{e}el state following Eq.~\eqref{eq:Neel}, and choose as time discretization step $dt=0.05$.

2. \textbf{State Evolution:} We compute the evolution of $U(t)$ following Eq.~\eqref{eq:QSD_Uevo}, resulting in the unnormalized coefficient matrix $\tilde{U}(t+ dt)$.  Before each time step, we compute $\langle \hat{n}_i\rangle_t$, which equals to the $i$-th diagonal elements of the correlation matrix $D(t)=U(t) U^\dag(t)$ defined in Eq.~\eqref{eq:CorM}. We compute the matrix multiplication employing the fourth-order Runge-Kutta method instead of exact diagonalization because numerically the latter is much more expensive.
We have checked that for $dt\leq 0.1$, and for all monitoring rates $\gamma$ we consider, the fourth-order Runge-Kutta method is always a good approximation.

3. \textbf{Normalization:} We use the QR decomposition $\tilde{U}(t+ dt)=U(t+ dt)R$, where $R$ is an upper-triangular matrix, to preserve the normalization condition $U^\dag U=\mathbf{1}_{N}$. The procedure ensures that the new state is correctly normalized $\langle \psi(t + dt)| \psi(t + dt)\rangle = 1$. This QR decomposition takes most of the computational time.

4. \textbf{Compute EE:} We compute the entanglement entropy and mutual information from the correlation matrix $D(t+ dt)=U(t+ dt) U^\dag(t+ dt)$. Note the state evolution only involves the updating of $U(t)$, so that this step does not affect the evolution, and we only choose some of time slices to compute the events we need.

5. \textbf{Repeat:} $2-4$ are repeated for each time step replacing $1$ by the state at the previous time step until the system reaches the steady state characterized by the saturation of the EE $\overline {S(\ell = L/2,t)}$ introduced in Sec~\ref{sec:model}.

The QSD measurement protocol combines both measurements and Hamiltonian dynamics at each time step of the dynamics. The main advantage of this method is that the purity of the Gaussian state is conserved by construction \cite{cao2019a}. Therefore, the time step $dt$ does not need to be very small to avoid non-physical errors.
The procedure for updating the state matrix $U$ under the QSD protocol is given by Algorithm~\ref{alg:QSD}. The probability distribution is computed from the data contained in $10^4 \sim 10^5$ quantum trajectories. The exact number of trajectories depends on the system size.

\begin{algorithm}
	\caption{Updating state matrix $U$ under QSD protocol}
	\SetAlgoLined
	\label{alg:QSD}
	\begin{algorithmic}[1]
		\STATE \textbf{Input:} {System size $L$, particle number $N=L/2$, monitoring rate $\gamma$, maximum time $t_f$, time step $dt$.}
		\STATE \textbf{Output:} Final state matrix $U(t_f)$.
		\STATE \textbf{Initialization:} {Time $t=0$. The state matrix $U_{ij}(t=0) = \delta_{2i-1,j}$ that characterizes the N\'{e}el state. The Hamitonian matrix $H_{ij}=\delta_{i,j\pm1}$.}
		\WHILE{$t\le t_f$}{
			\STATE Generate Gaussian random variable $\delta W_t^i \sim N(0,\sqrt{\gamma dt})$.
			\STATE Calculate the local occupation number $\langle \hat{n}_i \rangle = \sum_{j=1}^{N} U_{ij}U^*_{ij}$.
			\STATE Employ the 4th order Runge-Kutta algorithm to update the unrenormalized state matrix $\tilde{U}(t+dt)$ according to Eq~\eqref{eq:QSD_Uevo}
			\STATE QR decompose the state matrix $\tilde{U}(t+dt) = Q*R$ and get the new state matrix $U(t+dt) = Q$.
			\STATE 	$t=t+dt$.
		}\ENDWHILE	
		\RETURN $U(t_f)$
	\end{algorithmic}
\end{algorithm}

\subsection{Quantum jump (QJ) \label{app:QJ}}
In the QJ protocol, the state evolves continuously by a non-Hermitian  Hamiltonian but this continuous evolution is interrupted at certain times by sudden quantum jumps physically related to a measurement, which in our case is the observation of a particle $n_i =1$ at site $i$. The so-called jump operator that describes this effect is defined as $\hat{L}_j = \sqrt{\gamma}\hat{n}_j = \sqrt{\gamma} \hat{c}_j^\dagger \hat{c}_j$. The full dynamics including the quantum jumps can be modeled by the following equation \cite{minato2022,fuji2020}:
\begin{equation}
	d|\psi(t)\rangle = -i \hat{H} |\psi(t)\rangle dt + \sum_j \left[ \frac{\hat{L}_j |\psi(t)\rangle}{\sqrt{\langle \psi(t)|\hat{L}_j \hat{L}_j^\dagger|\psi(t)\rangle}} - |\psi(t)\rangle \right] dw_j(t), \label{eq.QJ}
\end{equation}
where $ dw_j(t) $ is the site-independent Poisson process. Numerically, $ dw_j(t) $ are discrete random variables that takes only $0$ or $1$. Its ensemble average satisfies $\langle dw_j(t) \rangle = \gamma dt$, where $ \gamma $ is the strength of the measurement that controls the rate at which the measurements are performed.

An efficient way to numerically simulate the time of evolution of $|\psi(t)\rangle$ Eq~\eqref{eq.QJ} is by employing the quantum trajectory algorithm, which can be summarized as follows \cite{minato2022,fuji2020,daley2014}:

1. \textbf{Initialization:} Start with the N\'{e}el state $|\psi(t=0)\rangle$ as an initial state at time $t = 0$, which means the system is half-filled.

2. \textbf{Waiting-time of quantum jump:} Determine the jump time $\tau$ by solving $\langle \psi(t+\tau^-)|\psi(t+\tau^-)\rangle = \eta$, where $\eta$ is a random number uniformly distributed between $(0, 1]$ and $\tau^-$ means the time right before the measurement. We note that the jump time can also be analytically determined by the expression $\tau = {\rm ln}(\eta)/(\gamma N)$ \cite{fuji2020}.

3. \textbf{Non-Hermitian time evolution:} Evolve the state according to the effective Hamiltonian $\hat{H}_{eff} = \hat{H} - i\gamma/2\sum_{j=1}^{L} \hat{c}_j^\dagger \hat{c}_j$:
\begin{equation}
	\frac{d}{dt}|\psi(t)\rangle = -i \hat{H}_{eff} |\psi(t)\rangle.
\end{equation}

4. \textbf{Quantum jump:} At time $t + \tau$, a quantum jump, $\hat{L}_j = \sqrt{\gamma}\hat{n}_j$ representing the detection of a particle ($n_j = 1$) at site $j$, occurs with probability $p_j$ given by:
\begin{equation}
	p_j = \frac{\gamma_j \langle \psi(t+\tau^-)|\hat{L}_j|\psi(t+\tau^-)\rangle}{\sum_k \gamma_k \langle \psi(t+\tau^-)|\hat{L}_k|\psi(t+\tau^-)\rangle}.
\end{equation}

5. \textbf{State update:} Update the state by applying the Lindblad operator $\hat{L}_j$:
\begin{equation}
	|\psi(t+\tau^+)\rangle \rightarrow \frac{\hat{L}_j |\psi(t+\tau^-)\rangle}{\sqrt{\langle \psi(t+\tau^-)|\hat{L}_j^\dagger \hat{L}_j|\psi(t+\tau^-)\rangle}}.
\end{equation}

6. \textbf{Orthonormalization:} Obtain the correlation matrix according to
\begin{equation}
	D_{i,i'} = \langle \psi' | \hat{c}_i^\dagger \hat{c}_{i'} | \psi' \rangle
	= \begin{cases}
		1, & i=i'=j \\
		0, & (i=j, i\ne i') \text{or} (i'=j, i\ne i') \\
		\langle \hat{c}_i^\dagger \hat{c}_{i'} \rangle - \frac{\langle \hat{c}_j^\dagger \hat{c}_{i'} \rangle \langle \hat{c}_i^\dagger \hat{c}_j \rangle}{\langle \hat{c}_j^\dagger \hat{c}_j \rangle}, & \text{otherwise}
	\end{cases}
	\label{eq:QJ_update}
\end{equation}
We then obtain the new orthonormal matrix $U$ from the correlation matrix through the SVD decomposition $ D = U S U^\dagger $, where $ S_{i,i} = 1 (1 \le i \le N ) $ and $ 0 (N + 1 \le i \le L) $. The new matrix $ U $ characterizes the new orthonormal state $ |\psi(t+\tau) \rangle $ at time $ t + \tau $.

7. \textbf{Repeat:} Go back to step 2 and repeat the process until the desired time is reached.

A detailed numerical algorithm for updating the state matrix under the QJ protocol is given in Algorithm~\ref{alg:QJ}. For the quantum jump protocol, depending on the system size and the measurement strength, we obtain between $10^3$ and $5\times 10^4$ quantum trajectories . Moreover, we collect the measurements in the time interval $[0.6L, 0.7L]$, when the average entanglement entropy has already reached it saturation value. As the figures show, this amount data is enough to obtain smooth distributions.

\begin{algorithm}
	\caption{Updating state matrix $U$ under QJ protocol}
	\SetAlgoLined
	\label{alg:QJ}
	\begin{algorithmic}[1]
		\STATE \textbf{Input:} {System size $L$, particle number $N=L/2$, monitoring rate $\gamma$, maximum time $t_f$.}
		\STATE \textbf{Output:} Final state matrix $U(t_f)$.
		\STATE \textbf{Initialization:} {Time $t=0$. The state matrix $U_{ij}(t=0) = \delta_{2i-1,j}$ that characterizes the N\'{e}el state. The Hamitonian matrix $H_{ij}=\delta_{i,j\pm1}$.}
		\WHILE{$t\le t_f$}{
			\STATE Generate a random value uniformly distributed $\eta \in (0,1]$ and get $\tau$ through $||exp(-i\hat{H}_{eff}\tau) U(t)||= \eta$.
			\STATE Update the matrix $U(t+\tau) = exp(-i\hat{H}_{eff}\tau) U(t)/||exp(-i\hat{H}_{eff}\tau) U(t)||$
			\STATE Obtain the correlation matrix $D = \sum_{k=1}^N U^*_{ik}(t+\tau)U_{jk}(t+\tau)$.
			\STATE Generate a random value uniformly distributed $p \in [0,1]$.
			\STATE Initiate $sum_{ni} = 0$;
			\FOR{$i = 1$ to $L$}
			\STATE $sum_{ni} = sum_{ni} +D_{ii}/N$ ~~~~~~\# select the measured site $i\in[1,L]$
			\IF{$ sum_{ni}\ge p$}
			\STATE break.
			\ENDIF
			\ENDFOR
			\STATE Update the correlation matrix $D$ following Eq.~\eqref{eq:QJ_update}.
			\STATE SVD decompose the correlation matrix $D=USU^\dagger$ and get the new state matrix $U(t+\tau)$.
			\STATE 	$t=t+\tau$.
		}\ENDWHILE	
		\RETURN $U(t_f)$
	\end{algorithmic}
\end{algorithm}

\subsection{Projective Measurements (PM)  \label{app:PM}}
This appendix introduces the PM protocol \cite{poboiko2023} which is closely related to the QJ protocol. The two main differences is that the state evolution between two measurements is governed by the Hermitian Hamiltonian and that the time between measurements $\tau$ is here a free parameter which we have decided to mimic that of the QJ protocol $\tau = {\rm ln}(\eta)/(\gamma N)$ , where $\eta$ is a random value uniformly distributed in $(0,1]$, $N=L/2$ is the particle number and $\gamma$ is the measuring rate per site.
At each measurement time $t+\tau$, a site $j$ is chosen randomly, and a projective measurement of the site occupation number $\hat{n}_j =\hat{c}_j^\dagger \hat{c}_j$ is conducted.
Upon performing a measurement at time $t+\tau$ with outcome $n_j$, the density matrix undergoes a discontinuous change:
\begin{equation}
	\hat{\rho}(t + \tau^+) = \hat{P}_{n_j}(j) \hat{\rho}(t + \tau^-) \hat{P}_{n_j}(j),
\end{equation}
where \(\hat{P}_{j}\) are the projection operators defined as:
\begin{equation}
	\hat{P}_0(j) = 1 - \hat{n}_j, \quad \hat{P}_1(j) = \hat{n}_j.
\end{equation}
The possible outcomes of this measurement are either $1$ or $0$ with equal probability, reflecting the fermionic nature of the particles. %
To keep the density matrix normalized along each quantum trajectory, we rescale $\hat{\rho}$ by the factor $1/\mathrm{tr}(\hat{\rho})$ after each measurement. %
We stress that between successive measurements, in this protocol the system undergoes unitary evolution. %
Similar to the quantum-jump protocol, we compute the evolution of the correlation matrix $D(t)$. The simulation is also initialized in the N\'{e}el state $|\psi(t=0)\rangle = |10101010\ldots\rangle$, and the corresponding element of the correlation matrix can be written as $D_{ij}(t=0) = \langle \psi(t=0)| \hat{c}_i^\dagger \hat{c}_j |\psi(t=0)\rangle $. For each time interval between two PM, apply the unitary evolution $D (t+\tau) = e^{-i\hat{H}\tau} D(t) e^{i\hat{H}\tau} $. At time $t+\tau$, we randomly select a site $j$, and compute $p_j = D_{j,j}$. We then generate a random value $p_c$. If $p_j< p_c$, we perform the projection operation $ \hat{P}_0(j) = 1 - \hat{n}_j $. The corresponding updated correlation matrix elements are given by \cite{schiro2024}:
\begin{equation}
	D_{i,i'} = \langle \psi' | \hat{c}_i^\dagger \hat{c}_{i'} | \psi' \rangle
	= \begin{cases}
		0, & (i=j) \text{or} (i'=j) \\
		\langle \hat{c}_i^\dagger \hat{c}_{i'} \rangle - \frac{\langle \hat{c}_j^\dagger \hat{c}_{i'} \rangle \langle \hat{c}_i^\dagger \hat{c}_j \rangle}{1 - \langle \hat{c}_j^\dagger \hat{c}_j \rangle}, & \text{otherwise}
	\end{cases}
	\label{eq:PM_update_0}
\end{equation}

If $p_j\ge p_c$, we perform the projection operators $ \hat{P}_1(j) = \hat{n}_j $. The updated correlation matrix is given by Eq~\eqref{eq:QJ_update} in Appendix~\ref{app:QJ}.

Following the similar calculation introduced in Appendix~\ref{app:QJ}, we apply the SVD decomposition $D = U S U^\dagger$ to restore the Gaussian state representation where the matrix $U$ characterizes the state after the PM measurement $ |\psi(t+\tau) \rangle$. We repeat the calculation until the steady state at which the EE reaches the saturation value. These steps are summarized in Algorithm~\ref{alg:PM}.
We calculate from $10^3$ to $5\times 10^4$ trajectories depending on the system size and the measurement strength. The measurements are collected in the time interval $[0.6L, 0.7L]$ so that the entanglement entropy has reached its saturation value. With this information, we compute the distribution of the change of entanglement entropy after each measurement.

\begin{algorithm}
	\caption{Updating state matrix $U$ under PM protocol}
	\SetAlgoLined
	\label{alg:PM}
	\begin{algorithmic}[1]
		\STATE \textbf{Input:} {System size $L$, particle number $N=L/2$, monitoring rate $\gamma$, maximum time $t_f$.}
		\STATE \textbf{Output:} Final state matrix $U(t_f)$.
		\STATE \textbf{Initialization:} {Time $t=0$. The state matrix $U_{ij}(t=0) = \delta_{2i-1,j}$ that characterizes the N\'{e}el state. The Hamitonian matrix $H_{ij}=\delta_{i,j\pm1}$.}
		\WHILE{$t\le t_f$}
		\STATE Generate a random value uniformly distributed $\eta \in (0,1]$ and get $\tau = \ln(\eta)/(\gamma N)$.
		\STATE Employ the 4th order Runge-Kutta algorithm to update the matrix $U(t+\tau) = exp(-iH\tau) U(t)$
		\STATE Obtain the correlation matrix $D = \sum_{k=1}^N U^*_{ik}(t+\tau)U_{jk}(t+\tau)$.
		\STATE Randomly select a site $i\in[1,L]$.
		\STATE Generate a random value uniformly distributed $p_i \in [0,1]$.
		
		\IF{$ D_{ii}\ge p_i$}
		\STATE Update the correlation matrix $D$ following Eq.~\eqref{eq:QJ_update}.
		\ELSIF{$ D_{ii}< p_i$}
		\STATE Update the correlation matrix $D$ following Eq.~\eqref{eq:PM_update_0}.
		\ENDIF
		\STATE SVD decompose the correlation matrix $D=USU^\dagger$ and get the new state matrix $U(t+\tau)$.
		\STATE 	$t=t+\tau$.
		
		\ENDWHILE	
		\RETURN $U(t_f)$
	\end{algorithmic}
\end{algorithm}

\newpage
\bibliography{monitoring.bib} 

\end{document}